\documentclass[11pt]{article}
\usepackage{}
\usepackage{amssymb}
\usepackage{amsfonts}
\usepackage{mathrsfs}
\usepackage{graphicx}
\usepackage{amsmath}
\usepackage{color}
\usepackage{amsthm}

\usepackage{pifont,bm}
\usepackage{tikz}
\usepackage{epstopdf}
\usepackage{enumerate}

\newtheorem{thm}{Theorem}[section]

\theoremstyle{definition}

\newtheorem{rem}[thm]{Remark}

\makeatletter
\def\@biblabel#1{[#1]}
\makeatother

\makeatletter \@addtoreset{equation}{section}
\renewcommand{\theequation}{\arabic{section}.\arabic{equation}}

\begin{document}

\begin{titlepage}
\title{\bf{Riemann-Hilbert approach for the NLSLab equation with nonzero boundary conditions
\footnote{
Corresponding author.\protect\\
\hspace*{3ex} \emph{E-mail addresses}: ts17080005a3@cumt.edu.cn (J.J. Mao) and sftian@cumt.edu.cn,
shoufu2006@126.com (S. F. Tian)}
}}
\author{Jin-Jin Mao and Shou-Fu Tian$^{*}$ \\
\small \emph{School of Mathematics and Institute of Mathematical Physics, China University of Mining } \\
\small \emph{and Technology, Xuzhou 221116, People's Republic of China}\\
\date{}}
\thispagestyle{empty}
\end{titlepage}
\maketitle

\vspace{-0.5cm}
\begin{center}
\rule{15cm}{1pt}\vspace{0.3cm}

\parbox{15cm}{\small
{\bf Abstract}\\
\hspace{0.5cm}  We consider the inverse scattering transform for the nonlinear Schr\"{o}dinger equation in laboratory coordinates (NLSLab equation) with nonzero boundary conditions (NZBCs) at infinity. In order to better deal with the scattering problem of NZBCs, we introduce the two-sheeted Riemann surface
of $\kappa$, then it convert into the standard complex $z$-plane. In the direct scattering problem, we study the analyticity, symmetries and asymptotic behaviors of the Jost function and the scattering matrix in detail. In addition, we establish the discrete spectrum, residual conditions, trace foumulae and theta conditions for the case of simple poles and double poles. The inverse problems of simple poles and double poles are from the Riemann-Hilbert problem (RHP). Finally, we obtain some soliton solutions
of the NLSLab equation, including stationary solitons, non-stationary solitons and multi-soliton
solutions.
Some features of these soliton solutions caused by the
influences of each parameters  are analyzed graphically in order to control such nonlinear phenomena.
}
\vspace{0.5cm}
\parbox{15cm}{\small{

\vspace{0.3cm} \emph{Key words:} The NLSLab equation; Nonzero boundary conditions; Direct scattering problem; Inverse scattering problem;  Simple poles; Double poles; Soliton solutions.\\

 } }
\end{center}
\vspace{0.3cm} \rule{15cm}{1pt} \vspace{0.2cm}

\section{Introduction}
The inverse scatter transform (IST) was first proposed to be applied to solve the well-known Korteweg-de Vries (KdV) equation with Lax pairs \cite{GGK-1967-19}. Later IST was extended to solve the increasingly important physical nonlinear wave equations with Lax pairs, such as modified KdV equation, nonlinear Schr\"{o}dinger (NLS) equation, Kadomtsev-Petviashvili (KP) equation, Benjamin-Ono (BO) equation, Davey-Stewartson (DS) equation, sine-Gordon equation, etc (see, Refs. \cite{AS-1981,AC-1991} and references therein). Among them, as a model with important physically meaning, the NLS equation and its extended equation appear in many nonlinear science fields, such as Bose-Einstein condensates, nonlinear optics, deep sea and even financial field (please refer to the reference \cite{A-2012}-\cite{Y-2015-47} and references therein).
As we well-known that  the
NLS reads
\begin{equation}\label{nls}
iq_{t'}=q_{x'x'}+2|q|^{2}q,
\end{equation}
which describes the slowly varying envelope for the modulations of an underlying carrier wave. Particularly,  Eq. \eqref{nls} can be also derived from the context of nonlinear optics, where     the light-cone variables $x$, $t$  and  the coordinates $z_{lab}$ and $t_{lab}$ are related under some transformation, $t'=z_{lab}$ and $x'=t_{lab}-z_{lab}/c_{g}$ in a laboratory frame. Here $z_{lab}$ and $t_{lab}$ denote the spatial and temporal coordinates and $c_{g}$ denotes the speed of light in the specific nonlinear medium \cite{Newell}. Eq. \eqref{nls} is  a   integrable system, and it admits  exact $N$-soliton solutions. The one-soliton solution reads
\begin{equation}
q(x',t')=Ae^{i(V^{2}-A^{2})t'-iVx'-i\psi}\mbox{sech}[A(x'-2Vt'-\delta)],
\end{equation}
where $A$, $V$, $\psi$ and $\delta$ are arbitrary constants.
In 1972,  Zakharov and Shabat  study the  integrability of Eq. \eqref{nls} via inverse scattering transform (IST) for the initial value problem (IVP)  on the infinite line $-\infty<x'<+\infty$ \cite{ZS-1972}.

In order to consider the boundary value problems (BVPs) for the NLS equation in a laboratory frame (NLSLab equation),
a more physically relevant group of problems to investigate is then perhaps those presented by
\begin{equation}\label{NLS-old}
iq_{t}+icq_{x}-q_{xx}-2|q|^{2}q=0,
\end{equation}
where $x$ and $t$ are new versions of $x'$ and $t'$,  and $c$ is some relevant group velocity.
 When $c=0$, these equation \eqref{NLS-1} reduce to   the defocusing NLS equation. It is remarking that  the additional  term $cu_{x}$ appears when we consider  the NLS equation from the context of both nonlinear optics   \cite{Newell} and deep water waves \cite{AS-1981}.   In \cite{LBK-2019-383},  Leisman, Biondini and  Kovacic studies the BVPs   for Eq. \eqref{NLS-old} on the half line $0<x<\infty$. They identify a class of linearizable BCs which reduces to $q_{x'}(0,t')+\alpha q(0,t')=0$  for $c=0$, and also
 construct a class of exact soliton solutions of the BVP with the corresponding behavior.

In this work, we consider the inverse scattering transform for the NLSLab equation with nonzero boundary conditions (NZBCs) at infinity, namely,
\begin{equation}\label{NLS-1}
iu_{t}+icu_{x}-u_{xx}-2(|u|^{2}-u_{0}^{2})u=0.
\end{equation}
The initial value problem of the NLSLab equation with the following NZBCs as $x\rightarrow\pm\infty$ reads
\begin{equation}\label{NLS-2}
\lim_{x\rightarrow\pm\infty}u(x,t)=u_{\pm},
\end{equation}
with $|u_{\pm}|=u_{0}\neq 0$. The boundary condition \eqref{NLS-2} are independent of time. According to the appropriate IST,  we can realize the initial value problem. And the additional item $2u_{0}^{2}u$ in equation \eqref{NLS-1} can also be eliminated by using a simple rescaling $q(x,t)=e^{2iu_{0}^{2}t}u(x,t)$, thereby converting equation \eqref{NLS-1} into the NLSLab equation \eqref{NLS-old}.

In \cite{ZS-1972}, Zakharov and Shabat first proposed the IST of the  focusing case with zero boundary conditions (ZBCs). One year later, in reference \cite{ZS-1973-37}, Zakharov and Shabat also proposed the IST of the  defocusing case with NZBCs.  Since then, both cases have been extensively studied \cite{PAB-2006-47}-\cite{ZCH-2017-24}. Among them, IST has almost no known results for the case of NZBCs. We believe that there are two primary reasons for this: (1) the technical difficulties caused by the NZBCs; (2) the existence of modulation instability (MI) (where is called  the Benjamin-Feir (BF) instability \cite{B-1967-299,BF-1967-27} in the context of water waves). We will discuss these above two questions successively in the following parts.

Regarding technical difficulties, we notice that there are still some problems about the IST with NZBCs that have not been solved even for the defocusing case \cite{BP-2014-132,DPM-2013-131}. The study of IST for the focusing case with NZBCs is reference \cite{M-1979-60}, but only part of the research results were included and no more solutions were given. Partial results were also obtained in reference \cite{GK-2012-45,Yangjj1,Yangjj2,Faneg,Faneg2}, which was used to study the stability of the Peregrine solitons under perturbations. Regarding the MI difficulties, we can get an overview of the the subject and a historical perspective from Zakharov and Ostrovsky's \cite{ZO-2009-238} article. From the linearized stability analysis, we observe that the uniform background is unstable for long-wavelength disturbances. It is believed that the MI is a mechanism for the generating of rogue waves which has recently attached renewed interest\cite{OOS-2006-96}. In the study of the NLS equation with periodic boundary conditions (BCs), we find that the underlying mechanism of the MI is related to the existence of the homoclinic solutions \cite{AH-1990-50,FL-1986-2}. However, it can provide a tool for the IST to the nonlinear stage for modulational instability instead of reducing the effectiveness of the IST or hindering the development of the IST.

The compatibility condition $X_{t}-T_{x}+[X,T]=0$ of the lax pair for equation \eqref{NLS-1} is
\begin{equation}\label{NLS-3}
\phi_{x}=X\phi,~~~X=-i\kappa\sigma_{3}+U,
\end{equation}
\begin{equation}\label{NLS-4}
\phi_{t}=T\phi,~~~T=-(2\kappa+c)(-i\kappa\sigma_{3}+U)+i(U_{x}+U^{2}+u_{0}^{2})\sigma_{3},
\end{equation}
where $\kappa$ means an spectral parameter, $\phi=\phi(x,t,\kappa)$ is a $2\times2$ matrix-valued eigenfunction, and the potential matrix $U$ is given by
\begin{align}\label{NLS-5}
 U=U(x,t)=\left(\begin{array}{cc}
    0 & u(x,t)\\
    -u^{\ast}(x,t) & 0\\
\end{array}\right),
\end{align}
with $u^{\ast}(x,t)$ standing for the complex conjugate of $u(x,t)$.

\begin{rem} In this work, the conjugate transposition and complex conjugate are represented by $\dag$ and $\ast$, respectively. And three Pauli matrices are defined as
\begin{align}\label{NLS-6}
 \sigma_{1}=\left(\begin{array}{cc}
    0 & 1\\
    1 & 0\\
\end{array}\right),\qquad  \sigma_{2}=\left(\begin{array}{cc}
    0 & -i\\
    i & 0\\
\end{array}\right), \qquad  \sigma_{3}=\left(\begin{array}{cc}
    1 & 0\\
    0 & -1\\
\end{array}\right),
\end{align}
and $e^{\alpha\widehat{\sigma}_{3}}U=e^{\alpha\sigma_{3}}Ue^{-\alpha\sigma_{3}}$ with $\alpha$ being a scalar variable.
\end{rem}

In this work, we will study the IST and soliton solutions of the NLSLab equation with NZBCs \eqref{NLS-2} from Lax \eqref{NLS-3} and \eqref{NLS-4}. The rest of this work is organized as follows: In Section 2, we study the asymptotic Lax pair of the NLSLab equation with NZBCs, and then the analyticity and symmetric of the Jost function and the scattering matrix are obtained by using spectral analysis. Moreover, discrete spectrum and residue conditions are given to research the inverse transformation process. In Section 3, the simple-pole solutions of the NLSLab equation with NZBCs are obtained by solving the matrix Riemann-Hilbert problem with the reflection-less potentials,  and their trace formulae and theta condition are obtained. In addition, their potential functions have been reconstructed. In Section 4, we solved the soliton solutions of the NLSLab equation and discuss the propagation behavior of different kinds of solutions by selecting different parameters with a brief analysis. In Section 5, we investigated the double-pole solutions of the NLSLab equation, and the construction of their trace formulae and theta condition. Some conclusions and discussions were presented in the last section.

\section{Direct scattering problem}

%

The asymptotic and the analyticity of the scattering eigenfunctions, asymptotics, and symmetries of the scattering matrix, residue conditions, and discrete spectrum can all be determined by using the direct scattering processes. Even if we have defined a suitable  uniformization variable \cite{FT-1987}, the two-sheeted Riemann surface for $\kappa$ can also be transformed into the standard complex $z$-plane, and the scattering problem is more conveniently handled on the standard complex $z$-plane.
\subsection{Preliminaries: Lax pair, Riemann surface, and uniformization coordinate}
Below we will considering the following asymptotic scattering problems $(x\rightarrow\pm\infty)$ of the Lax pairs \eqref{NLS-3} and \eqref{NLS-4}:
\begin{equation}\label{Pre-1}
\phi_{x}=X_{\pm}(\kappa)\phi,~~~X_{\pm}(\kappa)=\lim_{x\rightarrow\pm\infty}X=-i\kappa\sigma_{3}+U_{\pm},
\end{equation}
\begin{equation}\label{Pre-2}
\phi_{t}=T_{\pm}(\kappa)\phi,~~~T_{\pm}(\kappa)=\lim_{x\rightarrow\pm\infty}T=-(2\kappa+c)X_{\pm}(\kappa),
\end{equation}
where $U_{\pm}=\lim_{x\rightarrow\pm\infty}U(x,t)=\left(\begin{array}{cc}
    0 & u_{\pm}\\
    -u^{\ast}_{\pm} & 0\\
\end{array}\right)$.

We can get the eigenvalues of $X_{\pm}(\kappa)$ are $\pm i\sqrt{\kappa^{2}+u_{0}^{2}}$. It can be seen from the above that the eigenvalues of $X_{\pm}(\kappa)$ are doubly branched, so we need to introduce the two-sheeted Riemann surface defined by
\begin{equation}\label{Pre-3}
\lambda^{2}=\kappa^{2}+u_{0}^{2},
\end{equation}
such that $\lambda(\kappa)$ is a single-valued function on this surface. Here the branch point are the value of $\kappa$ for which $\kappa^{2}+u_{0}^{2}=0$, i.e., $\kappa=\pm iu_{0}$.  Letting $\kappa+iu_{0}=r_{1}e^{i\theta_{1}}$ and $\kappa-iu_{0}=r_{2}e^{i\theta_{2}}$, we can rewrite $\lambda(\kappa)=\sqrt{r_{1}r_{2}}e^{i(\frac{\theta_{1}+\theta_{2}}{2}+im\pi)}$, where $m=0,1$, respectively, on sheets I and sheets II. One take $\theta_{j}\in[-\frac{\pi}{2},\frac{3\pi}{2})$ for $j=1,2$.  According to these conventions, we obtained a discontinuity of $\lambda(\kappa)$ (which determines the position of the branch cut)  occurs on the segment $[-iu_{0},iu_{0}]$. We can then obtain the Riemann surface based on two copies of the complex plane along the cut. Along the real $\kappa$-axis, we can get $\lambda(\kappa)=\pm \sec\mbox{sign}(\kappa)\sqrt{\kappa^{2}+u_{0}^{2}}$, where the plus/minus sign applies to sheet I and II of the Riemann surface, respectively, and the square root sign represents the main branch of the real-valued square root function.

Below we will define the uniformization variable
\begin{equation}\label{Pre-4}
z=\kappa+\lambda.
\end{equation}
The inverse transformation is
\begin{equation}\label{Pre-5}
\kappa=\frac{1}{2}(z-\frac{u_{0}^{2}}{z}),~~~\lambda=\frac{1}{2}(z+\frac{u_{0}^{2}}{z}).
\end{equation}
Finally, in the complex $z$-plane, we take the circle of radius $u_{0}$ as $C_{0}$. Based on the above definition, we can get the branch cut on any sheet to be mapped to $C_{0}$.  In particular, $z(\pm iu_{0})=\pm iu_{0}$ from either sheet $z(0_{\textmd{I}}^{\pm})=\pm u_{0}$ and sheet $z(0_{\textmd{II}}^{\pm})=\mp u_{0}$; $C_{\textmd{I}}$ is mapped onto the exterior of $C_{0}$; $C_{\textmd{II}}$ is mapped onto the interior of $C_{0}$; in particular, on the $C_{\textmd{I}}$, $z=\infty$ as $k\rightarrow\infty$, while on the $C_{\textmd{II}}$, $z=0$ as $k\rightarrow\infty$; here we note that the first/second quadrants of the $C_{\textmd{I}}$ are mapped to the corresponding first/second quadrants outside $C_{0}$, respectively; the second/first quadrants of $C_{\textmd{II}}$ are mapped to the corresponding first/secondquadrants inside $C_{0}$, respectively. Also need to point out, $z_{\textmd{I}}z_{\textmd{II}}=(k+\sqrt{\kappa^{2}+u_{0}^{2}})(k-\sqrt{\kappa^{2}+u_{0}^{2}})=u_{0}^{2}$.

Here we can see that $\textmd{Im}\lambda$ is not determined by the sign in the upper half plane (UHP) and the lower half plane (LHP). Contrary to this, $\textmd{Im}\lambda>0$ in $D^{+}$, $\textmd{Im}\lambda<0$ in $D^{-}$, where
\begin{equation}\label{Pre-6}
D^{+}=\{z\in C:(|z|^{2}-u_{0}^{2})\textmd{Im}z>0\},~~~D^{-}=\{z\in C:(|z|^{2}-u_{0}^{2})\textmd{Im}z<0\}.
\end{equation}
In Figure 1, we show these two domains. As shown, the properties described above determines the analyticity regions of the Jost eigenfunction. For some of the above symbols, we will rewrite all $\kappa$ dependencies as dependency on $z$.
\\

\centerline{\begin{tikzpicture}
\path [fill=gray] (-5.25,0) -- (-0.25,0) to (-0.25,2.5) -- (-5.25,2.5);
\draw[-][thick](-5.25,0)--(-2.75,0);
\draw[fill] (-2.75,0) circle [radius=0.045];
\draw[->][thick](-2.75,0)--(-0.25,0)node[above]{$\textmd{Re}k$};
\draw[<-][thick](-2.75,2.5)node[right]{$\textmd{Im}k$}--(-2.75,1.5)node[right]{$iq_{0}$};
\draw[fill] (-2.75,1.5) circle [radius=0.045];
\draw[-][thick](-2.75,1.5)--(-2.75,0);
\draw[-][thick](-2.75,0)--(-2.75,-1.5)node[right]{$-iq_{0}$};
\draw[fill] (-2.75,-1.5) circle [radius=0.045];
\draw[-][thick](-2.75,-1.5)--(-2.75,-2.0);
\draw[fill] (-2.75,-0.3) node[right]{$0$};
\draw[fill] (-1.7,1) circle [radius=0.045] node[right]{$z^{*}_{n}$};
\draw[fill] (-1.7,-1) circle [radius=0.045] node[right]{$z_{n}$};
\path [fill=gray] (1.15,0) -- (6.35,0) to (6.35,2.5) -- (1.15,2.5);
\filldraw[white, line width=1](5.25,0) arc (0:180:1.5);
\filldraw[gray, line width=1](2.25,0) arc (-180:0:1.5);
\draw[->][thick](1.15,0)--(1.8,0);
\draw[-][thick](1.15,0)--(2.25,0);
\draw[<-][thick](2.75,0)--(4,0);
\draw[fill] (3.75,0) circle [radius=0.045];
\draw[-][thick](2.25,0)--(3.75,0);
\draw[<->][thick](4.5,0)--(5.75,0);
\draw[-][thick](3.75,0)--(6.35,0)node[above]{$\textmd{Re}z$};
\draw[-][thick](3.75,2.5)node[right]{$\textmd{Im}z$}--(3.75,0);
\draw[-][thick](3.75,0)--(3.75,-2.0);
\draw[fill] (3.8,-0.3) node[right]{$0$};
\draw[fill] (3.75,1.5) circle [radius=0.045];
\draw[fill] (3.75,-1.5) circle [radius=0.045];
\draw[fill] (3.80,1.7) node[right]{$iq_{0}$};
\draw[fill] (3.80,-1.7) node[right]{$-iq_{0}$};
\draw[fill][red] (5.25,1.8) circle [radius=0.045] node[right]{$z^{*}_{n}$};
\draw[fill] (5.25,-1.8) circle [radius=0.045] node[right]{$z_{n}$};
\draw[fill] (3.25,0.8) circle [radius=0.045] node[right]{$-\frac{q^{2}_{0}}{z^{*}_{n}}$};
\draw[fill][red] (3.25,-0.8) circle [radius=0.045] node[right]{$-\frac{q^{2}_{0}}{z_{n}}$};
\draw[-][thick](5.25,0) arc(0:360:1.5);
\draw[-<][thick](5.25,0) arc(0:30:1.5);
\draw[-<][thick](5.25,0) arc(0:150:1.5);
\draw[->][thick](5.25,0) arc(0:210:1.5);
\draw[->][thick](5.25,0) arc(0:330:1.5);
\end{tikzpicture}}
\noindent {\small \textbf{Figure 1.} (Color online) Left: Depicts the first sheet of the Riemann surface, showing different discrete spectral points in the spectral plane with $\textmd{Im}k>0$ (gray) and $\textmd{Im}k<0$ (white); Right: the discrete spectral points on the complex z-plane [zeros point of $s_{11}(z)$ (black) in the white region ($\textmd{Im}k<0$) and zeros point of $s_{22}(z)$ (red) in the gray region ($\textmd{Im}k>0$)], and also gives the orientation of the jump contours of the Riemann-Hilbert problem.}

\subsection{Jost solutions and analyticity}
On any sheet of the Riemann surface, we write the asymptotic eigenvector matrix as the following form
\begin{equation}\label{Pre-7}
E_{\pm}(z)=I-\frac{i}{z}\sigma_{3}U_{\pm}=\left(\begin{array}{cc}
    1 & -\frac{iu_{\pm}}{z}\\
    -\frac{iu^{\ast}_{\pm}}{z} & 1\\
\end{array}\right),
\end{equation}
where $I$ means the $2\times2$ identity matrix, so that
\addtocounter{equation}{1}
\begin{align}\label{cl-v-a}
&X_{\pm}E_{\pm}=-\frac{i}{2}(z+\frac{u_{0}^{2}}{z})E_{\pm}\sigma_{3}=-i\lambda E_{\pm}\sigma_{3},\tag{\theequation a}\\
\label{cl-v-b}
&T_{\pm}E_{\pm}=i\lambda(2\kappa+c)E_{\pm}\sigma_{3}.\tag{\theequation b}
\end{align}
\begin{rem} As $z=\pm iu_{0}$, $\det E_{\pm}\equiv 0$, whereas $z\neq\pm iu_{0}$, $\det E_{\pm}=1+\frac{u_{0}^{2}}{z^{2}}=\gamma(z)\neq0$ in which its inverse matrix exists, and its inverse matrix is \\
$~~~~~~~~~~~~~~~~~~~~~~~~~~~~~~~~~~~~~~~E_{\pm}^{-1}(z)=\frac{1}{\gamma(z)}\left(\begin{array}{cc}
    1 & \frac{iu_{\pm}}{z}\\
    \frac{iu^{\ast}_{\pm}}{z} & 1\\
\end{array}\right)$.
\end{rem}
In general, the continuum spectrum $\kappa$ is constructed by all values of $\kappa$ such that $\lambda(\kappa)\in R$,\cite{BK-2014-55} i.e., $\Sigma_{\kappa}=R\cup i[-u_{0},u_{0}]$. $\Sigma_{z}=R\cup C_{0}$ is the corresponding set of the complex $z$-plane. In the following expressions we will leave out the subscripts on $\Sigma$, where the expected meaning will be understood from the context.

\noindent \textbf{Theorem 1.} \emph{ For all of the above $z\in\Sigma$, we can define the solution of the Jost eigenfunction $\phi(x,t,z)$ as a simultaneous solutions of the two parts of the Lax pair, such that
\begin{equation}\label{Pre-8}
\phi(x,t,z)=E_{\pm}e^{i\theta(x,t,z)\sigma_{3}}, ~~~~ as ~~~~x\rightarrow\pm\infty,
\end{equation}
where
\begin{equation}\label{Pre-9}
\theta(x,t,z)=\lambda[-x+(2\kappa+c)t].
\end{equation}}

Here we need to subtract the asymptotic behavior of the potential and rewrite \eqref{NLS-3} as $(\phi_{\pm})_{x}=X_{\pm}\phi_{\pm}+\Delta U_{\pm}\phi_{\pm}$, where $\Delta U_{\pm}=U-U_{\pm}$. In the later calculation process, we need to introduce modified eigenfunctions by using the method of factorization for asymptotic exponential oscillation:
\begin{equation}\label{Pre-12}
\mu_{\pm}(x,t,z)=\phi_{\pm}(x,t,z)e^{-i\theta(x,t,z)\sigma_{3}},
\end{equation}
so that $\lim_{x\rightarrow\pm\infty}\mu_{\pm}(x,t,z)=E_{\pm}$. Then we formally integrated of the ODE of $\mu_{\pm}$ to get the linear integral equation of the modified eigenfunction:
\begin{equation}\label{cl3-v-a}
\mu_{-}(x,t,z)=E_{-}+\int_{-\infty}^{x}E_{-}e^{-i\lambda(x-y)\sigma_{3}}E_{-}^{-1}\Delta U_{-}(y,t)\mu_{-}(x,t,z)e^{i\lambda(x-y)\sigma_{3}}dy,
\end{equation}
By using the above integral equation, we can get the following theorem:

\noindent \textbf{Theorem 2.} \emph{Under the mild integrability conditions on the potential, we can extend the eigenfunction analysis to the following regions on the complex $z$-plane:
\begin{align}\label{Pre-13}
&D^{+}:~~~\mu_{-,1},~~~\mu_{+,2},\notag\\
&D^{-}:~~~\mu_{+,1},~~~\mu_{-,2},
\end{align}
where $\mu_{\pm}=(\mu_{\pm,1}, \mu_{\pm,2})$.}

\noindent \textbf{Proof:} We first rewriting the integral equation \eqref{cl3-v-a} that defines the Jost eigenfunction:
\begin{equation}\label{cl4-v-a}
\mu_{-}(x,z)=E_{-}\left[I+\int_{-\infty}^{x}e^{-i\lambda(x-y)\sigma_{3}}E_{-}^{-1}\Delta U_{-}(y)\mu_{-}(x,z)e^{i\lambda(x-y)\sigma_{3}}dy\right].
\end{equation}
The limits of integration means that $x-y$ is always negative for $\mu_{+}$ and always positive for $\mu_{-}$.
We should note that the matrix products in the RHS of \eqref{cl4-v-a} operate column-wise. In particular, letting $M_{\pm}(x,z)=E_{\pm}^{-1}\mu_{\pm}$ for the first column $m_{-,1}$ and the second column $m_{-,2}$ of $M_{-}$, one has
\addtocounter{equation}{1}
\begin{align}\label{cl5-v-a}
&m_{-,1}(x,z)=\left(\begin{aligned}
&1\\ &0\\
\end{aligned}\right)+\int_{-\infty}^{x}G_{1}(x-y,z)\Delta U_{-}(y)E_{\pm}(z)m_{-}(x,z)dy,\tag{\theequation a}\\
\label{cl5-v-b}
&m_{-,2}(x,z)=\left(\begin{aligned}
&0\\ &1\\
\end{aligned}\right)+\int_{-\infty}^{x}G_{2}(x-y,z)\Delta U_{-}(y)E_{\pm}(z)m_{-}(x,z)dy,\tag{\theequation b}
\end{align}
where
\addtocounter{equation}{1}
\begin{align}\label{cl6-v-a}
&G_{1}(x-y,z)=\textmd{diag}(1,e^{2i\lambda(x-y)})E_{-}^{-1}(z),\tag{\theequation a}\\
\label{cl6-v-b}
&G_{2}(x-y,z)=\textmd{diag}(e^{-2i\lambda(x-y)},1)E_{-}^{-1}(z),\tag{\theequation b}
\end{align}
with $e^{2i\lambda(x-y)}=e^{2i\textmd{Re}(\lambda(x-y))}e^{-2\textmd{Im}(\lambda(x-y))}$, $e^{-2i\lambda(x-y)}=e^{-2i\textmd{Re}(\lambda(x-y))}e^{2\textmd{Im}(\lambda(x-y))}$. Then, if we want to understand the analytic area of each column, we need to consider the situation of $\textmd{Im}(\lambda(x-y))>0$ and $\textmd{Im}(\lambda(x-y))<0$. In equation \eqref{cl4-v-a}, we obtain $(x-y)>0$, so the first column $\mu_{-,1}$ is analytic in the $z$-plane $D^{+}$ and the second column $\mu_{-,2}$ is analytic in the $z$-plane $D^{-}$. By using the same method, we obtain that the first column $\mu_{+,1}$ is analytic in the $z$-plane $D^{-}$ and the second column $\mu_{+,2}$ is analytic in the $z$-plane $D^{+}$.

The proof the above theorem.    $\Box$

\begin{rem} Here we will notice that the analytical properties of the columns for $\mu_{\pm}$ are basically the same as the analytical properties of $\phi_{\pm}$. Also, in the subsequent expression, we will superscript $\pm$ denote the regions $D^{\pm}$ of analyticity, and the subscript $\pm$ means the limiting value as $x\rightarrow\pm\infty$.
\end{rem}

\subsection{Scattering matrix}
According to Abel's theorem, we get any solution $\phi(x,t,z)$ of equations \eqref{Pre-1} and \eqref{Pre-2}, which has the property of $\partial_{x}(\textmd{det}\phi)=\partial_{t}(\textmd{det}\phi)=0$. Thus, since for all $z\in\Sigma$, \\ $\mathop{\textmd{lim}}_{x\rightarrow\infty}\mu_{\pm}(x,t,z)=\textmd{lim}_{x\rightarrow\infty}\phi_{\pm}(x,t,z)e^{-i\theta(x,t,z)\sigma_{3}}=E_{\pm}$, we can get
\begin{equation}\label{Pre-14}
\textmd{det}\mu_{\pm}(x,t,z)=\textmd{det}\phi_{\pm}(x,t,z)=\gamma(z)~~~~~x,t\in R, ~~~~z\in\Sigma.
\end{equation}
Letting $\Sigma_{0}=\Sigma\backslash\{\pm iu_{0}\}$, we have that $\forall z\in\Sigma_{0}$ both $\phi_{+}$ and $\phi_{-}$ are two fundamental matrix solutions for the scattering problem, and the two fundamental matrix solutions are linearly related, then we can relation the two matrices by $S(z)=(s_{ij})_{2\times2}$, namely:
\begin{equation}\label{Pre-15}
\phi_{+}(x,t,z)=\phi_{-}(x,t,z)S(z)~~~~~x,t\in R, ~~~~z\in\Sigma_{0}.
\end{equation}
From the relation \eqref{Pre-15}, we can get
\begin{align}\label{Pre-15-1}
&s_{11}(z)=\gamma^{-1}(z)|\phi_{+,1}(x,t,z),\phi_{-,2}(x,t,z)|,~~~
s_{22}(z)=\gamma^{-1}(z)|\phi_{-,1}(x,t,z),\phi_{+,2}(x,t,z)|,\notag\\
&s_{12}(z)=\gamma^{-1}(z)|\phi_{+,2}(x,t,z),\phi_{-,2}(x,t,z)|,~~~
s_{21}(z)=\gamma^{-1}(z)|\phi_{-,1}(x,t,z),\phi_{+,1}(x,t,z)|,
\end{align}
where $|\cdot,\cdot|=\det(\cdot,\cdot)$.

Furthermore, we can also get $\mu_{+}$ and $\mu_{-}$ are also linearly related, i.e.
\begin{equation}\label{Pre-15b}
\mu_{+}(x,t,z)=\mu_{-}(x,t,z)e^{i\theta(x,t,z)\sigma_{3}}S(z)e^{-i\theta(x,t,z)\sigma_{3}}~~~~~x,t\in R, ~~~~z\in\Sigma_{0}.
\end{equation}
Then we expand on the above formula by column, we can get
\begin{equation}\label{Pre-16}
\mu_{+,1}=s_{11}\mu_{-,1}+s_{21}\mu_{-,2}e^{-2i\theta(x,t,z)},~~~~~~~~\mu_{+,2}=s_{12}\mu_{-,1}e^{2i\theta(x,t,z)}+s_{22}\mu_{-,2}.
\end{equation}
From equation \eqref{Pre-14} and equation \eqref{Pre-15}, we can obtain
\begin{equation}\label{Pre-17}
\textmd{det}S(z)=1, ~~~~~~~~~~z\in\Sigma_{0}.
\end{equation}

\noindent \textbf{Theorem 3.} \emph{$s_{11}$ can be analytically extended to $D^{-}$, $s_{22}$ can be analytically extended to $D^{+}$. However, $s_{12}$ and $s_{21}$ are cannot be
extended off the $z$-plane in general.}

\noindent \textbf{Proof:} According to equation \eqref{Pre-14}, we know that $\mu_{\pm}$ is an invertible matrix, then the inverse matrix of $\mu_{\pm}$ is:
\begin{align}\label{Pre-17a}
 \mu_{\pm}^{-1}=\frac{1}{\textmd{det}\mu_{\pm}}\left(\begin{array}{cc}
    \mu_{\pm}^{22} & -\mu_{\pm}^{12}\\
    -\mu_{\pm}^{21} & \mu_{\pm}^{11}\\
\end{array}\right)=\frac{1}{\gamma}\left(\begin{array}{cc}
    \mu_{\pm}^{22} & -\mu_{\pm}^{12}\\
    -\mu_{\pm}^{21} & \mu_{\pm}^{11}\\
\end{array}\right)=\left(\begin{aligned}
&\widehat{\mu_{\pm,1}}\\ &\widehat{\mu_{\pm,2}}\\
\end{aligned}\right).
\end{align}
On the basis of theorem 2, we can get $\widehat{\mu_{+,1}}$ and $\widehat{\mu_{-,2}}$ are analytic in the $z$-plane $D^{+}$, and $\widehat{\mu_{-,1}}$ and $\widehat{\mu_{+,2}}$ are analytic in the $z$-plane $D^{-}$.

Form equation \eqref{Pre-15}, we can obtain
\begin{equation}\label{Pre-18}
e^{i\theta(x,t,z)\sigma_{3}}S(z)e^{-i\theta(x,t,z)\sigma_{3}}=\mu_{-}^{-1}(x,t,z)\mu_{+}(x,t,z).
\end{equation}
After further calculations, one can obtain
\begin{equation}\label{Pre-19}
e^{i\theta(x,t,z)\sigma_{3}}S(z)e^{-i\theta(x,t,z)\sigma_{3}}=\left(\begin{aligned}
&\widehat{\mu_{-,1}}\\ &\widehat{\mu_{-,2}}\\
\end{aligned}\right)(\mu_{+,1}, \mu_{+,2})=\left(\begin{array}{cc}
    \widehat{\mu_{-,1}}\mu_{+,1} & \widehat{\mu_{-,1}}\mu_{+,2}\\
    \widehat{\mu_{-,2}}\mu_{+,1} & \widehat{\mu_{-,2}}\mu_{+,2}\\
\end{array}\right).
\end{equation}
So $s_{11}$ is analytic in the $z$-plane $D^{-}$, and $s_{22}$ is analytic in the $z$-plane $D^{+}$.

The proof the above theorem.    $\Box$

\subsection{Symmetries}
The symmetries of the IST with NZBCs becomes intricate due to the following facts: (1) In the case of ZBCs, we only needs to process the map $\kappa\rightarrow\kappa^{\ast}$, but we also need to process the sheets of the Riemann surface; (2) Unlike the case of ZBCs, the Jost solution does not tend to the identity matrix after removing the asymptotic oscillation.

Recall $\lambda_{\textmd{I}}(\kappa)=-\lambda_{\textmd{II}}(\kappa)$, and consider some of the following conversions compatible with \eqref{Pre-3}: (1) when $z\rightarrow z^{\ast}$, implies $(\kappa,\lambda)\rightarrow(\kappa^{\ast},\lambda^{\ast})$ (here must be in the same sheet); (2) when $z\rightarrow-\frac{u_{0}^{2}}{z}$, indicates $(\kappa,\lambda)\rightarrow(\kappa,-\lambda)$ (here in the opposite sheets). The symmetries of the scattering problem has both of the above transformations. We will prove this in the following theorem.

\noindent \textbf{Theorem 4.} \emph{$\mu_{\pm}$ and $S(z)$ have the following symmetries:
\addtocounter{equation}{1}
\begin{align}\label{cl7-v-a}
&(1).~~~\mu_{\pm}(x,t,z)=-\sigma\mu_{\pm}^{\ast}(x,t,z^{\ast})\sigma,\tag{\theequation a}\\
\label{cl7-v-b}
&(2).~~~\phi_{\pm}(x,t,z)=-\sigma\phi_{\pm}^{\ast}(x,t,z^{\ast})\sigma,\tag{\theequation b}\\
\label{cl7-v-c}
&(3).~~~\mu_{\pm}(x,t,z)=-\frac{i}{z}\mu_{\pm}(x,t,-\frac{u_{0}^{2}}{z})\sigma_{3}U_{\pm},\tag{\theequation c}\\
\label{cl7-v-d}
&(4).~~~\phi_{\pm}(x,t,z)=-\frac{i}{z}\phi_{\pm}(x,t,-\frac{u_{0}^{2}}{z})\sigma_{3}U_{\pm},\tag{\theequation d}\\
\label{cl7-v-e}
&(5).~~~S^{\ast}(z^{\ast})=-\sigma S(z)\sigma,\tag{\theequation e}\\
\label{cl7-v-f}
&(6).~~~S(z)=(\sigma_{3}U_{-})^{-1}S(-\frac{u_{0}^{2}}{z})\sigma_{3}U_{+},\tag{\theequation f}\\
\label{cl7-v-g}
&(7).~~~S^{\ast}(z^{\ast})=-\sigma (\sigma_{3}U_{-})^{-1}S(-\frac{u_{0}^{2}}{z})\sigma_{3}U_{+}\sigma,\tag{\theequation g}
\end{align}}
where $\sigma=\left(\begin{array}{cc}
    0 & 1\\
    -1 & 0\\
\end{array}\right)$.

\noindent \textbf{Proof:} Here we only prove that (1) and (5). Other symmetries can be similarly proven.

$(1)$. Using $(\phi_{\pm})_{x}=X_{\pm}\phi_{\pm}+\Delta U_{\pm}\phi_{\pm}$ and equation \eqref{Pre-12}, we can get the lax pairs of $\mu_{\pm}$ as
\begin{equation}\label{Pre-20}
\left(E_{\pm}^{-1}(z)\mu_{\pm}(z)\right)_{x}-i\lambda(z)\left[E_{\pm}^{-1}(z)\mu_{\pm}(z),\sigma_{3}\right]=E_{\pm}^{-1}(z)\Delta U_{\pm}(z)\mu_{\pm}(z).
\end{equation}
In the calculation process, we first change $z$ to $z^{\ast}$, then complex conjugate the equation, and finally multiply $\sigma$ on both sides of the equation, one can get
\begin{equation}\label{Pre-21}
\left(\sigma E_{\pm}^{-1^{\ast}}(z^{\ast})\mu_{\pm}^{\ast}(z^{\ast})\sigma\right)_{x}+i\lambda^{\ast}(z^{\ast})\sigma\left[E_{\pm}^{-1^{\ast}}(z^{\ast})\mu_{\pm}^{\ast}(z^{\ast}),\sigma_{3}\right]\sigma=\sigma E_{\pm}^{-1^{\ast}}(z^{\ast})\Delta U_{\pm}^{\ast}(z^{\ast})\mu_{\pm}^{\ast}(z^{\ast})\sigma.
\end{equation}
Below we will add $\sigma\sigma$ (i.e. $\sigma \sigma=-I$) to the above equation, one can get
\begin{align}\label{Pre-22}
&\left(-\sigma E_{\pm}^{-1^{\ast}}(z^{\ast})\sigma \sigma\mu_{\pm}^{\ast}(z^{\ast})\sigma\right)_{x}+i\lambda^{\ast}(z^{\ast})\sigma\left(E_{\pm}^{-1^{\ast}}(z^{\ast})\sigma \sigma\mu_{\pm}^{\ast}(z^{\ast})\sigma \sigma\sigma_{3}\right.\notag\\
&\left.-\sigma_{3}\sigma \sigma E_{\pm}^{-1^{\ast}}(z^{\ast})\sigma \sigma\mu_{\pm}^{\ast}(z^{\ast})\right)\sigma=\sigma E_{\pm}^{-1^{\ast}}(z^{\ast})\sigma \sigma\Delta U_{\pm}^{\ast}(z^{\ast})\sigma \sigma\mu_{\pm}^{\ast}(z^{\ast})\sigma.
\end{align}
We noticed that $\sigma E_{\pm}^{-1^{\ast}}(z^{\ast})\sigma=-E_{\pm}^{-1}(z)$, $\lambda^{\ast}(z^{\ast})=\lambda(z)$, $\sigma\sigma_{3}\sigma=\sigma_{3}$, $\sigma\Delta U_{\pm}^{\ast}(z^{\ast})\sigma=-\Delta U_{\pm}(z)$. Then, we can further simplify the above equation, one can obtain
\begin{equation}\label{Pre-23}
\left(E_{\pm}^{-1}(z) \sigma\mu_{\pm}^{\ast}(z^{\ast})\sigma\right)_{x}-i\lambda(z)\sigma\left[E_{\pm}^{-1}(z) \sigma\mu_{\pm}^{\ast}(z^{\ast})\sigma,\sigma_{3}\right]=E_{\pm}^{-1}(z)\Delta U_{\pm}(z) \sigma\mu_{\pm}^{\ast}(z^{\ast})\sigma.
\end{equation}
We compare equation \eqref{Pre-20} with equation \eqref{Pre-23} and find that $\mu_{\pm}(z)$ and $\sigma\mu_{\pm}^{\ast}(z^{\ast})\sigma$ satisfy the same differential equation, and we are combining progressive analysis ($\lim_{x\rightarrow\pm\infty}\mu_{\pm}(x,t,z)=E_{\pm}$, $\lim_{x\rightarrow\pm\infty}-\sigma\mu_{\pm}^{\ast}(z^{\ast})\sigma=E_{\pm}$). We can get
\begin{equation}\label{Pre-24}
\mu_{\pm}(x,t,z)=-\sigma\mu_{\pm}^{\ast}(x,t,z^{\ast})\sigma.
\end{equation}
$(5)$. From \eqref{Pre-18}, one get
\begin{equation}\label{Pre-25}
S(z)=e^{-i\theta(x,t,z)\sigma_{3}}\mu_{-}^{-1}(x,t,z)\mu_{+}(x,t,z)e^{i\theta(x,t,z)\sigma_{3}}.
\end{equation}
Further calculations, we get
\begin{equation}\label{Pre-26}
S^{\ast}(z^{\ast})=e^{i\theta^{\ast}(x,t,z^{\ast})\sigma_{3}}\mu_{-}^{-1^{\ast}}(x,t,z^{\ast})\mu_{+}^{\ast}(x,t,z^{\ast})e^{-i\theta^{\ast}(x,t,z^{\ast})\sigma_{3}}.
\end{equation}
Below we will add $\sigma\sigma$ to the above equation, one can get
\begin{equation}\label{Pre-27}
\sigma S^{\ast}(z^{\ast})\sigma=-\sigma e^{i\theta^{\ast}(x,t,z^{\ast})\sigma_{3}}\sigma\sigma\mu_{-}^{-1^{\ast}}(x,t,z^{\ast})\sigma\sigma\mu_{+}^{\ast}(x,t,z^{\ast})\sigma\sigma e^{-i\theta^{\ast}(x,t,z^{\ast})\sigma_{3}}\sigma.
\end{equation}
From the conclusion of (1), we can obtain $\sigma e^{i\theta^{\ast}(x,t,z^{\ast})\sigma_{3}}\sigma=-e^{-i\theta(x,t,z)\sigma_{3}}$, $\sigma e^{-i\theta^{\ast}(x,t,z^{\ast})\sigma_{3}}\sigma=-e^{i\theta(x,t,z)\sigma_{3}}$, $\mu_{-}^{-1}(x,t,z)=-\sigma\mu_{-}^{-1^{\ast}}(x,t,z^{\ast})\sigma$ and $\mu_{+}(x,t,z)=-\sigma\mu_{+}^{\ast}(x,t,z^{\ast})\sigma$, thus simplifying the above equation \eqref{Pre-27}, one can obtain
\begin{equation}\label{Pre-28}
\sigma S^{\ast}(z^{\ast})\sigma=-\mu_{-}^{-1}(x,t,z)\mu_{+}(x,t,z)=-S(z).
\end{equation}
The proof the above theorem.    $\Box$

We can expand the above theorem 4 to get the following theorem:

\noindent \textbf{Theorem 5.} \emph{$\mu_{\pm}$ and $S(z)$ have the following symmetries:
\addtocounter{equation}{1}
\begin{align}\label{cl8-v-a}
&(1).~~~\mu_{\pm,1}(x,t,z)=\sigma\mu_{\pm,2}^{\ast}(x,t,z^{\ast}),~~~~~~~~~\mu_{\pm,2}(x,t,z)=-\sigma\mu_{\pm,1}^{\ast}(x,t,z^{\ast})\tag{\theequation a}\\
\label{cl8-v-b}
&(2).~~~\mu_{\pm,1}(x,t,z)=-\frac{iu_{\pm}^{\ast}}{z}\mu_{\pm,2}(x,t,-\frac{u_{0}^{2}}{z}),~\mu_{\pm,2}(x,t,z)=-\frac{iu_{\pm}}{z}\mu_{\pm,1}(x,t,-\frac{u_{0}^{2}}{z}),\tag{\theequation b}\\
\label{cl8-v-c}
&(3).~~~s^{\ast}_{11}(z^{\ast})=s_{22},~~~~~~~~~~~~~~~~~~~~~~~~~~~s^{\ast}_{12}(z^{\ast})=-s_{21},\tag{\theequation c}\\
\label{cl8-v-d}
&(4).~~~s_{11}(z)=\frac{u_{+}^{\ast}}{u_{-}^{\ast}}s_{22}(-\frac{u_{0}^{2}}{z}),~~~~~~~~~~~~~~~~s_{12}(z)=\frac{u_{+}}{u_{-}^{\ast}}s_{21}(-\frac{u_{0}^{2}}{z}).\tag{\theequation d}\\
\label{cl8-v-e}
&(5).~~~s_{21}(z)=\frac{u_{+}^{\ast}}{u_{-}}s_{12}(-\frac{u_{0}^{2}}{z}),~~~~~~~~~~~~~~~~s_{22}(z)=\frac{u_{+}}{u_{-}}s_{11}(-\frac{u_{0}^{2}}{z}).\tag{\theequation e}\\
\label{cl8-v-f}
&(6).~~~s^{\ast}_{11}(z^{\ast})=\frac{u_{+}}{u_{-}}s_{11}(-\frac{u_{0}^{2}}{z}),~~~~~~~~~~~~~~~s^{\ast}_{12}(z^{\ast})=-\frac{u^{\ast}_{+}}{u_{-}}s_{12}(-\frac{u_{0}^{2}}{z}).\tag{\theequation f}\\
\label{cl8-v-g}
&(7).~~~s^{\ast}_{21}(z^{\ast})=-\frac{u_{+}}{u^{\ast}_{-}}s_{21}(-\frac{u_{0}^{2}}{z}),~~~~~~~~~~~~s^{\ast}_{22}(z^{\ast})=\frac{u^{\ast}_{+}}{u^{\ast}_{-}}s_{22}(-\frac{u_{0}^{2}}{z}).\tag{\theequation g}
\end{align}}
\noindent \textbf{Proof:} For the proof process of theorem 5, we will not explain in detail. $\Box$

Note that:

(i) Although the above theorem 4 and 5 are only effective for $z\in\Sigma$,  whenever the scattering coefficients and individual columns are analytic, they can be extended to the appropriate regions of the $z$-plane by using the Schwartz reflection principle.

(ii) Since the continuum spectrum is not only a subset of the real $z$-axis, then it is different the case of ZBCs, and the symmetries of the non-analytic scattering coefficients also needs to involve the map $z\rightarrow z^{\ast}$.

(iii) For the first involution $z\rightarrow z^{\ast}$, they are the same as ZBCs. For the second  involution $z\rightarrow-\frac{u_{0}^{2}}{z}$, they represents the switch from one sheet to the other sheet. But since this transformation does not affect $\kappa$, so we consider $f(\kappa)$ is any single-valued function of $\kappa$, one have $f_{\textmd{I}}(\kappa)=f_{\textmd{II}}(\kappa)$. In other words, when expressed is a function of $z$, $f(z)$ satisfies the symmetry $f(z)=f(-\frac{u_{0}^{2}}{z})$. This is because $f$ does not depends on $z$ directly, but only through the combination $k=\frac{z-\frac{u_{0}^{2}}{z}}{2}$. In general, \eqref{cl7-v-b} and \eqref{cl8-v-d} relate to the values of the Jost eigenfunction and the scattering coefficient on the opposite slice of the Riemann surface.

\subsection{Discrete spectrum and residue conditions}
The set of all values $z\in C\backslash\Sigma$ constitutes the discrete spectrum of the scattering problem, such that the eigenfunction exists in $L^{2}(R)$. We will next prove that these values are zeros of $s_{11}$ in $D^{-}$ and  those of $s_{22}$ in $D^{+}$. It is worth noting that in the case of ZBCs produce the so-called real spectral singularities\cite{Z-1989-42}. Now we only consider potentials without spectral singularities. However, in later section, since the discrete eigenvalues tend to $\Sigma$, we will consider the limit form of a soliton solution. And we will prove that this limit is well defined and produces an ordinary solution.

For all $z\in D^{-}$, $\mu_{+,1}\rightarrow0$ as $x\rightarrow+\infty$, and $\mu_{-,2}\rightarrow0$ as $x\rightarrow-\infty$. We the hypothesis that $s_{11}(z)$ has a finite number $N$ of simple zeros $z_{1}, z_{2}, \ldots, z_{N}$ in $D^{-}\cap\{z\in C: \textmd{Im}z<0\}$. That is, let $s_{11}(z_{n})=0$ and $s_{11}'(z_{n})\neq0$, where $s_{11}'(z_{n})$ means differential of $s_{11}(z_{n})$ with respect to $z$, and with $|z_{n}|>u_{0}$ and $\textmd{Im}z<0$ for $n=1,2,\ldots,N$. According to the theorem 5, we can get
\begin{equation}\label{Pre-29}
s_{11}(z_{n})=0\Leftrightarrow s_{22}(z^{\ast}_{n})=0\Leftrightarrow s_{22}(-\frac{u_{0}^{2}}{z_{n}})=0\Leftrightarrow s_{11}(-\frac{u_{0}^{2}}{z^{\ast}_{n}})=0.
\end{equation}
Therefore, for each $n=1,2,\ldots,N$,  we have a quartet of discrete eigenvalues, i.e. the discrete spectrum is the set
\begin{equation}\label{Pre-30}
Z=\{z_{n}, z^{\ast}_{n}, -\frac{u_{0}^{2}}{z_{n}}, -\frac{u_{0}^{2}}{z^{\ast}_{n}}\}_{n=1}^{N}.
\end{equation}
In the next process, we will derive the residual conditions needed for the inverse problem.

From theorem 2 and theorem 3, we know that $\mu_{-,1}$, $\mu_{+,2}$,  $s_{22}$ are analytic in the $z$-plane $D^{+}$, and $\mu_{+,1}$, $\mu_{-,2}$, $s_{11}$ are analytic in the $z$-plane $D^{-}$. Thus we convert equation \eqref{Pre-16} into the following form:
\addtocounter{equation}{1}
\begin{align}\label{cl9-v-a}
&\mu_{+,1}(z_{n})=s_{21}(z_{n})\mu_{-,2}(z_{n})e^{-2i\theta(z_{n})},\tag{\theequation a}\\
\label{cl9-v-b}
&\mu_{+,2}(z^{\ast}_{n})=s_{12}(z^{\ast}_{n})\mu_{-,1}(z^{\ast}_{n})e^{2i\theta(z^{\ast}_{n})}.\tag{\theequation b}
\end{align}
Furthermore, we have
\addtocounter{equation}{1}
\begin{align}\label{cl9-v-1}
&\phi_{+,1}(z_{n})=A_{-}(z_{n})\phi_{-,2}(z_{n}),\tag{\theequation a}\\
\label{cl9-v-2}
&\phi_{+,2}(z^{\ast}_{n})=A_{+}(z^{\ast}_{n})\phi_{-,1}(z^{\ast}_{n}).\tag{\theequation b}
\end{align}
where $A_{-}(z_{n})$ and $A_{+}(z^{\ast}_{n})$ are all norming constants.

Thus, one obtain
\addtocounter{equation}{1}
\begin{align}\label{cl10-v-a}
&\mathop{\textmd{Res}}_{z=z_{n}}\left(\frac{\mu_{+,1}(z)}{s_{11}(z)}\right)=\frac{\mu_{+,1}(z_{n})}{s_{11}'(z_{n})}=
\frac{s_{21}(z_{n})e^{-2i\theta(z_{n})}}{s_{11}'(z_{n})}\mu_{-,2}(z_{n})
=b_{n}\mu_{-,2}(z_{n}),\tag{\theequation a}\\
\label{cl10-v-b}
&\mathop{\textmd{Res}}_{z=z^{\ast}_{n}}\left(\frac{\mu_{+,2}(z)}{s_{22}(z)}\right)=\frac{\mu_{+,2}(z^{\ast}_{n})}{s_{22}'(z^{\ast}_{n})}=
\frac{s_{12}(z^{\ast}_{n})e^{2i\theta(z^{\ast}_{n})}}{s_{22}'(z^{\ast}_{n})}\mu_{-,1}(z^{\ast}_{n})
=c_{n}\mu_{-,1}(z^{\ast}_{n}),\tag{\theequation b}
\end{align}
\addtocounter{equation}{1}
\begin{align}\label{cl10-v-1}
&\mathop{\textmd{Res}}_{z=z_{n}}\left(\frac{\phi_{+,1}(z)}{s_{11}(z)}\right)=\frac{\phi_{+,1}(z_{n})}{s_{11}'(z_{n})}=
\frac{A_{-}(z_{n})}{s_{11}'(z_{n})}\phi_{-,2}(z_{n})
=K_{-}(z_{n})\phi_{-,2}(z_{n}),\tag{\theequation a}\\
\label{cl10-v-2}
&\mathop{\textmd{Res}}_{z=z^{\ast}_{n}}\left(\frac{\phi_{+,2}(z)}{s_{22}(z)}\right)=\frac{\phi_{+,2}(z^{\ast}_{n})}{s_{22}'(z^{\ast}_{n})}=
\frac{A_{+}(z^{\ast}_{n})}{s_{22}'(z^{\ast}_{n})}\phi_{-,1}(z^{\ast}_{n})
=K_{+}(z^{\ast}_{n})\phi_{-,1}(z^{\ast}_{n}),\tag{\theequation b}
\end{align}
where $b_{n}=\frac{s_{21}(z_{n})e^{-2i\theta(z_{n})}}{s_{11}'(z_{n})}$, $c_{n}=\frac{s_{12}(z^{\ast}_{n})e^{2i\theta(z^{\ast}_{n})}}{s_{22}'(z^{\ast}_{n})}$, $K_{-}(z_{n})=\frac{A_{-}(z_{n})}{s_{11}'(z_{n})}$ and $K_{+}(z^{\ast}_{n})=\frac{A_{+}(z^{\ast}_{n})}{s_{22}'(z^{\ast}_{n})}$.

According to $(3)$ in the previous Theorem 5, we can easily get $c_{n}^{\ast}=-b_{n}$.

Finally, we discuss the remaining two points of the eigenvalue quartet by using the similarity method, we can obtain
\addtocounter{equation}{1}
\begin{align}\label{cl11-v-a}
&\mathop{\textmd{Res}}_{z=-\frac{u_{0}^{2}}{z^{\ast}_{n}}}\left(\frac{\mu_{+,1}(z)}{s_{11}(z)}\right)=\frac{\mu_{+,1}(-\frac{u_{0}^{2}}{z^{\ast}_{n}})}{s_{11}'(-\frac{u_{0}^{2}}{z^{\ast}_{n}})}=\frac{s_{21}(-\frac{u_{0}^{2}}{z^{\ast}_{n}})e^{-2i\theta(-\frac{u_{0}^{2}}{z^{\ast}_{n}})}}{s_{11}'(-\frac{u_{0}^{2}}{z^{\ast}_{n}})}\mu_{-,2}(-\frac{u_{0}^{2}}{z^{\ast}_{n}})=b_{N+n}\mu_{-,2}(-\frac{u_{0}^{2}}{z^{\ast}_{n}}),\tag{\theequation a}\\
\label{cl11-v-b}
&\mathop{\textmd{Res}}_{z=-\frac{u_{0}^{2}}{z_{n}}}\left(\frac{\mu_{+,2}(z)}{s_{22}(z)}\right)=\frac{\mu_{+,2}(-\frac{u_{0}^{2}}{z_{n}})}{s_{22}'(-\frac{u_{0}^{2}}{z_{n}})}=\frac{s_{12}(-\frac{u_{0}^{2}}{z_{n}})e^{2i\theta(-\frac{u_{0}^{2}}{z_{n}})}}{s_{22}'(-\frac{u_{0}^{2}}{z_{n}})}\mu_{-,1}(-\frac{u_{0}^{2}}{z_{n}})=c_{N+n}\mu_{-,1}(-\frac{u_{0}^{2}}{z_{n}}),\tag{\theequation b}
\end{align}
where $b_{N+n}=\frac{s_{21}(-\frac{u_{0}^{2}}{z^{\ast}_{n}})e^{-2i\theta(-\frac{u_{0}^{2}}{z^{\ast}_{n}})}}{s_{11}'(-\frac{u_{0}^{2}}{z^{\ast}_{n}})}$ and $c_{N+n}=\frac{s_{12}(-\frac{u_{0}^{2}}{z_{n}})e^{2i\theta(-\frac{u_{0}^{2}}{z_{n}})}}{s_{22}'(-\frac{u_{0}^{2}}{z_{n}})}$. According to $(3)$ in the previous theorem 5, we can easily get $c_{N+n}^{\ast}=-b_{N+n}$.

Moreover, differentiating $(4)$, $(5)$ and $(6)$ in Theorem 5, using $(3)$ in Theorem 5, and evaluating at $z=z_{n}$ or $z=z^{\ast}_{n}$, we have
\addtocounter{equation}{1}
\begin{align}\label{cl12-v-a}
&s_{11}'(-\frac{u_{0}^{2}}{z^{\ast}_{n}})=(\frac{z^{\ast}_{n}}{u_{0}})^{2}(\frac{u_{-}}{u_{+}})(s_{11}'(z_{n}))^{\ast},\tag{\theequation a}\\
\label{cl12-v-b}
&s_{22}'(-\frac{u_{0}^{2}}{z_{n}})=(\frac{z_{n}}{u_{0}})^{2}(\frac{u^{\ast}_{-}}{u^{\ast}_{+}})(s_{22}'(z^{\ast}_{n}))^{\ast}.\tag{\theequation b}
\end{align}
After further calculation, we then have
\addtocounter{equation}{1}
\begin{align}\label{cl13-v-a}
&b_{N+n}=(\frac{u_{0}}{z^{\ast}_{n}})^{2}(\frac{u^{\ast}_{+}}{u_{+}})c_{n},\tag{\theequation a}\\
\label{cl13-v-b}
&c_{N+n}=(\frac{u_{0}}{z_{n}})^{2}(\frac{u_{+}}{u^{\ast}_{+}})b_{n}.\tag{\theequation b}
\end{align}

\subsection{Asymptotics as $z\rightarrow\infty$ and $z\rightarrow0$}
Here we needed the asymptotic properties of the scattering matrix and the eigenfunction to define the inverse problem. In addition, the next-to-leading-order behavior for the eigenfunctions will help us rebuild the potential from the solution of the Riemann-Hilbert problem.

We may discover that the asymptotics with NZBCs is more intricate than the asymptotics with ZBCs, but the calculations are lessened in the uniformization variable. It is worth noting that limit $\kappa\rightarrow\infty$ corresponds to $z\rightarrow\infty$ in $C_{\textmd{I}}$ and to $z\rightarrow0$ in $C_{\textmd{II}}$, and we will need both limits. Consider the following formal of extension:
\addtocounter{equation}{1}
\begin{align}\label{cl14-v-a}
&~~~~~~~~~~~~~~~~~~~~~~~~~~~~~\mu_{-}(x,t,z)=\sum_{n=0}^{\infty}\mu^{n}(x,t,z),\tag{\theequation a}\\
\label{cl14-v-b}
&~~~~~~~~~~~~~~~~~~~~~~~~~~~~~~~with~~\mu^{0}(x,t,z)=E_{-},\tag{\theequation b}\\
\label{cl14-v-c}
&\mu^{n+1}(x,t,z)=\int_{-\infty}^{x}E_{-}e^{-i\lambda(x-y)\sigma_{3}}E_{-}^{-1}\Delta U_{-}(y,t)\mu^{n}(x,t,z)e^{i\lambda(x-y)\sigma_{3}}dy.\tag{\theequation c}
\end{align}
Let $A_{d}$ and $A_{o}$ means the diagonal and the off-diagonal parts of a matrix $A$, respectively.

\noindent \textbf{Theorem 6.} \emph{Equations \eqref{cl14-v-a}-\eqref{cl14-v-c} can provides an asymptotic extension of the columns for $\mu_{-}(x,t,z)$ as $z\rightarrow\infty$ in the suitable region of the z-plane. Generally speaking, as long as potential is a continuous derivative
\begin{equation}\label{Pre-31}
\mu^{2m}_{d}O(\frac{1}{z^{m}}),~~~\mu^{2m}_{o}=O(\frac{1}{z^{m+1}}),~~~\mu^{2m+1}_{d}=O(\frac{1}{z^{m+1}}),~~~\mu^{2m+1}_{o}=O(\frac{1}{z^{m+1}}),
\end{equation}
for all $m\in N$.}

\noindent \textbf{Proof:} From equation \eqref{Pre-7} and \eqref{cl14-v-b}, we can easily get
\begin{equation}\label{Pre-32}
\mu^{0}_{d}(x,z)=I,~~~\mu^{0}_{o}(x,z)=-\frac{i}{z}\sigma_{3}U_{-}.
\end{equation}
We first substituting equation \eqref{Pre-32} into equation \eqref{cl14-v-c}, and then separating the diagonal and off-diagonal parts, we can get
\begin{align}\label{Pre-34}
&\mu^{n+1}_{d}(x,z)=\frac{1}{1+\frac{u_{0}^{2}}{z^{2}}}\left[\int_{-\infty}^{x}\left(\Delta U_{-}(y)\mu^{n}_{o}(x,z)+\frac{i\sigma_{3}U_{-}}{z}\Delta U_{-}(y)\mu^{n}_{d}(x,z)\right)dy\right.\notag\\
&~~~~~~~~~~~~\left.-\frac{i\sigma_{3}U_{-}}{z}\int_{-\infty}^{x}e^{-i\lambda(x-y)\sigma_{3}}\left(\Delta U_{-}(y)\mu^{n}_{d}(x,z)+\frac{i\sigma_{3}U_{-}}{z}\Delta U_{-}(y)\mu^{n}_{o}(x,z)\right)e^{i\lambda(x-y)\sigma_{3}}dy\right],\notag\\
&\mu^{n+1}_{o}(x,z)=\frac{1}{1+\frac{u_{0}^{2}}{z^{2}}}\left[-\frac{i\sigma_{3}U_{-}}{z}\int_{-\infty}^{x}\left(\Delta U_{-}(y)\mu^{n}_{o}(x,z)+\frac{i\sigma_{3}U_{-}}{z}\Delta U_{-}(y)\mu^{n}_{d}(x,z)\right)dy\right.\notag\\
&~~~~~~~~~~~~~\left.+\int_{-\infty}^{x}e^{-i\lambda(x-y)\sigma_{3}}\left(\Delta U_{-}(y)\mu^{n}_{d}(x,z)+\frac{i\sigma_{3}U_{-}}{z}\Delta U_{-}(y)\mu^{n}_{o}(x,z)\right)e^{i\lambda(x-y)\sigma_{3}}dy\right].
\end{align}
where we have temporarily ignored $\mu^{l}$ for time dependence for the convenience of calculation. As $z\rightarrow\infty$, the four terms in the RHS of the first expression for equation \eqref{Pre-34} are, respectively,
\begin{equation}\label{Pre-35}
O(\mu^{n}_{o}),~~~O(\frac{\mu^{n}_{d}}{z}),~~~O(\frac{\mu^{n}_{d}}{z^{2}}),~~~O(\frac{\mu^{n}_{o}}{z^{3}}),
\end{equation}
where the last two estimates are obtained by using partial integration, taking advantage for the differentiability of $e^{-i\lambda(x-y)\sigma_{3}}$ ( with $\lambda=\frac{1}{2}(z+\frac{u_{0}^{2}}{z})$ as $z\rightarrow\infty$).

By using the same method, as $z\rightarrow\infty$, the four terms in the RHS of the second expression for equation \eqref{Pre-34} are, respectively,
\begin{equation}\label{Pre-36}
O(\frac{\mu^{n}_{o}}{z}),~~~O(\frac{\mu^{n}_{d}}{z^{2}}),~~~O(\frac{\mu^{n}_{d}}{z}),~~~O(\frac{\mu^{n}_{o}}{z^{2}}).
\end{equation}
Finally, when we make $n$ is either odd or even number, look for the dominant contribution, and one can get
\begin{equation}\label{Pre-37}
\mu^{2m}_{d}=O(\frac{1}{z^{m}}),~~~\mu^{2m}_{o}=O(\frac{1}{z^{m+1}}),~~~\mu^{2m+1}_{d}=O(\frac{1}{z^{m+1}}),~~~\mu^{2m+1}_{o}=O(\frac{1}{z^{m+1}}).
\end{equation}
The proof the above theorem.    $\Box$

After observation, we can see that the above expression is always $\textmd{Im}z\leq0$ for the first column and $\textmd{Im}z\geq0$ for the second column. We can use the same method to prove that $\mu_{+}(x,t,z)$ also has the same results hold.

Below we will discuss the situation of asymptotic as $z\rightarrow0$. Using a similar method, we can get the following theorem.

\noindent \textbf{Theorem 7.} \emph{Equations $(2.49)$ can provides an asymptotic extension of the columns for $\mu_{-}(x,t,z)$ as $z\rightarrow0$ in the suitable region of the $z$-plane, with
\begin{equation}\label{Pre-38}
\mu^{2m}_{d}=O(z^{m}),~~~\mu^{2m}_{o}=O(z^{m-1}),~~~\mu^{2m+1}_{d}=O(z^{m}),~~~\mu^{2m+1}_{o}=O(z^{m}),
\end{equation}
for all $m\in N$.}

Based on theorems 6 and 7, We can show that as $z\rightarrow\infty$,
\begin{equation}\label{Pre-39}
\mu_{-}(x,t,z)=I-\frac{i}{z}\sigma_{3}U+O(\frac{1}{z^{2}}),
\end{equation}
as $z\rightarrow0$,
\begin{equation}\label{Pre-40}
\mu_{-}(x,t,z)=-\frac{i}{z}\sigma_{3}U+O(1).
\end{equation}

Based on equation \eqref{Pre-39} and \eqref{Pre-40}, we will reconstruct the scattering potential $U(x,t)$ from the perspective of the solution for the inverse problem.

Finally, we take the resulting Jost eigenfunctions \eqref{Pre-39} into \eqref{Pre-19}, one shows that, as $z\rightarrow\infty$ in the appropriate areas of the $z$-plane,
\begin{equation}\label{Pre-41}
S(z)=I+O(\frac{1}{z}).
\end{equation}
Based on the above results, we can obtain the above estimate holds with $\textmd{Im}z\leq0$ and $\textmd{Im}z\geq0$ for $s_{11}$ and $s_{22}$, respectively, and with $\textmd{Im}z=0$ for $s_{12}$ and $s_{21}$. By using the same method, we shows that, as $z\rightarrow0$ in the appropriate areas of the $z$-plane,
\begin{equation}\label{Pre-42}
S(z)=\textmd{diag}(-\frac{u_{-}}{u_{+}},-\frac{u_{+}}{u_{-}})+O(z).
\end{equation}

\section{Inverse scattering problem: simple poles}

\subsection{Riemann-Hilbert problem}
As before, the formulation of the inverse problem begins from equation \eqref{Pre-15}, which we consider as the relation between the eigenfunction analysis in $D^{-}$ and the eigenfunction resolution in $D^{+}$. Therefore, we present the sectionally meromorphic matrices
\begin{equation}\label{Ip-1}
M^{-}(x,t,z)=\left(\frac{\mu_{+,1}}{s_{11}},\mu_{-,2}\right),~~~~~M^{+}(x,t,z)=\left(\mu_{-,1},\frac{\mu_{+,2}}{s_{22}}\right).
\end{equation}
From \eqref{Pre-16}, we can get the jump condition
\begin{equation}\label{Ip-2}
M^{+}(x,t,z)=M^{-}(x,t,z)(I-G(x,t,z)),~~z\in\Sigma,
\end{equation}
where
\begin{equation}\label{Ip-3}
G(x,t,z)=\left(\begin{array}{cc}
    0 & -e^{2i\theta(x,t,z)}\widetilde{\rho}(z)\\
    e^{-2i\theta(x,t,z)}\rho(z) & \rho(z)\widetilde{\rho}(z)\\
\end{array}\right),
\end{equation}
with the reflection coefficients is
\begin{equation}\label{Ip-4}
\rho(z)=\frac{s_{21}}{s_{11}},~~~~\widetilde{\rho}(z)=\frac{s_{12}}{s_{22}}.
\end{equation}
Equations \eqref{Ip-1}-\eqref{Ip-3} define a matrix, multiplicative, and homogeneous Riemann-Hilbert problem (RHP), respectively.

As before, in order to complete the formulation of the RHP, we need a normalization condition. In this case, we get the asymptotic behavior of $M^{\pm}$ to be $z\rightarrow\infty$. Combining the asymptotic behavior of the Jost eigenfunctions and scattering coefficients, we can easily obtain
\begin{equation}\label{Ip-5}
M^{\pm}=I+O\left(\frac{1}{z}\right),~~~~z\rightarrow\infty.
\end{equation}
Using the same method, we can get
\begin{equation}\label{Ip-6}
M^{\pm}=-\frac{i}{z}\sigma_{3}U_{\pm}+O(1),~~~~z\rightarrow0.
\end{equation}
Therefore, as with the NLSLab equation with NZBCs, in addition to considering the behavior at $z=\infty$ and the poles from the discrete spectrum, we need to subtract the pole at $z=0$ to facilitate the regular RHP.

In order to solve the RHP, we needs to normalize it by reducing the asymptotic behavior and the pole contributions. As we have seen earlier that discrete eigenvalues appear in symmetric quartets [cf. \eqref{Pre-29}]. Then, in order to facilitate our calculations in the following, we define $\zeta_{n}=z_{n}$ and $\zeta_{N+n}=-\frac{u_{0}^{2}}{z^{\ast}_{n}}$ for $n=1,2,\ldots,N$, and rewrite the above equation \eqref{Ip-2} as
\begin{align}\label{Ip-7}
&~~~~~M^{+}-I+\frac{i}{z}\sigma_{3}U_{+}
-\sum_{n=1}^{2N}\frac{\textmd{Res}_{\zeta^{\ast}_{n}}M^{+}}{z-\zeta^{\ast}_{n}}
-\sum_{n=1}^{2N}\frac{\textmd{Res}_{\zeta_{n}}M^{-}}{z-\zeta_{n}}\notag\\
&=M^{-}-I+\frac{i}{z}\sigma_{3}U_{+}
-\sum_{n=1}^{2N}\frac{\textmd{Res}_{\zeta_{n}}M^{-}}{z-\zeta_{n}}
-\sum_{n=1}^{2N}\frac{\textmd{Res}_{\zeta^{\ast}_{n}}M^{+}}{z-\zeta^{\ast}_{n}}-M^{-}G.
\end{align}
From equation \eqref{Ip-7}, we can get that the left-hand side of equation \eqref{Ip-7} is analytic in $D^{+}$, and $O(\frac{1}{z})$ as $z\rightarrow\infty$, whereas the sum of the first four terms of the right-hand side of equation \eqref{Ip-7} is analytic in $D^{-}$, and $O(\frac{1}{z})$ as $z\rightarrow\infty$. Finally, the asymptotic behavior of the off-diagonal scattering coefficients means that $G(x,t,z)$ is $O(\frac{1}{z})$ as $z\rightarrow\pm\infty$, and $O(z)$ as $z\rightarrow0$ along the real axis. Then, we introduced the Cauchy projector $Q_{\pm}[f](z)$ over $\Sigma$:
\begin{equation}\label{Ip-8}
Q_{\pm}[f](z)=\frac{1}{2\pi i}\int\frac{f(s)}{s-(z\pm i0)}ds,
\end{equation}
where $\int_{\Sigma}$ represents the integral along the oriented contour shown in Figure 1, and the notation $z\pm i0$ represents that when $z\in\Sigma$, the limit is chosen from the left/right of it. Now we need to combine the Plemelj's formulae: if $f^{\pm}$ are  analytic in $D^{\pm}$, and are $O(\frac{1}{z})$ as $z\rightarrow\pm\infty$, one has $Q^{\pm}f^{\pm}=\pm f^{\pm}$ and $Q^{+}f^{-}=Q^{-}f^{+}=0$. Applying $Q^{+}$ and $Q^{-}$ to equation \eqref{Ip-7}, we can get
\begin{align}\label{Ip-9}
&M(x,t,z)=I-\frac{i}{z}\sigma_{3}U_{+}
+\sum_{n=1}^{2N}\frac{\textmd{Res}_{\zeta^{\ast}_{n}}M^{+}}{z-\zeta^{\ast}_{n}}
+\sum_{n=1}^{2N}\frac{\textmd{Res}_{\zeta_{n}}M^{-}}{z-\zeta_{n}}\notag\\
&~~~~~~~~~~~~~~~~+\frac{1}{2\pi i}\int_{\Sigma}\frac{M^{-}(x,t,s)}{s-z}G(x,t,s)ds,~~~~z\in C\setminus\Sigma.
\end{align}
Here, the expressions of $M^{+}$ and $M^{-}$ have the same meaning as before, except that the integrals appearing in the right-hand side represent $Q^{+}$ and $Q^{-}$ projectors, respectively.

\subsection{Residue conditions and reconstruction formula}
In order to close the system, we need to get the expression of the residues appearing in equation \eqref{Ip-9}. Among the residue relations \eqref{cl10-v-a} and \eqref{cl11-v-a}, we know that only the first column of $M^{-}$ has poles at $z=z_{n}$ and $z=-\frac{u_{0}^{2}}{z^{\ast}_{n}}$, and the pole at which the residual is proportional to the second column of $M^{-}$. On the other hand, in the residual relations \eqref{cl10-v-b} and \eqref{cl11-v-b}, we know that only the second column of $M^{+}$ has poles at $z=z^{\ast}_{n}$ and $z=-\frac{u_{0}^{2}}{z_{n}}$, and its residue is proportional to the first column of $M^{+}$ at that point. Specifically,
\addtocounter{equation}{1}
\begin{align}\label{Ip1-v-a}
&\textmd{Res}_{\zeta_{n}}M^{-}=\textmd{Res}_{\zeta_{n}}(\frac{\mu_{+,1}(\zeta_{n})}{s_{11}},\mu_{-,2}(\zeta_{n}))
=(b_{n}\mu_{-,2}(\zeta_{n}),0),~~~n=1,2,\ldots,2N,\tag{\theequation a}\\
\label{Ip1-v-b}
&\textmd{Res}_{\zeta^{\ast}_{n}}M^{+}=\textmd{Res}_{\zeta^{\ast}_{n}}(\mu_{-,1}(\zeta^{\ast}_{n}),\frac{\mu_{+,2}(\zeta^{\ast}_{n})}{s_{22}})
=(0,c_{n}\mu_{-,1}(\zeta^{\ast}_{n})),~~~n=1,2,\ldots,2N.\tag{\theequation b}
\end{align}
Therefore, we can assess the second column of equation \eqref{Ip-9} at $z=z_{n}$ and $z=-\frac{u_{0}^{2}}{z^{\ast}_{n}}$, and get
\begin{align}\label{Ip-10}
&\mu_{-,2}(x,t,\zeta_{n})=\left(\begin{aligned}
&-\frac{i}{\zeta_{n}}u_{+}\\ &~~~~~1\\
\end{aligned}\right)
+\sum_{n=1}^{2N}\frac{c_{n}\mu_{-,1}(x,t,\zeta^{\ast}_{n})}{\zeta_{n}-\zeta^{\ast}_{n}}+\frac{1}{2\pi i}\int_{\Sigma}\frac{M^{-}(x,t,s)}{s-\zeta_{n}}G(x,t,s)ds,\notag\\
&~~~~~~~~~~~~~~~~~~~~~~~~~~~~~~~~~~~~~~~~~~~~~~n=1,2,\ldots,2N.
\end{align}
Using the same method, we know that when $z=z^{\ast}_{n}$ or $z=-\frac{u_{0}^{2}}{z_{n}}$, $M$ is resolved on $D^{+}$, thus:
\begin{align}\label{Ip-11}
&\mu_{-,1}(x,t,\zeta^{\ast}_{n})=\left(\begin{aligned}
&~~~~~~~1\\ &-\frac{i}{\zeta^{\ast}_{n}}u^{\ast}_{+}\\
\end{aligned}\right)
+\sum_{n=1}^{2N}\frac{b_{n}\mu_{-,2}(x,t,\zeta_{n})}{\zeta^{\ast}_{n}-\zeta_{n}}+\frac{1}{2\pi i}\int_{\Sigma}\frac{M^{-}(x,t,s)}{s-\zeta^{\ast}_{n}}G(x,t,s)ds,\notag\\
&~~~~~~~~~~~~~~~~~~~~~~~~~~~~~~~~~~~~~~~~~~~~~~n=1,2,\ldots,2N.
\end{align}
Finally, based on equation \eqref{Ip-10} and equation \eqref{Ip-11}, we can obtain $M^{-}(x,t,z)$ and by using equation \eqref{Ip-9} for $z\in\Sigma$, and obtain a closed linear system for algebraic-integral equations of the solution for the RHP. We hope that the solvability conditions of the RHP can be established by using techniques similar to Ref.\cite{B-1984-37}.

Next, we will reconstruct the potential form for the solution of the RHP. From equation \eqref{Ip-9}, we can get the asymptotic behavior of $M^{\pm}$ as
\begin{align}\label{Ip-12}
&M(x,t,z)=I+\frac{1}{z}\left[-i\sigma_{3}U_{+}
+\sum_{n=1}^{2N}\textmd{Res}_{\zeta^{\ast}_{n}}M^{+}
+\sum_{n=1}^{2N}\textmd{Res}_{\zeta_{n}}M^{-}\right.\notag\\
&\left.~~~~~~~~~~~~~~~~-\frac{1}{2\pi i}\int_{\Sigma}M^{-}(x,t,s)G(x,t,s)ds\right]+O(\frac{1}{z^{2}}),~~~~z\rightarrow\infty,
\end{align}
where $\textmd{Res}_{\zeta_{n}}M^{-}$ and $\textmd{Res}_{\zeta^{\ast}_{n}}M^{+}$ are given by \eqref{Ip1-v-a} and \eqref{Ip1-v-b}, respectively. We take $M=M^{\pm}$  and contrasting the 1, 2 element of equation \eqref{Ip-12} with \eqref{Pre-39}, and then we can get the reconstruction formula of the potential as:
\begin{equation}\label{Ip-12a}
u(x,t)=u_{+}+i\sum_{n=1}^{2N}c_{n}\mu_{-,11}(\zeta^{\ast}_{n})
-\frac{1}{2\pi}\int_{\Sigma}\left(M^{-}(x,t,s)G(x,t,s)\right)_{12}ds.
\end{equation}
Because of the previous knowledge, we know that the Jost eigenfunction is obtained by using simultaneous solution of both parts of the Lax pair, so the time dependence of the solution is automatically considered.

\subsection{Trace formulae and ``theta" condition}
From theorem 3, we can obtain $s_{11}$ and $s_{22}$ are analytic in $D^{-}$ and in $D^{+}$, respectively. Moreover, in the previous calculations, it is also proved that the discrete spectrum is composed of quartets: $z_{n}$, $z^{\ast}_{n}$, $-\frac{u_{0}^{2}}{z_{n}}$, $-\frac{u_{0}^{2}}{z^{\ast}_{n}}$ ($\forall n=1,2,\ldots,N$). Then the functions
\begin{equation}\label{Ip-13}
\beta^{-}(z)=s_{11}(z)\prod_{n=1}^{N}\frac{(z-z^{\ast}_{n})(z+\frac{u_{0}^{2}}{z_{n}})}{(z-z_{n})(z+\frac{u_{0}^{2}}{z^{\ast}_{n}})},~~~~~
\beta^{+}(z)=s_{22}(z)\prod_{n=1}^{N}\frac{(z-z_{n})(z+\frac{u_{0}^{2}}{z^{\ast}_{n}})}{(z-z^{\ast}_{n})(z+\frac{u_{0}^{2}}{z_{n}})},
\end{equation}
are analytic in $D^{-}$ and in $D^{+}$, respectively. But, $\beta^{-}(z)$ and $\beta^{+}(z)$ are not the same as $s_{11}(z)$ and $s_{22}(z)$ because they have no zeros. In addition,  $\beta^{\pm}(z)$ as $z\rightarrow\infty$ in the appropriate domains. Finally, for all $z\in\Sigma$, we can get $\beta^{-}(z)\beta^{+}(z)=s_{11}(z)s_{22}(z)$, and by \eqref{Pre-17} we can get $\frac{1}{s_{11}(z)s_{22}(z)}=1-\rho(z)\widetilde{\rho}(z)=1+\rho(z)\rho^{\ast}(z^{\ast})$, which means
\begin{equation}\label{Ip-14}
\beta^{-}(z)\beta^{+}(z)=\frac{1}{1-\rho(z)\widetilde{\rho}(z)}
=\frac{1}{1+\rho(z)\rho^{\ast}(z^{\ast})},~~~~z\in\Sigma.
\end{equation}
We notice that since $\Sigma$ is not just the real $z$-axis, the above expression \eqref{Ip-14} is not reduced to $1+|\rho(z)|^{2}$. Equation \eqref{Ip-14} can be seen as a jump condition of a scalar, multiplicative and  homogenous RHP. Taking the logarithms and using the Cauchy projectors, we can get
\begin{equation}\label{Ip-15}
\textmd{ln}\beta^{\pm}(z)=\mp\frac{1}{2\pi i}\int_{\Sigma}\frac{\textmd{ln}(1+\rho(s)\rho^{\ast}(s^{\ast}))}{s-z}ds,~~~~z\in D^{\pm}.
\end{equation}
Substituting \eqref{Ip-15} into \eqref{Ip-13}, we can obtain the so-called ``trace" formula:
\addtocounter{equation}{1}
\begin{align}\label{Ip2-v-a}
&s_{11}(z)=\exp\left[\frac{1}{2\pi i}\int_{\Sigma}\frac{\textmd{ln}(1+\rho(s)\rho^{\ast}(s^{\ast}))}{s-z}ds\right]
\prod_{n=1}^{N}\frac{(z-z_{n})(z+\frac{u_{0}^{2}}{z^{\ast}_{n}})}{(z-z^{\ast}_{n})(z+\frac{u_{0}^{2}}{z_{n}})},
~~~~~z\in D^{-},\tag{\theequation a}\\
\label{Ip2-v-b}
&s_{22}(z)=\exp\left[-\frac{1}{2\pi i}\int_{\Sigma}\frac{\textmd{ln}(1+\rho(s)\rho^{\ast}(s^{\ast}))}{s-z}ds\right]
\prod_{n=1}^{N}\frac{(z-z^{\ast}_{n})(z+\frac{u_{0}^{2}}{z_{n}})}{(z-z_{n})(z+\frac{u_{0}^{2}}{z^{\ast}_{n}})},
~~~z\in D^{+},\tag{\theequation b}
\end{align}
which the analytical scattering coefficient is expressed by the reflection coefficient and the discrete eigenvalues. Therefore, in the special case for reflectionless solution, $s_{12}(z)=s_{21}(z)\equiv 0$ $(\forall z\in\Sigma)$, and the integrals in \eqref{Ip2-v-a} and \eqref{Ip2-v-b} are all zero.

Next, we can derive the asymptotic phase difference of the boundary values $u_{+}$ and $u_{-}$(also referred to as ``theta" condition) by using the obtained trace formulae. From equation \eqref{Pre-42}, we obtain $s_{11}(z)=-\frac{u_{-}}{u_{+}}$. In equation \eqref{Ip2-v-a}, we take $z\rightarrow 0$, and we can obtain
\begin{equation}\label{Ip-16}
\textmd{arg}(-\frac{u_{-}}{u_{+}})=4\sum_{n=1}^{N}\textmd{arg}z_{n}+\frac{1}{2\pi }\int_{\Sigma}\frac{\textmd{ln}(1+\rho(s)\rho^{\ast}(s^{\ast}))}{s}ds,
\end{equation}
which relates the phase difference between the asymptotic values for the potential to the reflection coefficient and the discrete spectrum. Note that
\begin{equation}\label{Ip-17}
\int_{u_{0}}^{\infty}\frac{\textmd{ln}(1+|\rho(s)|^{2})}{s}ds=-\int_{-u_{0}}^{0}\frac{\textmd{ln}(1+|\rho(s)|^{2})}{s}ds,
\end{equation}
because $|\rho(s)|=|\rho(\frac{-u_{0}^{2}}{s})|$ as a result of the symmetry \eqref{cl8-v-e} and \eqref{cl8-v-f}. We will find that a similar relation holds between the integral form $0$ to $u_{0}$ and from $-\infty$ to $-u_{0}$. Finally, we can get
\begin{equation}\label{Ip-18}
\int_{C_{0}^{+}}\frac{\textmd{ln}(1+|\rho(s)|^{2})}{s}ds=-\int_{C_{0}^{-}}\frac{\textmd{ln}(1+|\rho(s)|^{2})}{s}ds,
\end{equation}
where $C_{0}^{+}$ and $C_{0}^{-}$ denote semicircles of the upper half and the lower half of the radius $u_{0}$, respectively. However, since the orientation of the contributions of these individuals are additive to each other and do not cancel each other out, this means that the reflection coefficient contribute to the asymptotic phase difference in principle.

\subsection{Reflectionless potentials}
We now see the potentials $u(x,t)$ of which the reflection coefficient $\rho(z)$ vanishes identically. We also notice that in this case there is no jump from $M^{+}$ to $M^{-}$ thwart the continuous spectrum, so that the inverse problem can be reduced to an algebraic system, in which such a solution can obtain the soliton solution of the integrable nonlinear equation.

In order to simplify the calculation, we introduce the quantities
\begin{equation}\label{Ip-19}
d_{k}(z)=\frac{b_{k}(z)}{z-\zeta_{k}},~~~~k=1,2,\ldots,2N.
\end{equation}
Further, we can obtain
\addtocounter{equation}{1}
\begin{align}\label{Ip3-v-a}
&d_{k}(\zeta^{\ast}_{n})=\frac{b_{k}(\zeta^{\ast}_{n})}{\zeta^{\ast}_{n}-\zeta_{k}},
~~~~k=1,2,\ldots,2N,\tag{\theequation a}\\
\label{Ip3-v-b}
&d^{\ast}_{k}(\zeta^{\ast}_{n})=\left[\frac{b_{k}(\zeta^{\ast}_{n})}{\zeta^{\ast}_{n}-\zeta_{k}}\right]^{\ast}
=\frac{b^{\ast}_{k}(\zeta^{\ast}_{n})}{\zeta_{n}-\zeta^{\ast}_{k}}
=-\frac{c_{k}(\zeta_{n})}{\zeta_{n}-\zeta^{\ast}_{k}},
~~~~k=1,2,\ldots,2N.\tag{\theequation b}
\end{align}
In equation \eqref{Ip-12a}, we will find that only the first component $\mu_{-,11}$ of the eigenfunction is demanded in the reconstruction formula. Then the algebraic system get  from the perspective of the inverse problem is the represented as
\addtocounter{equation}{1}
\begin{align}\label{Ip4-v-a}
&\mu_{-,12}(\zeta_{j})
=-\frac{i}{\zeta_{j}}u_{+}-\sum_{k=1}^{2N}d^{\ast}_{k}(\zeta^{\ast}_{j})\mu_{-,11}(\zeta^{\ast}_{k}),
~~~~j=1,2,\ldots,2N,\tag{\theequation a}\\
\label{Ip4-v-b}
&\mu_{-,11}(\zeta^{\ast}_{n})
=1+\sum_{j=1}^{2N}d_{j}(\zeta^{\ast}_{n})\mu_{-,12}(\zeta_{j}),
~~~~~~~~~~~n=1,2,\ldots,2N,\tag{\theequation b}
\end{align}
where we have omitted the $x$ and $t$ dependencies for the simplicity of writing. Substituting \eqref{Ip4-v-a} into \eqref{Ip4-v-b}, and one get
\begin{equation}\label{Ip-20}
\mu_{-,11}(\zeta^{\ast}_{n})
=1-iu_{+}\sum_{j=1}^{2N}\frac{d_{j}(\zeta^{\ast}_{n})}{\zeta_{j}}
-\sum_{j=1}^{2N}\sum_{k=1}^{2N}d_{j}(\zeta^{\ast}_{n})d^{\ast}_{k}(\zeta^{\ast}_{j})\mu_{-,11}(\zeta^{\ast}_{k}),
~~~~n=1,2,\ldots,2N.
\end{equation}
Next, we will rewrite this system in matrix form. First, let's introducing $\textbf{X}=(X_{1},X_{2},\ldots,X_{2N})^{T}$ and $\textbf{B}=(B_{1},B_{2},\ldots,B_{2N})^{T}$, where
\begin{equation}\label{Ip-21}
X_{n}=\mu_{-,11}(\zeta^{\ast}_{n}),~~~B_{n}=1-iu_{+}\sum_{j=1}^{2N}\frac{d_{j}(\zeta^{\ast}_{n})}{\zeta_{j}},
~~~~n=1,2,\ldots,2N.
\end{equation}
and the $2N\times 2N$  matrix $A=(A_{n,k})$, where
\begin{equation}\label{Ip-22}
A_{n,k}=\sum_{j=1}^{2N}d_{j}(\zeta^{\ast}_{n})d^{\ast}_{k}(\zeta^{\ast}_{j}),
~~~~n,k=1,2,\ldots,2N.
\end{equation}
Next we convert equation \eqref{Ip-20} to $\textbf{M}\textbf{X}=\textbf{B}$, where $\textbf{M}=I+A=(M_{1},M_{2},\ldots,M_{2N})$. Below we will use the Cramer's Rule to get the solution of the system to $X_{n}=\frac{\textmd{det}M_{n}^{rep}}{\textmd{det}\textbf{M}}$ for $n=1,2,\ldots,2N$, where $M_{n}^{rep}=(M_{1},\ldots,M_{n-1},\textbf{B},M_{n+1},\ldots,M_{2N})$.

Finally, when we substituting the above result $X_{1},X_{2},\ldots,X_{2N}$ into the reconstruction formula, we can compactly written the resulting expression for the potential into the following form as
\begin{equation}\label{Ip-23}
u(x,t)=u_{+}+i\sum_{n=1}^{2N}c_{n}\mu_{-,11}(\zeta^{\ast}_{n})
=u_{+}+i\sum_{n=1}^{2N}c_{n}\frac{\textmd{det}M_{n}^{rep}}{\textmd{det}\textbf{M}}
=u_{+}+i\frac{\textmd{det}\textbf{M}^{aug}}{\textmd{det}\textbf{M}},
\end{equation}
where the matrix $\textbf{M}^{aug}$ is a $(2N+1)\times(2N+1)$ and is given by
\begin{equation}\label{Ip-24}
\textbf{M}^{aug}=\left(\begin{array}{cc}
    0 & \textbf{Y}\\
    \textbf{B} & \textbf{M}\\
\end{array}\right),~~~~~~\textbf{Y}=(c_{1},c_{2},\ldots,c_{2N}).
\end{equation}
In the calculation process, we found that even if the discrete eigenvalues arise in the quartets of the NZBCs case rather than in the ZBCs case, the number of unknowns contained in the inverse problem are still the same as the unknowns in the ZBCs case.

\section{Soliton solutions}
In Ref.\cite{LBK-2019-383}, the NLSLab equation has been solved for the soliton solution. Next, we should note that in view of these soliton solutions, the relationship between MI and rogue waves, it is found that these soliton solutions may be related to the rogue waves in water wave as well as optics.

\subsection{Stationary solitons}
The first thing, we discussed was the one-soliton solution: $N=1$. According to the previous symmetry description, we can get the NLSLab equation possesses a scaling symmetry. That is, if $u(x,t)$ is a solution of the NLSLab equation, then for any $a\in R$, $au(ax,a^{2}t)$ is also the solution of the NLSLab equation. Therefore, in the later calculation process, we can assume $u_{0}=1$.

The first thing we discuss the case of purely imaginary eigenvalue. Let $z_{1}=\frac{3i}{2}$, and $b_{1}=e^{-2i\theta(z_{1})}$, with $\xi,\varphi\in R$. From equation \eqref{Ip-16}, we can get a corresponding progressive phase difference is $2\pi$,  in other words,  no  asymptotic phase difference in this case. From the general $N$ solution formula \eqref{Ip-23}, we can get
\begin{equation}\label{Ss-1}
u(x,t)=-\frac{\frac{169}{225}e^{\eta_{1}}e^{\eta_{2}}-\frac{16}{45}ie^{\eta_{1}}
+\frac{9}{5}ie^{\eta_{2}}+1}{\frac{169}{225}e^{\eta_{1}}e^{\eta_{2}}-\frac{4}{5}ie^{\eta_{1}}
+\frac{4}{5}ie^{\eta_{2}}+1},
\end{equation}
where
\begin{equation}\label{Ss-2}
\eta_{1}=-\frac{5}{6}x+(\frac{1}{12}+\frac{65}{36}i)t,~~~
\eta_{2}=-\frac{5}{6}x+(\frac{1}{12}-\frac{65}{36}i)t,
\end{equation}
which is an expression of one-soliton solutions (also referred to as a time-periodic breather solutions) showed in Fig.2(a). We can the fixed $c=0.1$ first and then choose another NZBCs. For example, for $u_{+}=-0.5$ and $u_{+}=-0.05$, we can get the corresponding breather solutions of the NLSLab equation (see Figs.2(b) and (c)). We can see from the above Figs.2(a)-(c) that when the non-zero parameter $u_{+}$ becomes larger, the periodic behavior of the breather will only appears in its top part, and the maximal amplitude in the background will also gradually decreases. In particular, when $u_{0}\rightarrow0$, the breather solutions the NLSLab equation of NZBCs becomes the bright soliton of the NLSLab equation for ZBCs.(see Fig.2(d)).

\noindent
{\rotatebox{0}{\includegraphics[width=3.5cm,height=3.0cm,angle=0]{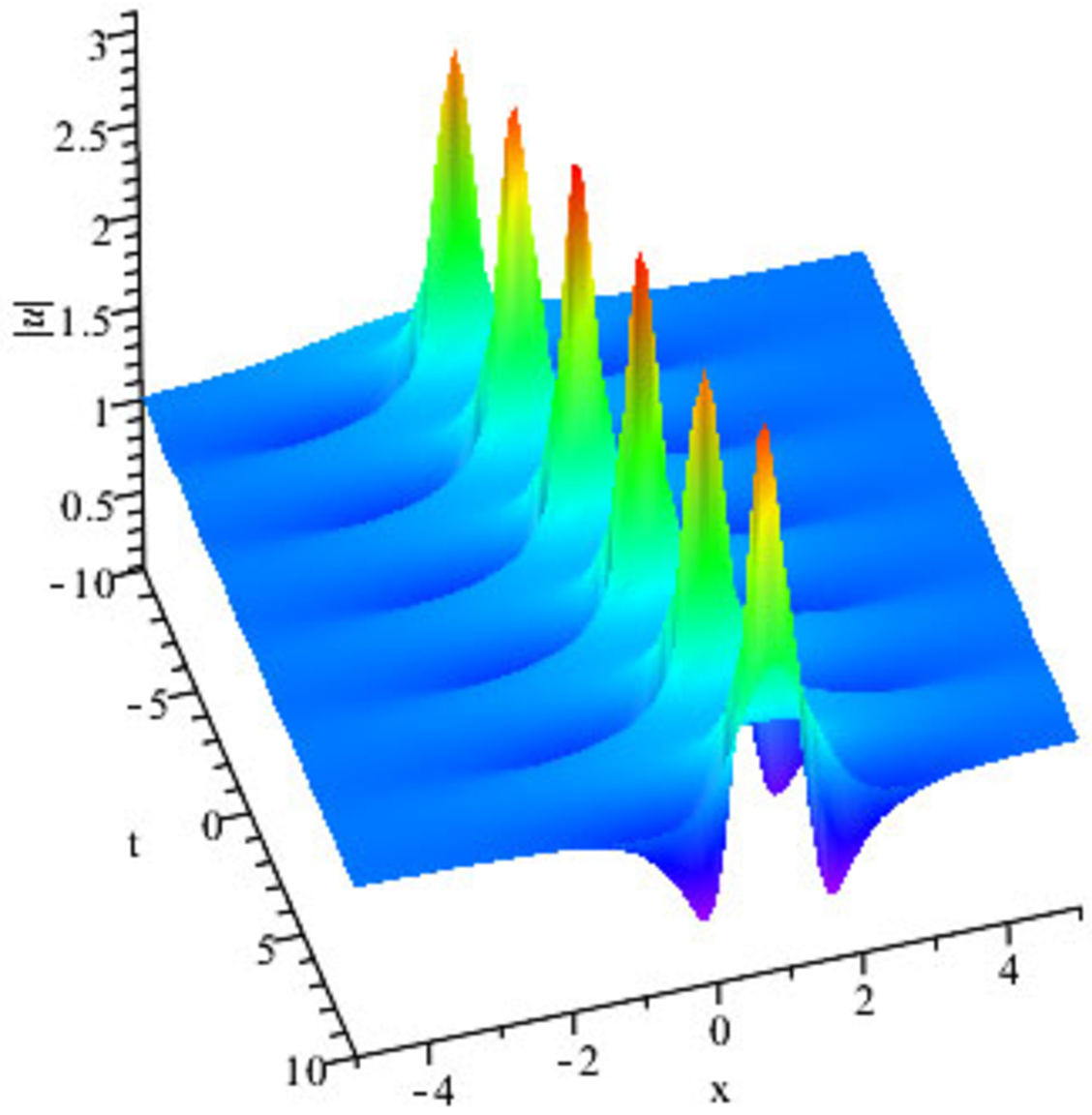}}}
~~~~
{\rotatebox{0}{\includegraphics[width=3.5cm,height=3.0cm,angle=0]{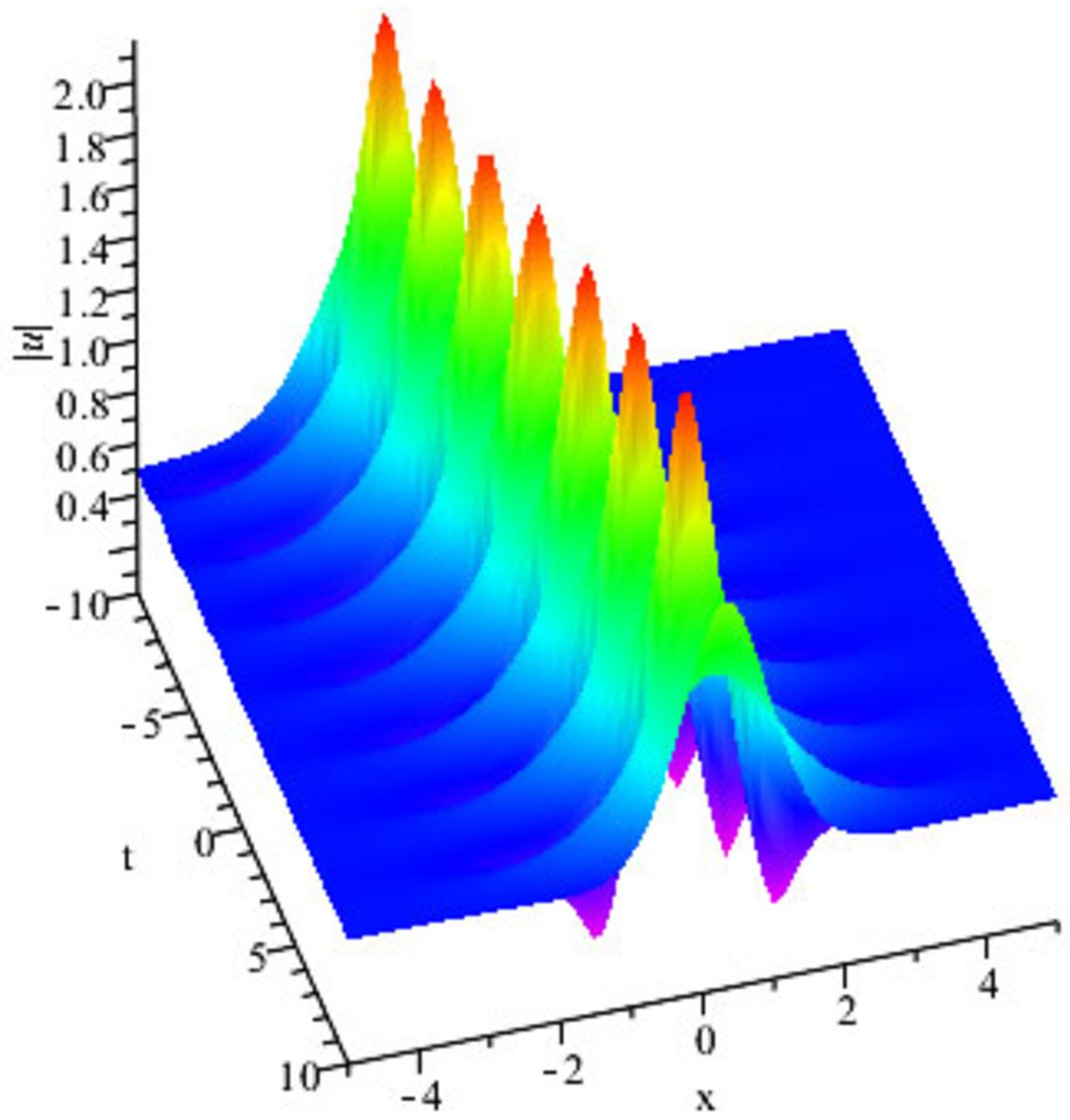}}}
~~~~
{\rotatebox{0}{\includegraphics[width=3.5cm,height=3.0cm,angle=0]{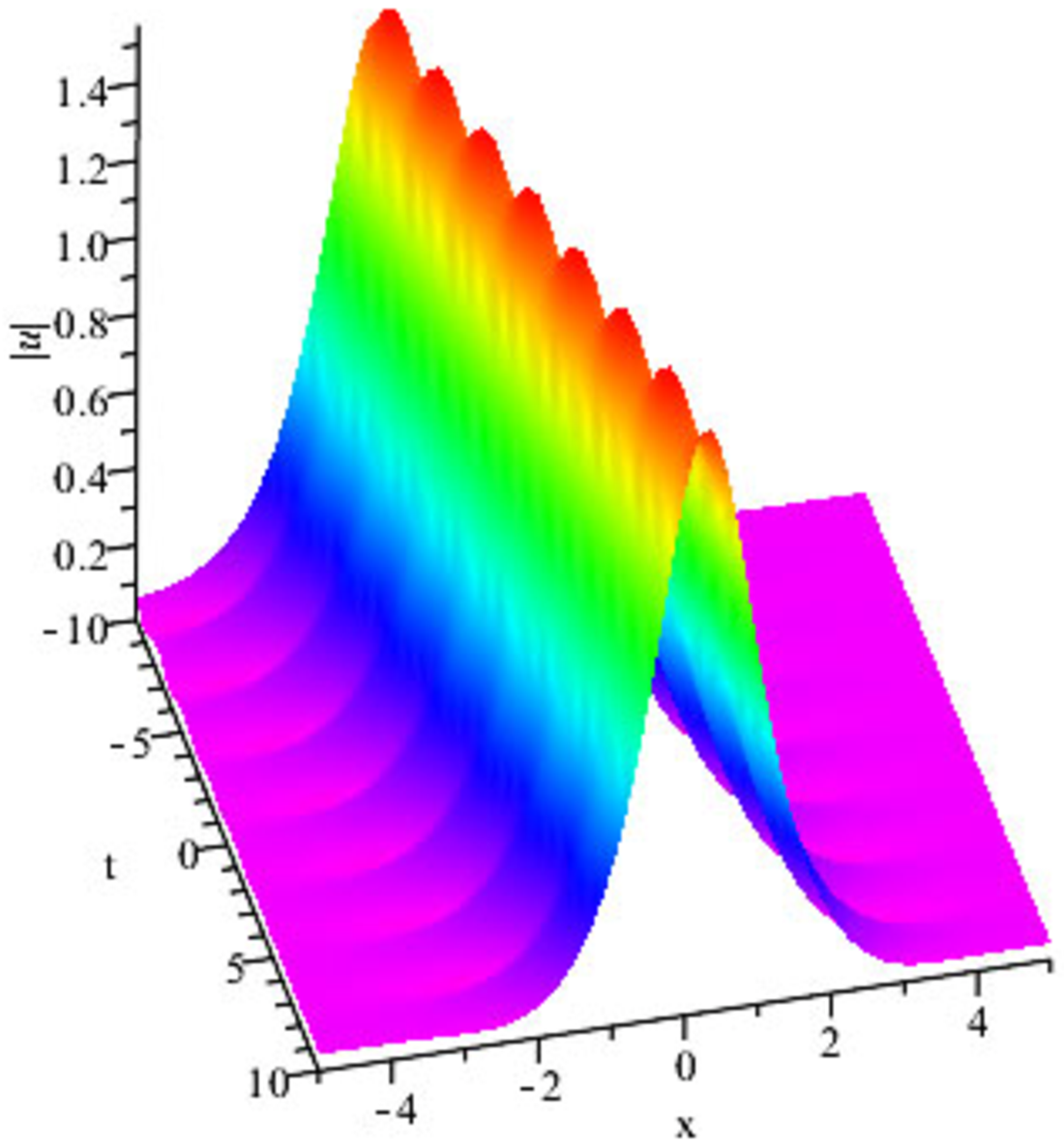}}}
~~~~
{\rotatebox{0}{\includegraphics[width=3.5cm,height=3.0cm,angle=0]{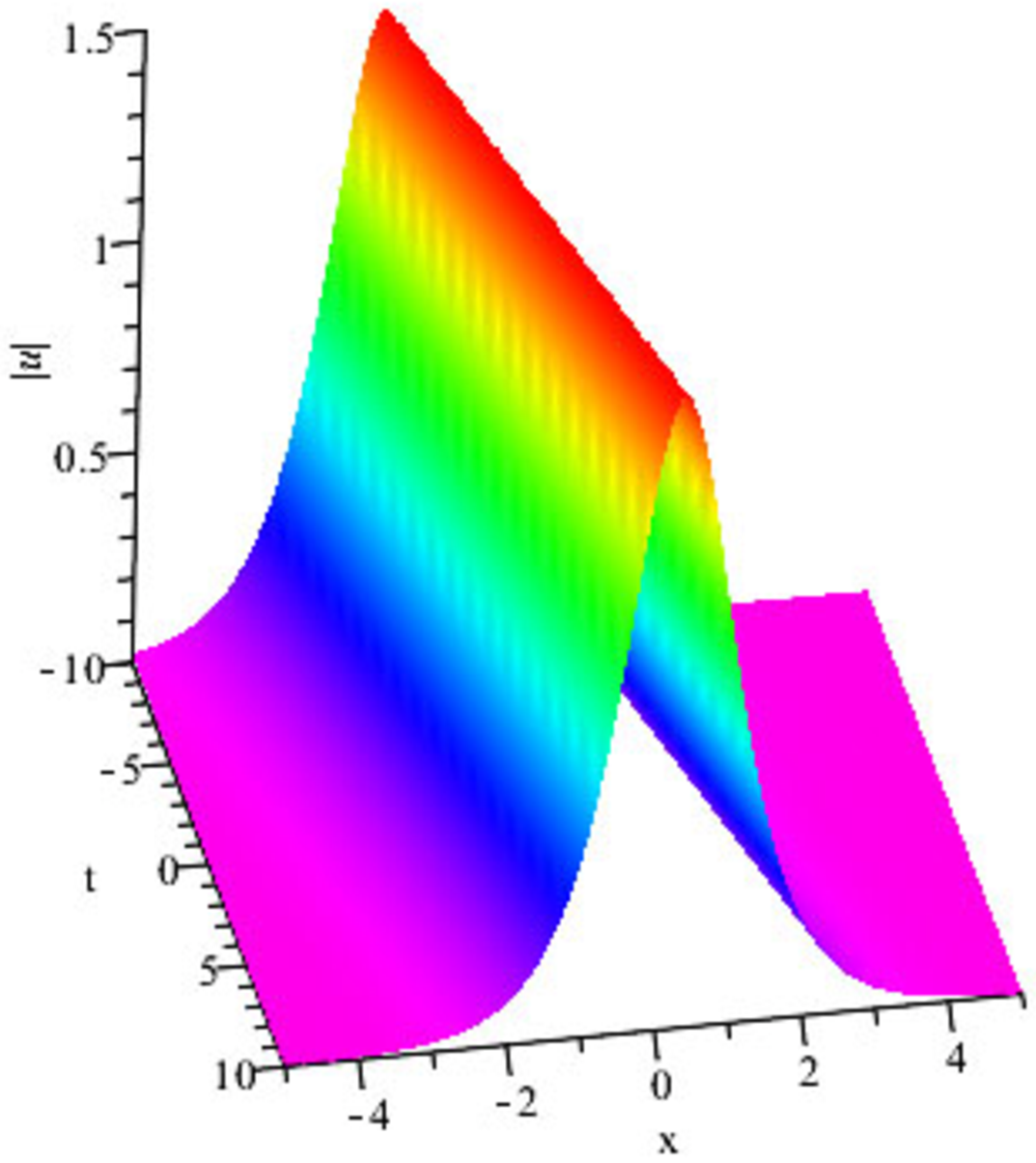}}}
$~~~~~~~~~~~(\textbf{a})~~
~~~~~~~~~~~~~~~~~~~~~~~~(\textbf{b})
~~~~~~~~~~~~~~~~~~~~~~~~~~~~~~(\textbf{c})
~~~~~~~~~~~~~~~~~~~~~~~~~~~(\textbf{d})$\\
\noindent { \small \textbf{Figure 2.} (Color online) Simple-pole stationary one-soliton solutions $u(x,t)$ of the NLSLab equation \eqref{NLS-1}  by choosing suitable parameters:
 $c=0.1, N=1, z_{1}=\frac{3i}{2}, b_{1}=e^{-2i\theta(z_{1})}.$
 (a): the breather solutions $u(x,t)$ with NZBCs and parameters $u_{+}=-1$; (b): the breather solution $u(x,t)$ with NZBCs and parameters $u_{+}=-0.5$; (c): the breather solutions $u(x,t)$ with NZBCs and parameters $u_{+}=-0.05$; (d): the breather solutions $u(x,t)$ with NZBCs and parameters $u_{+}\rightarrow0$.}

\subsection{Non-stationary solitons}
Next, we will discuss the one-soliton solutions ($N=1$) obtained for a generic position of discrete eigenvalues. By using the same method, we set $u_{0}=1$ by the scaling invariance of the NLSLab equation then without loss of generality. For $z_{1}=ae^{i\upsilon}$ with $a>1$ and $\upsilon\in(0, \frac{\pi}{2})\cup(\frac{\pi}{2}, \pi)$, the asymptotic phase difference is $\textmd{arg}(-\frac{u_{-}}{u_{+}})=4\upsilon\in(0, 2\pi)$. For example, for $z_{1}=2e^{i\frac{\pi}{4}}$, and $b_{1}=e^{-2i\theta(z_{1})}$, with $\xi,\varphi\in R$, we can get the non-stationary solitons of the NLSLab equation with NZBCs as
\begin{equation}\label{Ss-3}
u(x,t)=\frac{25e^{\eta_{1}}e^{\eta_{2}}-32e^{\eta_{1}}+2e^{\eta_{2}}-128ie^{\eta_{1}}
-8ie^{\eta_{2}}-136}{25e^{\eta_{1}}e^{\eta_{2}}+32e^{\eta_{1}}+32e^{\eta_{2}}-8ie^{\eta_{1}}
+8ie^{\eta_{2}}+136},
\end{equation}
where
\begin{align}\label{Ss-4}
&\eta_{1}=(\frac{1}{4}-i)(-\sqrt{2}ix+\sqrt{2}x+4it+t+\frac{\sqrt{2}i}{10}t-\frac{\sqrt{2}}{10}t),\notag\\
&\eta_{2}=(-\frac{1}{4}-i)(-\sqrt{2}ix-\sqrt{2}x+4it-t+\frac{\sqrt{2}i}{10}t+\frac{\sqrt{2}}{10}t),
\end{align}
and the asymptotic phase difference is $\textmd{arg}(-\frac{u_{-}}{u_{+}})=\pi$.
\\

\noindent
{\rotatebox{0}{\includegraphics[width=4.2cm,height=3.5cm,angle=0]{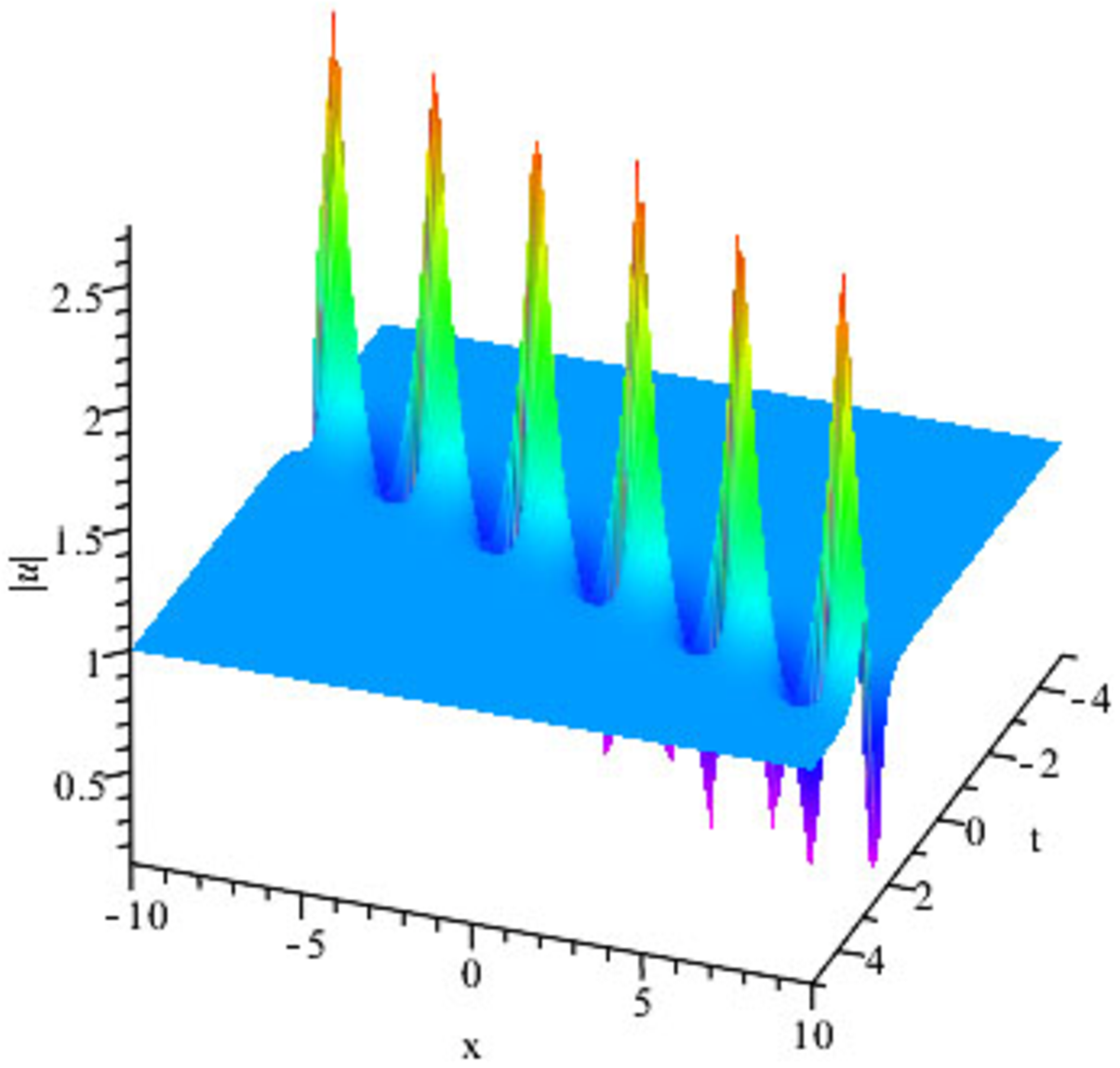}}}
~~~~
{\rotatebox{0}{\includegraphics[width=4.2cm,height=3.5cm,angle=0]{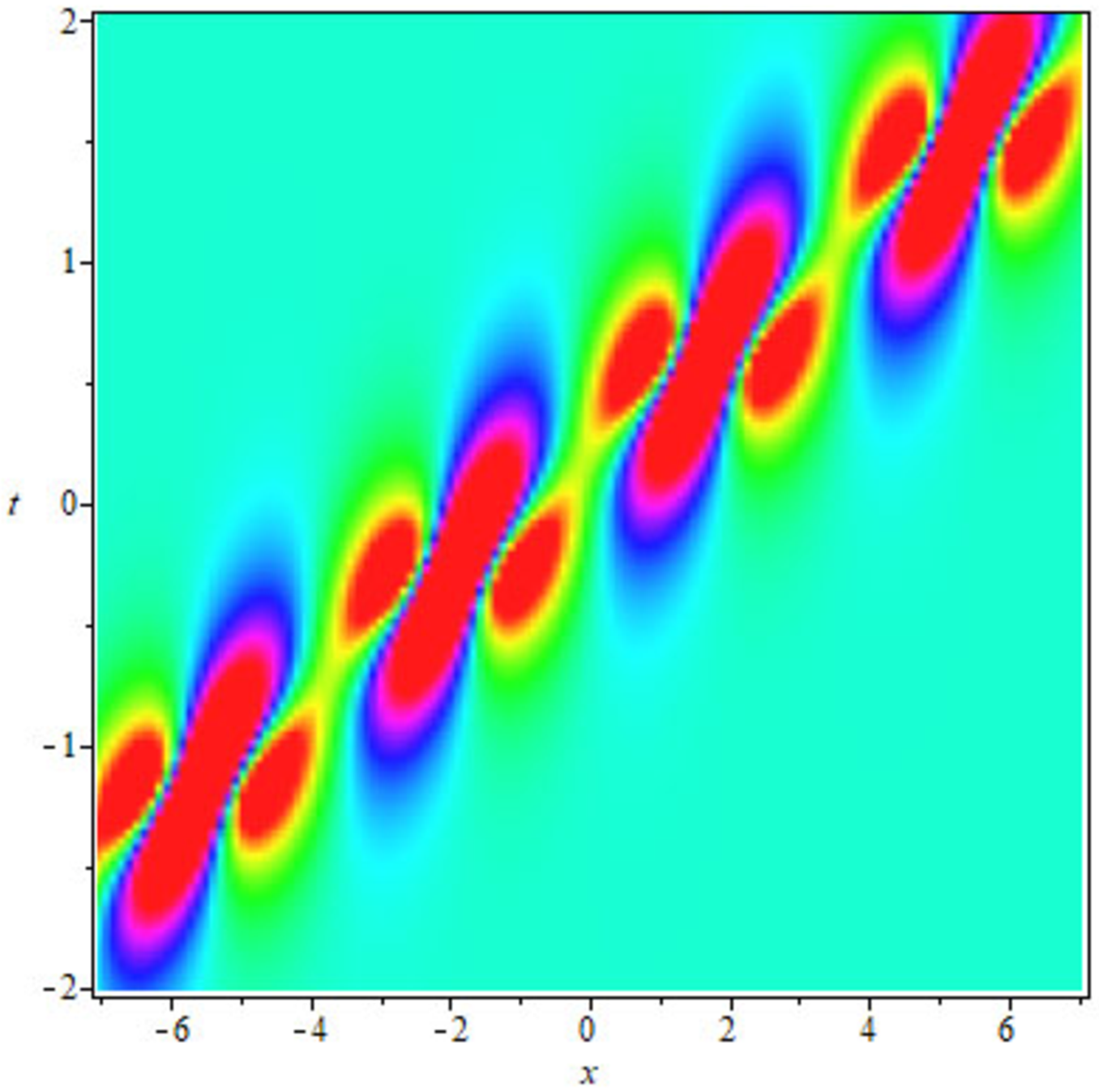}}}
~~~~
{\rotatebox{0}{\includegraphics[width=4.2cm,height=3.5cm,angle=0]{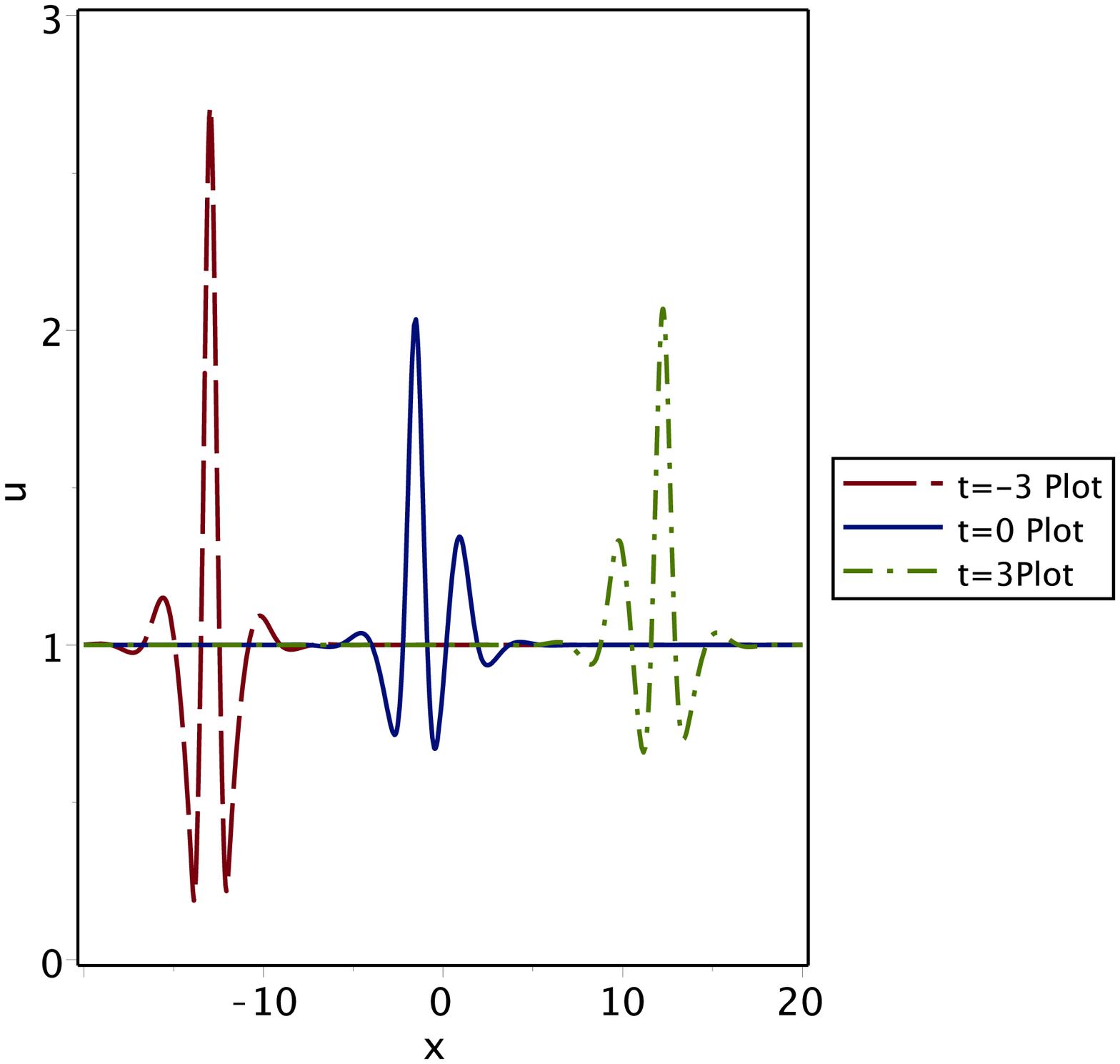}}}
$~~~~~~~~~~~~~~~(\textbf{a1})~~
~~~~~~~~~~~~~~~~~~~~~~~~~~~~~~~~~~~(\textbf{a2})
~~~~~~~~~~~~~~~~~~~~~~~~~~~~~~~~(\textbf{a3})$\\

\noindent
{\rotatebox{0}{\includegraphics[width=4.2cm,height=3.5cm,angle=0]{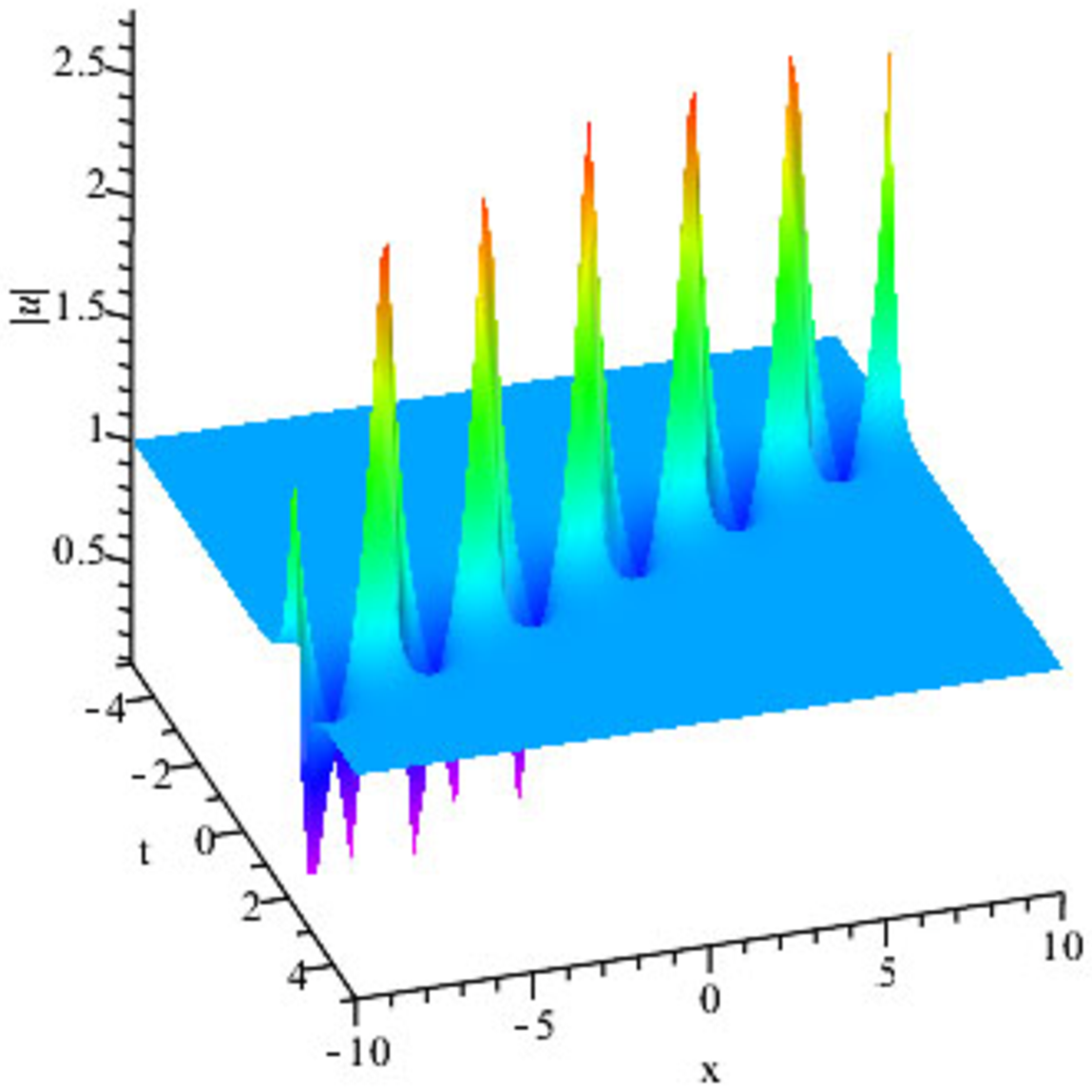}}}
~~~~
{\rotatebox{0}{\includegraphics[width=4.2cm,height=3.5cm,angle=0]{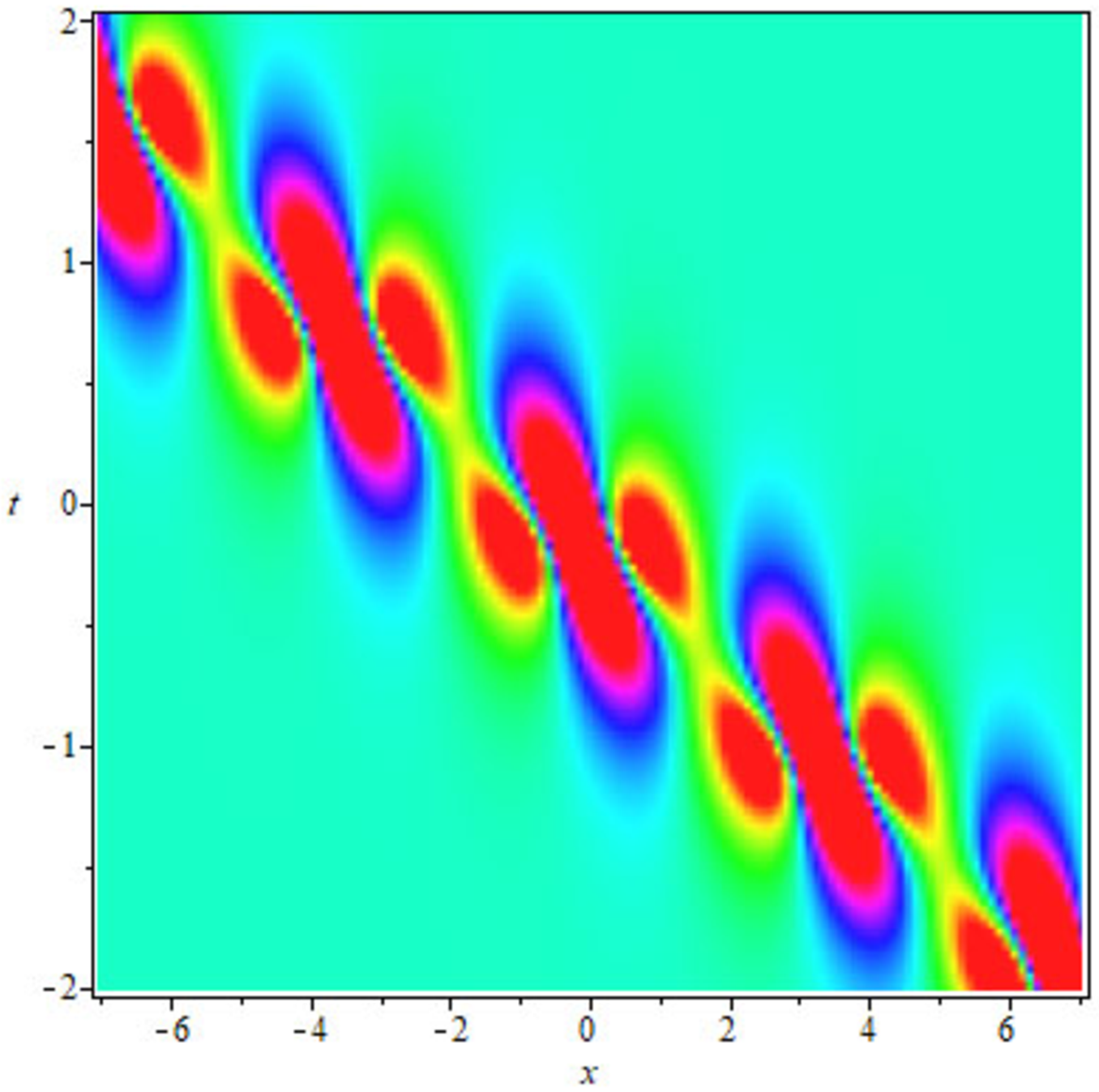}}}
~~~~
{\rotatebox{0}{\includegraphics[width=4.2cm,height=3.5cm,angle=0]{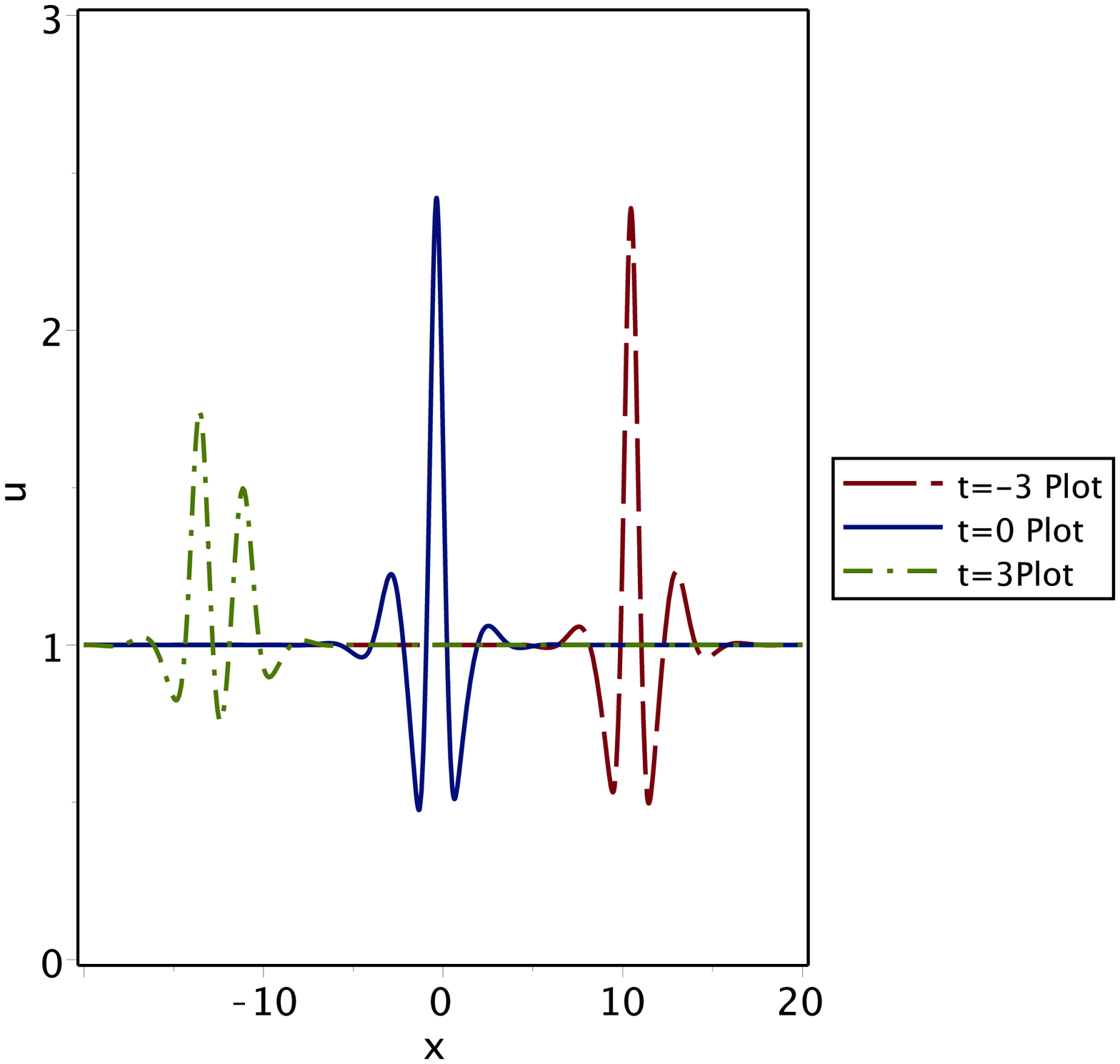}}}
$~~~~~~~~~~~~~~~(\textbf{b1})~~
~~~~~~~~~~~~~~~~~~~~~~~~~~~~~~~~~~~(\textbf{b2})
~~~~~~~~~~~~~~~~~~~~~~~~~~~~~~~~(\textbf{b3})$\\
\noindent { \small \textbf{Figure 3.} (Color online) Simple-pole non-stationary breather solutions $u$ of the NLSLab equation \eqref{NLS-1}  by choosing suitable parameters:
 $c=0.1, N=1, u_{0}=1, u_{+}=-1, b_{1}=e^{-2i\theta(z_{1})}$.
($\textbf{(a1)}, \textbf{(a2)},\textbf{(a3)}$): the breather-breather solutions with NZBCs and parameter $z=2e^{\frac{i\pi}{4}}$;
($\textbf{(b1)}, \textbf{(b2)},\textbf{(b3)}$): the breather-breather solutions with NZBCs and parameter $z=2e^{\frac{3i\pi}{4}}$.}

\subsection{Multi-soliton solutions}
In the previous proof process, we have obtained the explicit solutions \eqref{Ip-23} of an arbitrary number of solitons of equation \eqref{NLS-1}. Now, we will use the two-soliton solution: $N=2$ as an example. For $N=2$, we have the weak interactions of breather-breather solutions of the NLSLab equation with NZBCs. (see Figures 4 (a1)-(a3) for $u_{+}=-1$, Figures 4 (b1)-(b3) for $u_{+}=-0.5$ and Figures 4 (c1)-(c3) for $u_{+}=-0.05$, where $z_{1}=\frac{1}{5}+\frac{3i}{2}, z_{2}=-\frac{1}{5}+\frac{3i}{2}$). As can be seen from Figs. 4 (a1)-(a3) to Figs. 4 (c1)-(c3), as the parameter $u_{+}$ increases, the periodicity of  the simple-pole breather-breather solutions will gradually shift to the top parts. In particular, when $u_{+}\rightarrow0$, the weak interaction of the simple-pole bright-bright solitons of the NLSLab equation with the ZBCs is observed (see Fig.s 4(d1)-(d3)).
\\

\noindent
{\rotatebox{0}{\includegraphics[width=4.2cm,height=3.5cm,angle=0]{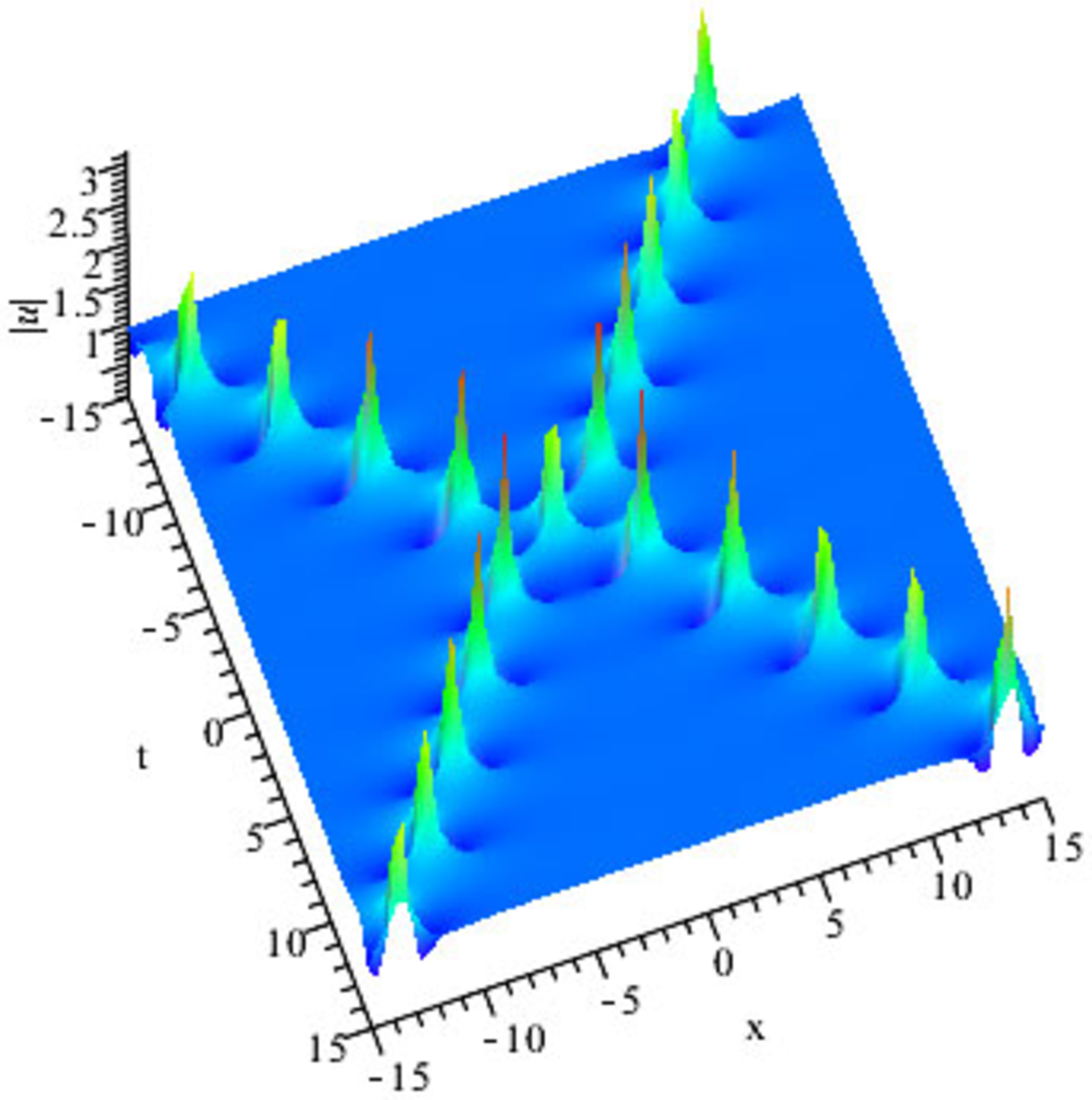}}}
~~~~
{\rotatebox{0}{\includegraphics[width=4.2cm,height=3.5cm,angle=0]{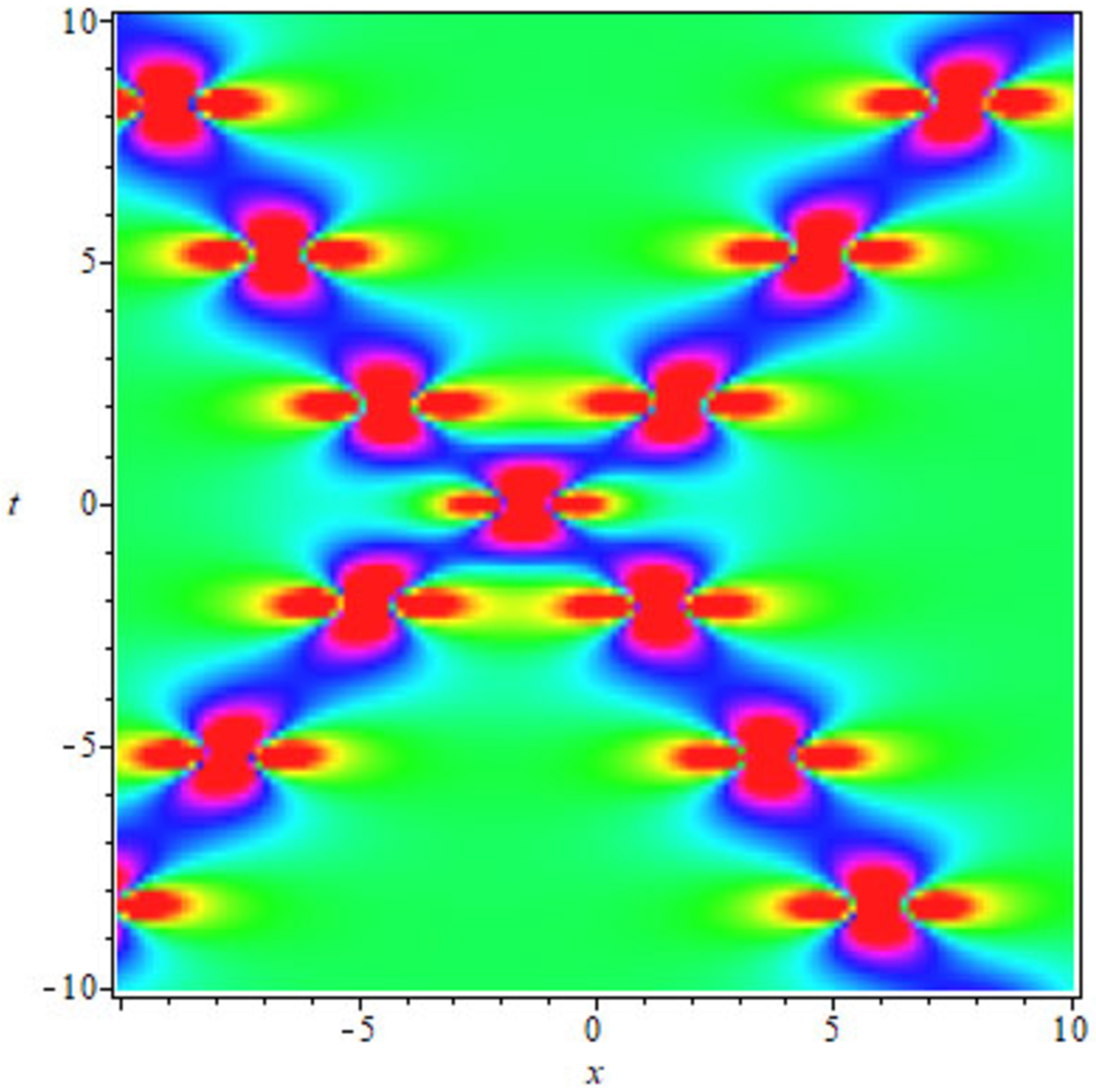}}}
~~~~
{\rotatebox{0}{\includegraphics[width=4.2cm,height=3.5cm,angle=0]{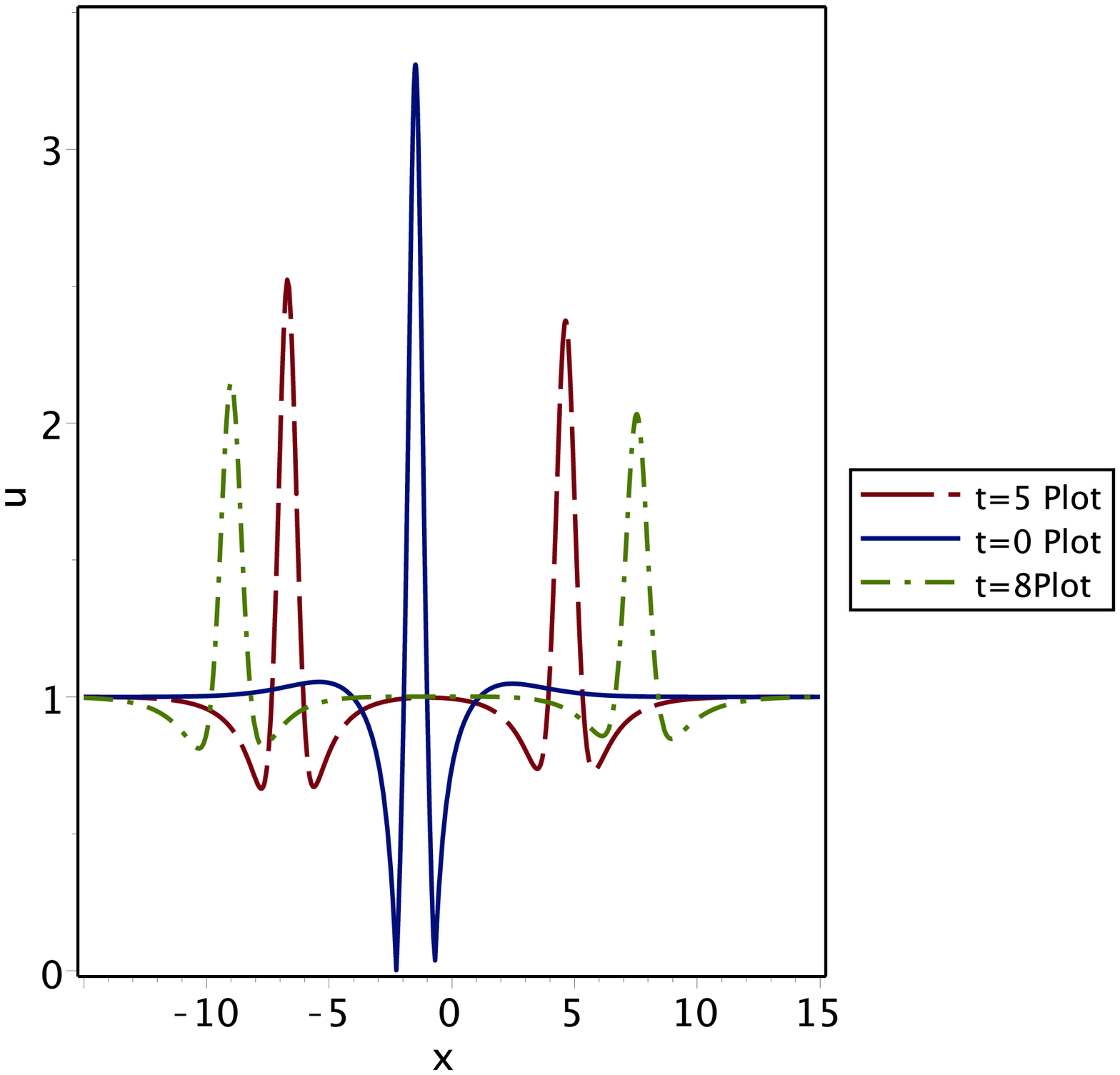}}}
$~~~~~~~~~~~~~~~(\textbf{a1})~~
~~~~~~~~~~~~~~~~~~~~~~~~~~~~~~~~~~~(\textbf{a2})
~~~~~~~~~~~~~~~~~~~~~~~~~~~~~~~~(\textbf{a3})$\\

\noindent
{\rotatebox{0}{\includegraphics[width=4.2cm,height=3.5cm,angle=0]{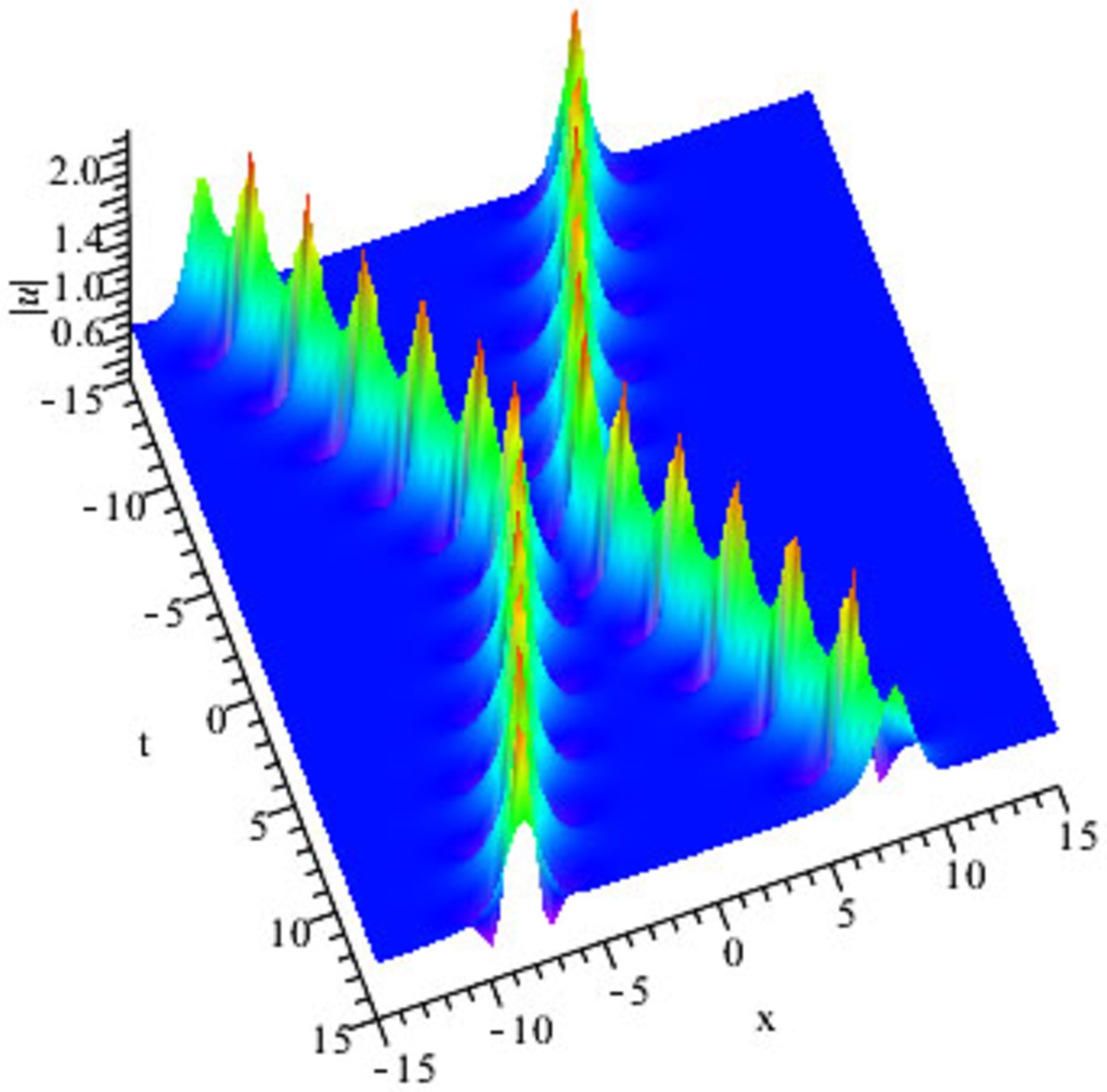}}}
~~~~
{\rotatebox{0}{\includegraphics[width=4.2cm,height=3.5cm,angle=0]{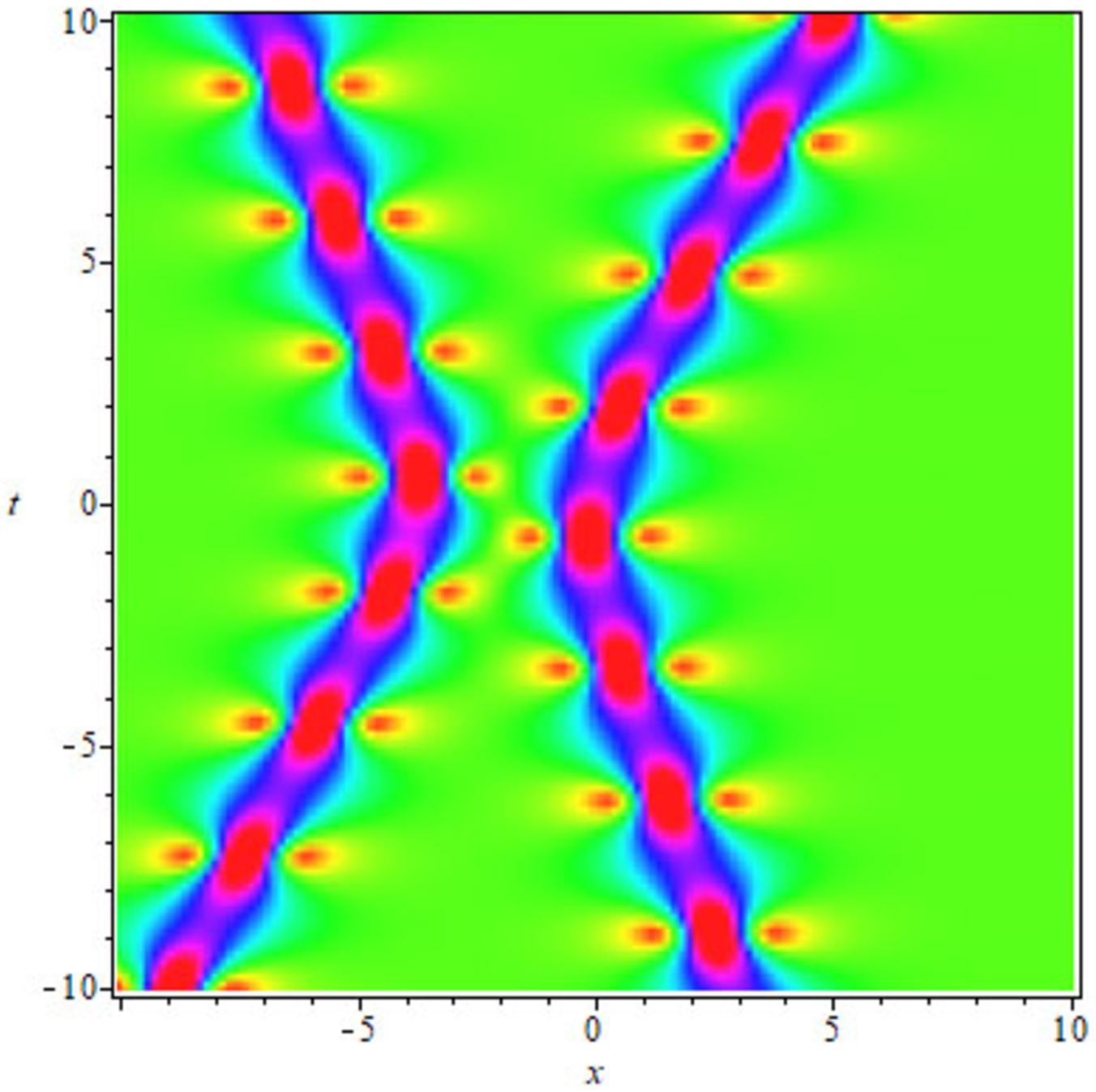}}}
~~~~
{\rotatebox{0}{\includegraphics[width=4.2cm,height=3.5cm,angle=0]{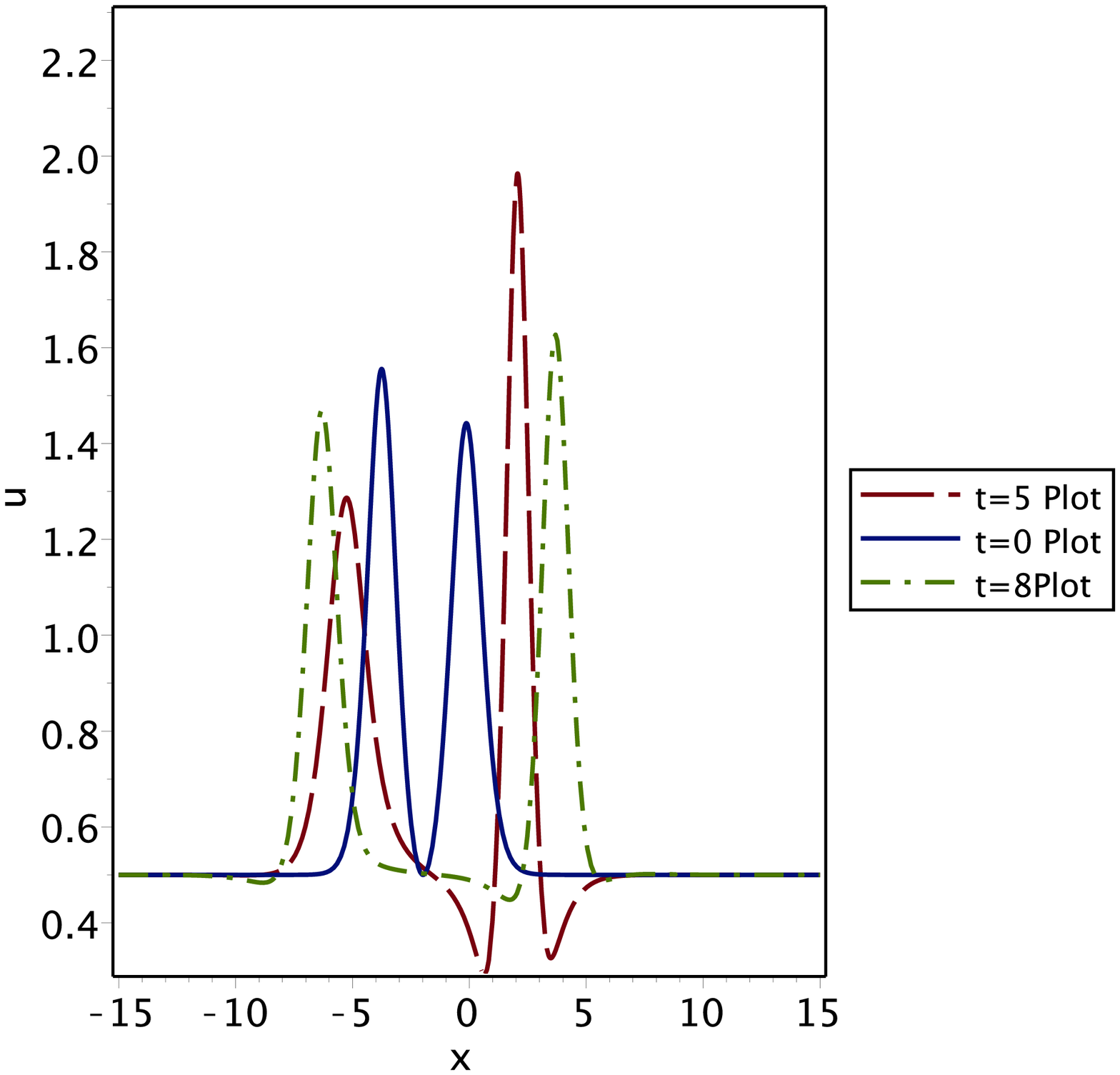}}}
$~~~~~~~~~~~~~~~(\textbf{b1})~~
~~~~~~~~~~~~~~~~~~~~~~~~~~~~~~~~~~~(\textbf{b2})
~~~~~~~~~~~~~~~~~~~~~~~~~~~~~~~~(\textbf{b3})$\\

\noindent
{\rotatebox{0}{\includegraphics[width=4.2cm,height=3.5cm,angle=0]{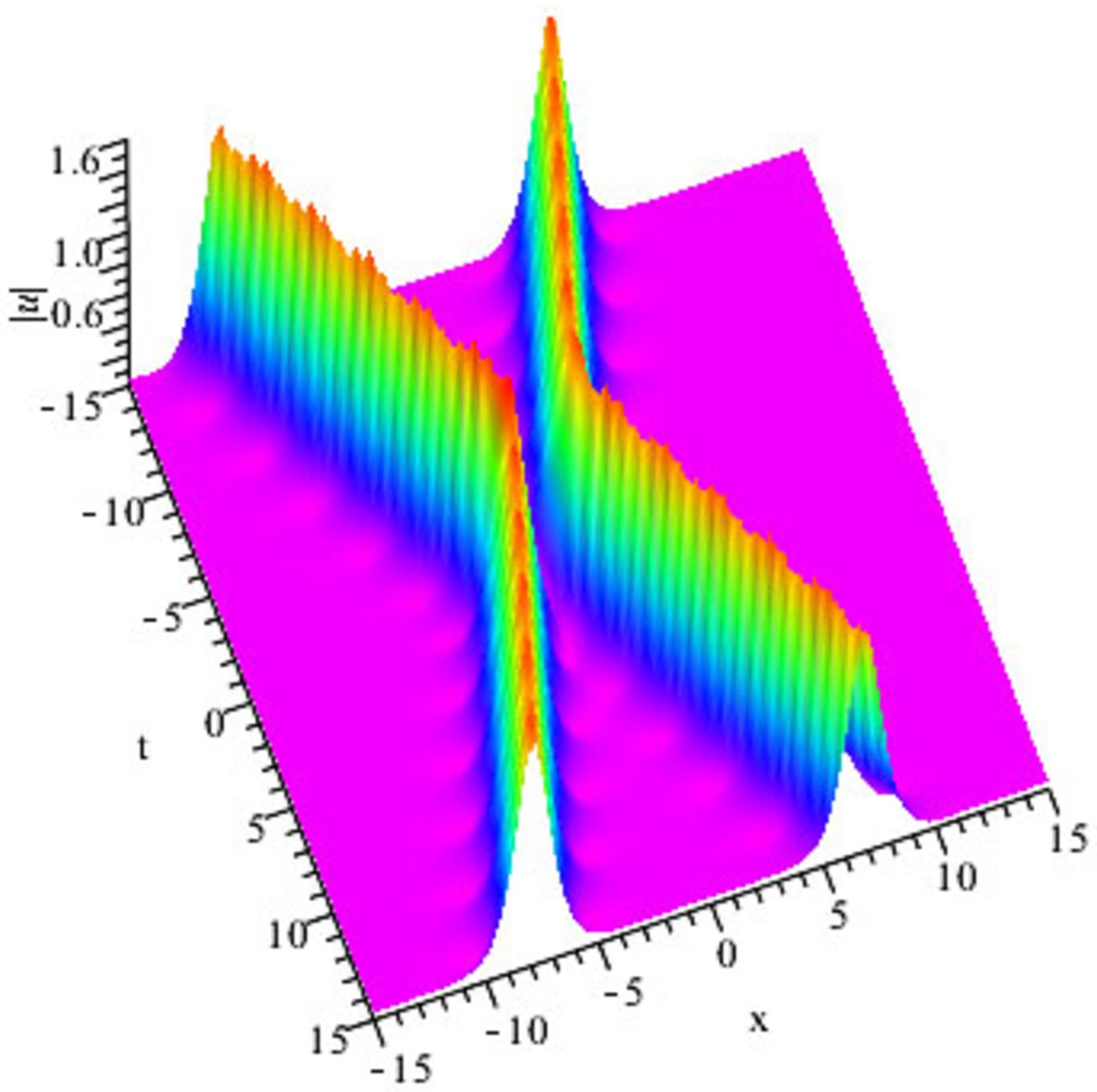}}}
~~~~
{\rotatebox{0}{\includegraphics[width=4.2cm,height=3.5cm,angle=0]{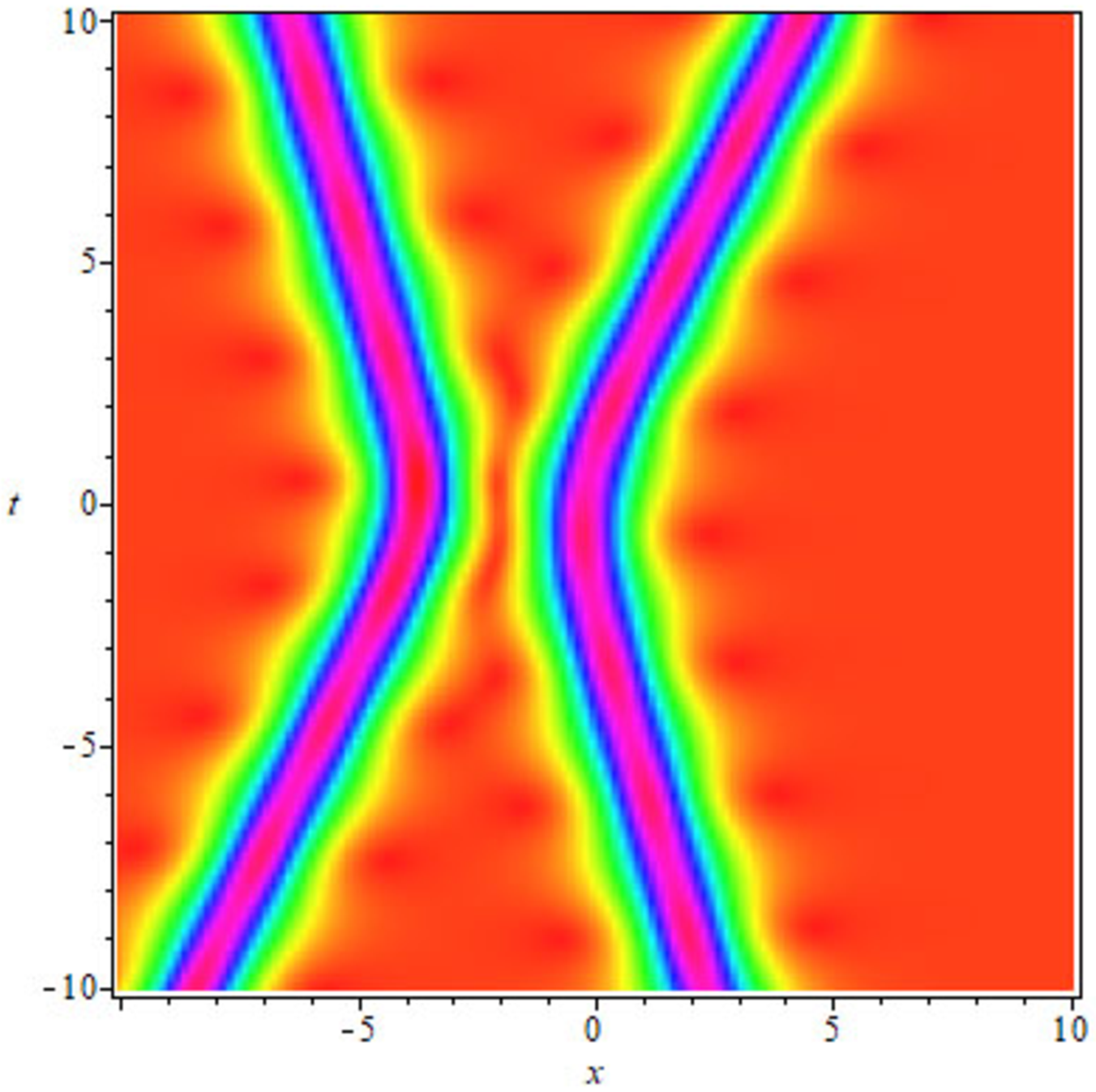}}}
~~~~
{\rotatebox{0}{\includegraphics[width=4.2cm,height=3.5cm,angle=0]{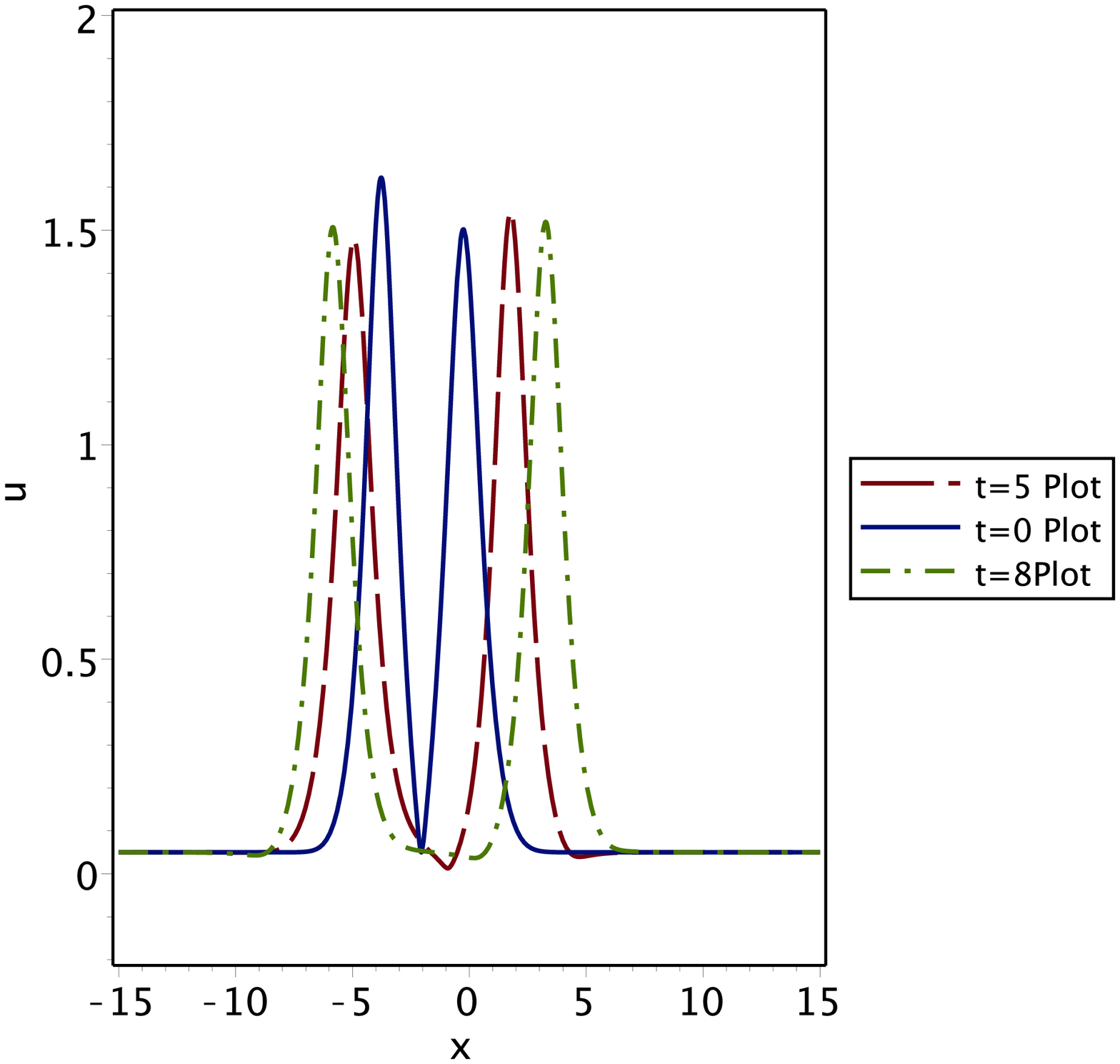}}}
$~~~~~~~~~~~~~~~(\textbf{c1})~~
~~~~~~~~~~~~~~~~~~~~~~~~~~~~~~~~~~~(\textbf{c2})
~~~~~~~~~~~~~~~~~~~~~~~~~~~~~~~~(\textbf{c3})$\\

\noindent
{\rotatebox{0}{\includegraphics[width=4.2cm,height=3.5cm,angle=0]{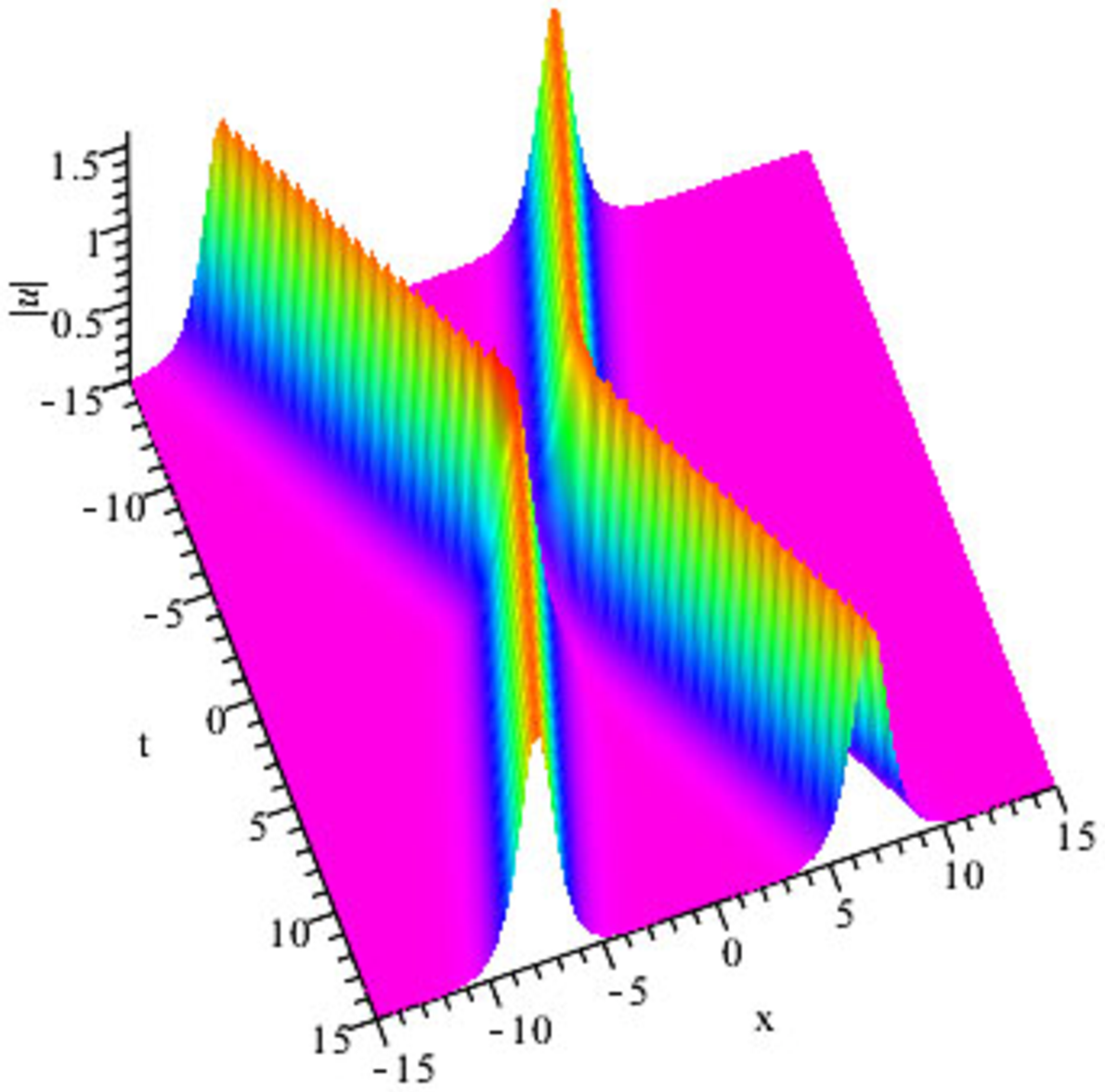}}}
~~~~
{\rotatebox{0}{\includegraphics[width=4.2cm,height=3.5cm,angle=0]{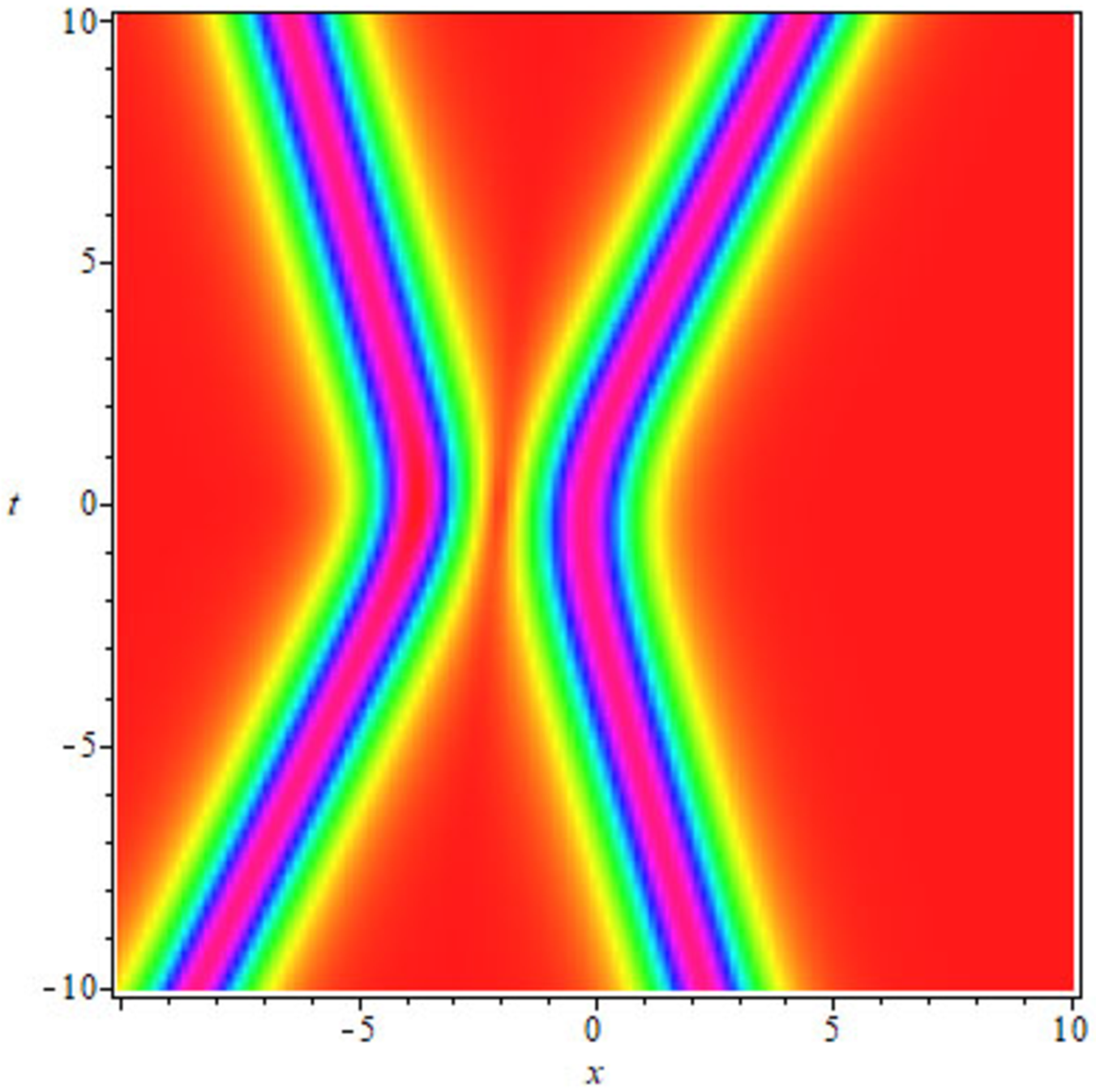}}}
~~~~
{\rotatebox{0}{\includegraphics[width=4.2cm,height=3.5cm,angle=0]{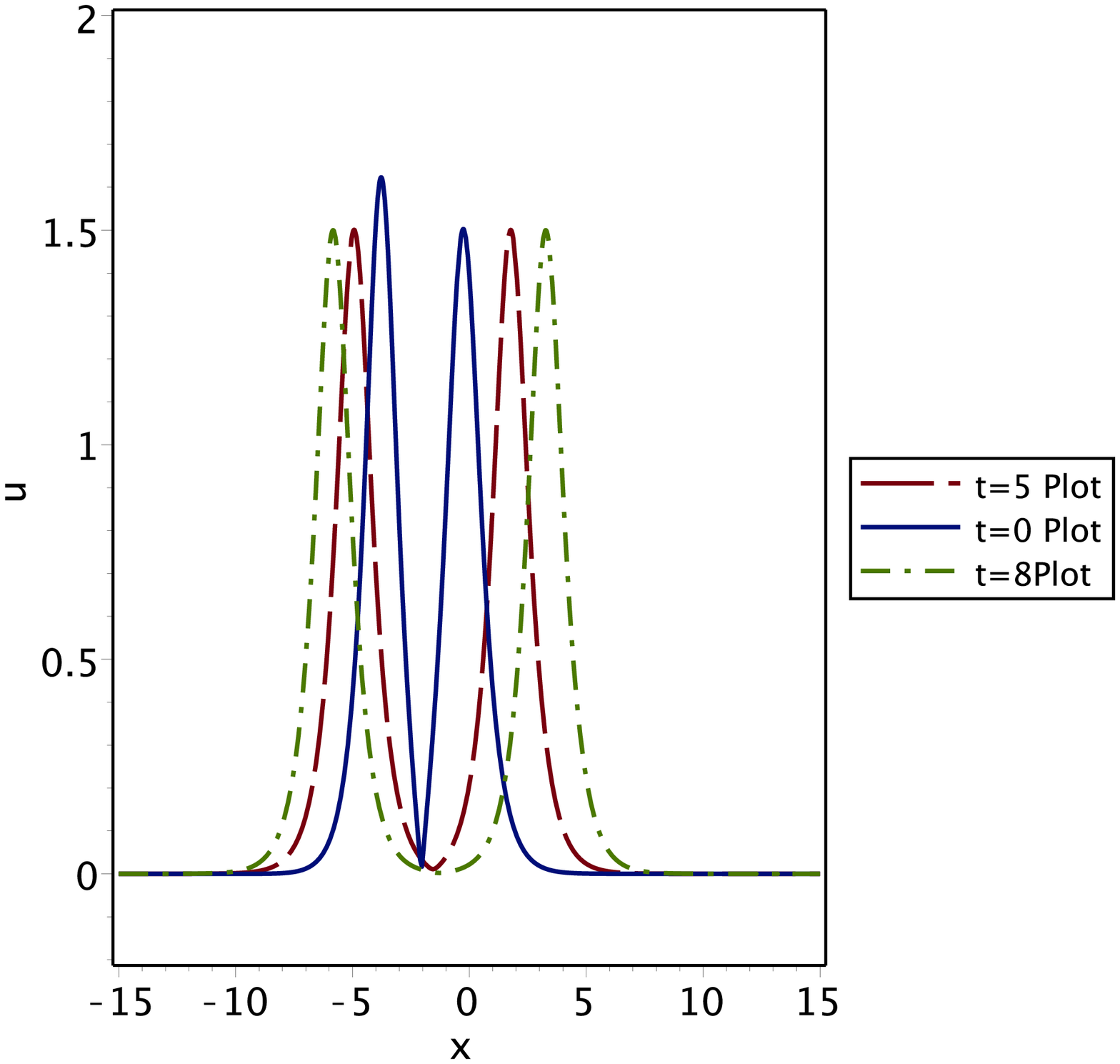}}}
$~~~~~~~~~~~~~~~(\textbf{d1})~~
~~~~~~~~~~~~~~~~~~~~~~~~~~~~~~~~~~~(\textbf{d2})
~~~~~~~~~~~~~~~~~~~~~~~~~~~~~~~~(\textbf{d3})$\\
\noindent { \small \textbf{Figures 4.} (Color online) Weak interactions of simple-pole two-soliton solutions $u(x,t)$ of the NLSLab equation \eqref{NLS-1}  by choosing suitable parameters:
 $c=0.1, N=2, z_{1}=\frac{1}{5}+\frac{3i}{2}, z_{2}=-\frac{1}{5}+\frac{3i}{2}, b_{1}=e^{-2i\theta(z_{1})}, b_{2}=e^{-2i\theta(z_{2})}$.
($\textbf{(a1)}, \textbf{(a2)},\textbf{(a3)}$): the simple-pole two-soliton solutions with NZBCs and parameter $u_{+}=-1$; ($\textbf{(b1)}, \textbf{(b2)},\textbf{(b3)}$): the simple-pole two-soliton solutions with NZBCs and parameter $u_{+}=-0.5$; ($\textbf{(c1)}, \textbf{(c2)},\textbf{(c3)}$): the simple-pole two-soliton solutions with NZBCs and parameter $u_{+}=-0.05$; ($\textbf{(d1)}, \textbf{(d2)},\textbf{(d3)}$): the simple-pole two-soliton solutions with NZBCs and parameter $ u_{+}\rightarrow0$.}

\section{The NLSLab equation with NZBCs: double poles}
Although the direct and inverse scattering problems in the case of double poles are similar to the forward and reverse scatter problems in the case of simple poles, there is a certain difference between them, that is, they can make the meromorphic matrix function $M(x,t,z)$ in Riemann-Hilbert problem. The problem shows distinct types of poles (i.e., simple poles and double poles), and the matrix function $M(x,t,z)$ can produce different potentials for the simple poles and double poles. In addition to this, the resulting trace formulae and theta conditions will change ( the focusing NLS equation with double poles, see [30]).

\subsection{Residue conditions}
Next, we will assume that the discrete spectral points $Z$ are the double zeros of the scattering coefficients $s_{11}(z)$ and $s_{22}(z)$, i.e. $z_{n}$ ($z_{n}\in Z\cap D^{-}$) is a double zero of $s_{11}(z)$ and $z^{\ast}_{n}$ ($z^{\ast}_{n}\in Z\cap D^{+}$) is a double zero of $s_{22}(z)$, than $s_{11}(z_{n})=s_{11}'(z_{n})=0$ along with $s_{11}''(z_{n})\neq0$ and $s_{22}(z^{\ast}_{n})=s_{22}'(z^{\ast}_{n})=0$ along with $s_{22}''(z^{\ast}_{n})\neq0$. In the following calculation process, we first give a simple proposition\cite{PB-2017-82}:

\noindent \textbf{Proposition 1.} If the functions $f(z)$ and $g(z)$ are analytic in a certain complex domain $\Omega$, and $z_{0}\in\Omega$ represents a double zeros of the functions $g(z_{0})$ and $f(z_{0})\neq0$, then the residue condition $\mathrm{Res}[f(z)/g(z)]$ and the coefficient $P_{-2}[f(z)/g(z)]$ of $f(z)/g(z)$ can be given by Laurent expansion at $z=z_{0}$, namely
\begin{equation}\label{dp-1}
\mathop{P_{-2}}_{z=z_{0}}\left[\frac{f(z)}{g(z)}\right]=\frac{2f(z_{0})}{g''(z_{0})},~~~
\mathop{\mathrm{Res}}_{z=z_{0}}\left[\frac{f(z)}{g(z)}\right]=2\left(\frac{f'(z_{0})}{g''(z_{0})}-\frac{f(z_{0})g'''(z_{0})}{3[g''(z_{0})]^{2}}\right).
\end{equation}
The calculation here is very similar to the calculation process of \cite{PB-2017-82}. For the NLS equation with a double pole, there is $s_{11}(z_{n})=s_{11}'(z_{n})=0$, $s_{11}''(z_{n})\neq0$ in $\forall z_{n}\in Z\cap D^{-}$ and $s_{22}(z^{\ast}_{n})=s_{22}'(z^{\ast}_{n})=0$, $s_{22}''(z^{\ast}_{n})\neq0$ in $\forall z^{\ast}_{n}\in Z\cap D^{+}$. Obviously it can be found that equations \eqref{cl9-v-a} and \eqref{cl9-v-b} are still holds, but for which the coefficient $s_{21}(z_{n})$ and $s_{12}(z^{\ast}_{n})$ may not be the same as before. Equations \eqref{Pre-15-1} can be rewritten as
\begin{align}\label{dp-2}
&s_{11}(z)\gamma(z)=|\phi_{+,1}(x,t,z),\phi_{-,2}(x,t,z)|,~~~
s_{22}(z)\gamma(z)=|\phi_{-,1}(x,t,z),\phi_{+,2}(x,t,z)|,\notag\\
&s_{12}(z)\gamma(z)=|\phi_{+,2}(x,t,z),\phi_{-,2}(x,t,z)|,~~~
s_{21}(z)\gamma(z)=|\phi_{-,1}(x,t,z),\phi_{+,1}(x,t,z)|.
\end{align}
At this time, we derive the first order  partial derivative with respect to $z$ on the two equations in the first row of equations \eqref{dp-2}, one  can get
\begin{align}\label{dp-3}
&[s_{11}(z)\gamma(z)]'=|\phi_{+,1}'(x,t,z),\phi_{-,2}(x,t,z)|+|\phi_{+,1}(x,t,z),\phi_{-,2}'(x,t,z)|,\notag\\
&[s_{22}(z)\gamma(z)]'=|\phi_{-,1}'(x,t,z),\phi_{+,2}(x,t,z)|+|\phi_{-,1}(x,t,z),\phi_{+,2}'(x,t,z)|.
\end{align}
Let $z=z_{n}\in Z\cap D^{-}$ in equation \eqref{dp-3} and by using $s_{11}(z_{n})=s_{11}'(z_{n})=0$, $s_{22}(z^{\ast}_{n})=s_{22}'(z^{\ast}_{n})=0$ and equations \eqref{cl9-v-1} and \eqref{cl9-v-2}, one get
\begin{align}\label{dp-4}
&|\phi_{+,1}'(x,t,z_{n})-A_{-}(z_{n})\phi_{-,2}'(x,t,z_{n}),\phi_{-,2}(x,t,z_{n})|=0,\notag\\
&|\phi_{-,1}(x,t,z^{\ast}_{n}),\phi_{+,2}'(x,t,z^{\ast}_{n})-A_{+}(z^{\ast}_{n})\phi_{-,1}'(x,t,z^{\ast}_{n})|=0,
\end{align}
which means that there exists another constant $B_{-}(z_{n})$ and $B_{+}(z^{\ast}_{n})$ that makes
\begin{align}\label{dp-5}
&\phi_{+,1}'(x,t,z_{n})=B_{-}(z_{n})\phi_{-,2}(x,t,z_{n})+ A_{-}(z_{n})\phi_{-,2}'(x,t,z_{n}),\notag\\
&\phi_{+,2}'(x,t,z^{\ast}_{n})=B_{+}(z^{\ast}_{n})\phi_{-,1}(x,t,z^{\ast}_{n})+
A_{+}(z^{\ast}_{n})\phi_{-,1}'(x,t,z^{\ast}_{n}).
\end{align}
According to equations \eqref{cl9-v-1}, \eqref{dp-1} and \eqref{dp-5}, we have
\begin{align}\label{dp-6}
&\mathop{P_{-2}}_{z=z_{n}}\left[\frac{\phi_{+,1}(x,t,z)}{s_{11}(z)}\right]=\frac{2\phi_{+,1}(x,t,z_{n})}{s_{11}''(z_{n})}=
\frac{2A_{-}(z_{n})}{s_{11}''(z_{n})}\phi_{-,2}(x,t,z_{n})\notag\\
&~~~~~~~~~~~~~~~~~~~~~~~~~=K_{-}(z_{n})\phi_{-,2}(x,t,z_{n}),\notag\\
&\mathop{\mathrm{Res}}_{z=z_{n}}\left[\frac{\phi_{+,1}(x,t,z)}{s_{11}(z)}\right]=
\frac{2\phi_{+,1}'(x,t,z_{n})}{s_{11}''(z_{n})}-\frac{2\phi_{+,1}(x,t,z_{n})s_{11}'''(z_{n})}
{3(s_{11}''(z_{n}))^{2}}\notag\\
&~~~~~~~~~~~~~~~~~~~~~~~~~=\frac{2A_{-}(z_{n})}{s_{11}''(z_{n})}\left[\phi_{-,2}'(x,t,z_{n})+
\left(\frac{B_{-}(z_{n})}{A_{-}(z_{n})}-\frac{s_{11}'''(z_{n})}{3s_{11}''(z_{n})}\right)\phi_{-,2}(x,t,z_{n})\right]\notag\\
&~~~~~~~~~~~~~~~~~~~~~~~~~=K_{-}(z_{n})[\phi_{-,2}'(x,t,z_{n})+J_{-}(z_{n})\phi_{-,2}(x,t,z_{n})],
\end{align}
where $K_{-}(z_{n})=\frac{2A_{-}(z_{n})}{s_{11}''(z_{n})}$ and $J_{-}(z_{n})=\frac{B_{-}(z_{n})}{A_{-}(z_{n})}-\frac{s_{11}'''(z_{n})}{3s_{11}''(z_{n})}$.

Similarly, according to equations \eqref{cl9-v-2}, \eqref{dp-1} and \eqref{dp-5}, we can obtain
\begin{align}\label{dp-7}
&\mathop{P_{-2}}_{z=z^{\ast}_{n}}\left[\frac{\phi_{+,2}(x,t,z)}{s_{22}(z)}\right]=\frac{2\phi_{+,2}(x,t,z^{\ast}_{n})}{s_{22}''(z_{n})}=
\frac{2A_{+}(z^{\ast}_{n})}{s_{22}''(z^{\ast}_{n})}\phi_{-,1}(x,t,z^{\ast}_{n})\notag\\
&~~~~~~~~~~~~~~~~~~~~~~~~~=K_{+}(z^{\ast}_{n})\phi_{-,1}(x,t,z^{\ast}_{n}),\notag\\
&\mathop{\mathrm{Res}}_{z=z^{\ast}_{n}}\left[\frac{\phi_{+,2}(x,t,z)}{s_{22}(z)}\right]=
\frac{2\phi_{+,2}'(x,t,z^{\ast}_{n})}{s_{22}''(z^{\ast}_{n})}-\frac{2\phi_{+,2}(x,t,z^{\ast}_{n})s_{22}'''(z^{\ast}_{n})}
{3(s_{22}''(z^{\ast}_{n}))^{2}}\notag\\
&~~~~~~~~~~~~~~~~~~~~~~~~~=\frac{2A_{+}(z^{\ast}_{n})}{s_{22}''(z^{\ast}_{n})}\left[\phi_{-,1}'(x,t,z^{\ast}_{n})+
\left(\frac{B_{+}(z^{\ast}_{n})}{A_{+}(z^{\ast}_{n})}-\frac{s_{22}'''(z^{\ast}_{n})}{3s_{22}''(z^{\ast}_{n})}\right)\phi_{-,1}(x,t,z^{\ast}_{n})\right]\notag\\
&~~~~~~~~~~~~~~~~~~~~~~~~~=K_{+}(z^{\ast}_{n})[\phi_{-,1}'(x,t,z^{\ast}_{n})+J_{+}(z^{\ast}_{n})\phi_{-,1}(x,t,z^{\ast}_{n})],
\end{align}
where $K_{+}(z^{\ast}_{n})=\frac{2A_{+}(z^{\ast}_{n})}{s_{22}''(z^{\ast}_{n})}$ and $J_{+}(z^{\ast}_{n})=\frac{B_{+}(z^{\ast}_{n})}{A_{+}(z^{\ast}_{n})}-\frac{s_{22}'''(z^{\ast}_{n})}{3s_{22}''(z^{\ast}_{n})}$.

From (2) and (4) of theorem 4, we can get
\begin{align}\label{dp-8}
&\phi_{\pm,1}(x,t,z)=\sigma\phi_{\pm,2}^{\ast}(x,t,z^{\ast}),
~~~~~~~~~~~~\phi_{\pm,2}(x,t,z)=-\sigma\phi_{\pm,1}^{\ast}(x,t,z^{\ast}),\notag\\
&\phi_{\pm,1}(x,t,z)=-\frac{iu^{\ast}_{\pm}}{z}\phi_{\pm,2}(x,t,-\frac{u_{0}^{2}}{z}),
~~~\phi_{\pm,2}(x,t,z)=-\frac{iu_{\pm}}{z}\phi_{\pm,1}(x,t,-\frac{u_{0}^{2}}{z}).
\end{align}
After further calculations, we can get the following symmetries (Please refer to \cite{PB-2017-82,ZCY-2020-80} for the details of the certification process.)
\begin{align}\label{dp-9}
&K_{-}(z_{n})=-K_{+}^{\ast}(z^{\ast}_{n})=(\frac{z_{n}}{u_{0}})^{4}(\frac{u^{\ast}_{+}}{u_{+}})
K_{+}(-\frac{u_{0}^{2}}{z_{n}})\notag\\
&~~~~~~~~~~=-(\frac{z_{n}}{u_{0}})^{4}(\frac{u^{\ast}_{+}}{u_{+}})
K_{-}^{\ast}(-\frac{u_{0}^{2}}{z^{\ast}_{n}}),~~z_{n}\in Z^{f}\cap D^{-}\notag\\
&J_{-}(z_{n})=J_{+}^{\ast}(z^{\ast}_{n})=(\frac{u_{0}}{z_{n}})^{2}
J_{+}(-\frac{u_{0}^{2}}{z_{n}})+\frac{2}{z_{n}}\notag\\
&~~~~~~~~~=(\frac{u_{0}}{z_{n}})^{2}J_{-}^{\ast}(-\frac{u_{0}^{2}}{z^{\ast}_{n}})+\frac{2}{z_{n}},
~~z_{n}\in Z^{f}\cap D^{-}.
\end{align}
It follows from equations \eqref{Pre-12}, \eqref{Ip-1}, \eqref{dp-6} and \eqref{dp-7} that
\begin{align}\label{dp-10}
&\mathop{P_{-2}}_{z=z_{n}}M^{-}_{1}(x,t,z)=\mathop{P_{-2}}_{z=z_{n}}\left[\frac{\mu_{+,1}(x,t,z)}{s_{11}(z)}\right]
=K_{-}(z_{n})e^{-2i\theta(x,t,z_{n})}\mu_{-,2}(x,t,z_{n}),\notag\\
&\mathop{P_{-2}}_{z=-\frac{u_{0}^{2}}{z_{n}}}M^{+}_{2}(x,t,z)=\mathop{P_{-2}}_{z=-\frac{u_{0}^{2}}{z_{n}}}\left[\frac{\mu_{+,2}(x,t,z)}{s_{22}(z)}\right]
=K_{+}(-\frac{u_{0}^{2}}{z_{n}})e^{2i\theta(x,t,-\frac{u_{0}^{2}}{z_{n}})}
\mu_{-,1}(x,t,-\frac{u_{0}^{2}}{z_{n}}),\notag\\
&\mathop{\mathrm{Res}}_{z=z_{n}}M^{-}_{1}(x,t,z)=\mathop{\mathrm{Res}}_{z=z_{n}}
\left[\frac{\mu_{+,1}(x,t,z)}{s_{11}(z)}\right]\notag\\
&~~~~~~=K_{-}(z_{n})e^{-2i\theta(x,t,z_{n})}
\left\{\mu_{-,2}'(x,t,z_{n})+\left[J_{-}(z_{n})-2i\theta'(x,t,z_{n})\right]
\mu_{-,2}(x,t,z_{n})\right\},\notag\\
&\mathop{\mathrm{Res}}_{z=-\frac{u_{0}^{2}}{z_{n}}}M^{+}_{2}(x,t,z)
=\mathop{\mathrm{Res}}_{z=-\frac{u_{0}^{2}}{z_{n}}}\left[\frac{\mu_{+,2}(x,t,z)}{s_{22}(z)}\right]\notag\\
&~~~~~~=K_{+}(-\frac{u_{0}^{2}}{z_{n}})e^{2i\theta(x,t,-\frac{u_{0}^{2}}{z_{n}})}
\left\{\mu_{-,1}'(x,t,-\frac{u_{0}^{2}}{z_{n}})
+\left[J_{+}(-\frac{u_{0}^{2}}{z_{n}})+2i\theta'(x,t,-\frac{u_{0}^{2}}{z_{n}})\right]
\mu_{-,1}(x,t,-\frac{u_{0}^{2}}{z_{n}})\right\}.
\end{align}

\subsection{RH problem}
For the case of the double pole, equations \eqref{Ip-2}, \eqref{Ip-5} and \eqref{Ip-6} are still valid. Therefore, in order to solve such a Riemann-Hilbert problem, we must eliminate the singularity contributions and the asymptotic values as $z\rightarrow\infty$ and $z\rightarrow0$
\begin{align}\label{dp-11}
&M_{dp}(x,t,z)=I-\frac{i}{z}\sigma_{3}U_{+}
+\sum_{n=1}^{2N}M_{dp}^{(n)},\notag\\
&M_{dp}^{(n)}=\frac{\mathop{P_{-2}}_{z=z_{n}}M^{-}}{(z-z_{n})^{2}}+\frac{\mathop{\mathrm{Res}}_{z=z_{n}}M^{-}}{z-z_{n}}
+\frac{\mathop{P_{-2}}_{z=-\frac{u_{0}^{2}}{z_{n}}}M^{+}}{(z+\frac{u_{0}^{2}}{z_{n}})^{2}}+
\frac{\mathop{\mathrm{Res}}_{z=-\frac{u_{0}^{2}}{z_{n}}}M^{+}}{z+\frac{u_{0}^{2}}{z_{n}}},
\end{align}
from the jump condition \eqref{Ip-2}, we get
\begin{equation}\label{dp-12}
M^{+}(x,t,z)-M_{dp}(x,t,z)=M^{-}(x,t,z)-M_{dp}(x,t,z)-M^{-}(x,t,z)G(x,t,z).
\end{equation}
From equation \eqref{dp-12}, we can get that the left-hand side of equation \eqref{dp-12} is analytic in $D^{+}$, whereas the sum for the first four terms of the right-hand side of equation \eqref{dp-12} is analytic in $D^{-}$. Moreover, the asymptotic behavior of the off-diagonal scattering coefficients means that $G(x,t,z)$ is $O(\frac{1}{z})$ as $z\rightarrow\pm\infty$, and $O(z)$ as $z\rightarrow0$ along the real axis. Finally, we use the Plemelj's formula and Cauchy projector to solve equation \eqref{dp-12} to yield
\begin{equation}\label{dp-13}
M(x,t,z)=M_{dp}(x,t,z)+\frac{1}{2\pi i}\int_{\Sigma}\frac{M^{-}(x,t,s)}{s-z}G(x,t,s)ds,~~~~z\in C\setminus\Sigma,
\end{equation}
where $\int_{\Sigma}$ represents the integral of the oriented contours shown in Figure 1 (right).

Therefore, based on equations \eqref{dp-10}, we obtain the parts of $P_{-2}(\cdot)$ and $\mathrm{Res}(\cdot)$ in equation \eqref{dp-13} as
\begin{equation}\label{dp-14}
M_{dp}^{(n)}=\left(V_{n}(z_{n})\left[\mu_{-,2}'(z_{n})+(W_{n}+\frac{1}{z-z_{n}})\mu_{-,2}(z_{n})\right],
\widehat{V_{n}}(z_{n})\left[\mu_{-,1}'(\widehat{z_{n}})
+(\widehat{W_{n}}+\frac{1}{z-\widehat{z_{n}}})\mu_{-,1}(\widehat{z_{n}})\right]\right),
\end{equation}
where
\begin{align}\label{dp-15}
&V_{n}(z_{n})=\frac{K_{-}(z_{n})}{z-z_{n}}e^{-2i\theta(x,t,z_{n})},
~~~~~~~~~~W_{n}=J_{-}(z_{n})-2i\theta'(x,t,z_{n}),\notag\\
&\widehat{V_{n}}(z_{n})=\frac{K_{+}(\widehat{z_{n}})}{z-\widehat{z_{n}}}e^{2i\theta(x,t,\widehat{z_{n}})},
~~\widehat{W_{n}}=J_{+}(\widehat{z_{n}})+2i\theta'(x,t,\widehat{z_{n}}),
~~\widehat{z_{n}}=-\frac{u_{0}^{2}}{z_{n}}.
\end{align}

\subsection{Reconstruction formula for the potential}
\noindent \textbf{Theorem 8.} \emph{The potential with double poles of the NLSLab equation with NZBCs will be given by
\begin{equation}\label{dp-16}
u(x,t)=u_{+}+i\sum_{n=1}^{2N}K_{+}(\widehat{\eta_{n}})e^{2i\theta(\widehat{\eta_{n}})}
\left(\mu_{-,11}'(\widehat{\eta_{n}})
+\widehat{W_{n}}\mu_{-,11}(\widehat{\eta_{n}})\right)
-\frac{1}{2\pi}\int_{\Sigma}\left(M^{-}(s)G(s)\right)_{12}ds,
\end{equation}
where $K_{+}(\widehat{\eta_{n}})=\frac{2A_{+}(\widehat{\eta_{n}})}{s_{22}''(\widehat{\eta_{n}})}$, $\widehat{W_{n}}=J_{+}(\widehat{\eta_{n}})+2i\theta'(x,t,\widehat{\eta_{n}})$, $J_{+}(\widehat{\eta_{n}})=\frac{B_{+}(\widehat{\eta_{n}})}{A_{+}(\widehat{\eta_{n}})}
-\frac{s_{22}'''(\widehat{\eta_{n}})}{3s_{22}''(\widehat{\eta_{n}}}$, and $\mu_{-,11}'(\widehat{\eta_{n}})$ and $\mu_{-,11}(\widehat{\eta_{n}})$, $n=1,2,\cdots,2N$ are given by equations \eqref{dp-21}.}

\noindent \textbf{Proof:} Before proving the above theorem, we make
\begin{align}\label{dp-17}
&\eta_{n}=\left\{ \begin{aligned}
&z_{n}, ~~~~~~~~~~~n=1,2,\cdots,N\\
&-\frac{u_{0}^{2}}{z^{\ast}_{n-N}}, ~~~n=N+1,N+2,\cdots,2N,
\end{aligned} \right.
\end{align}
and $\widehat{\eta_{n}}=-\frac{u_{0}^{2}}{\eta_{n}}$. Then $Z=\{\eta_{n}, \widehat{\eta_{n}}\}_{n=1}^{2N}.$
 In order to determine $\mu_{-,1}'(\widehat{\eta_{n}}), \mu_{-,1}(\widehat{\eta_{n}}), \mu_{-,2}'(\eta_{n})$ and $\mu_{-,2}(\eta_{n})$ in equation \eqref{dp-14}, for $z=\eta_{n} (n=1,2,\cdots, 2N)$, and it follows from the second column of $M(x,t;z)$ given by equation \eqref{dp-13} with equation \eqref{dp-14} that
\begin{align}\label{dp-18}
\mu_{-,2}(z)=&\left(\begin{aligned}
&-\frac{i}{z}u_{+}\\ &~~~~~1\\
\end{aligned}\right)
+\sum_{n=1}^{2N}\widehat{V_{n}}(\widehat{z})\left[\mu_{-,1}'(\widehat{\eta_{n}})
+(\widehat{W_{n}}+\frac{1}{z-\widehat{\eta_{n}}})\mu_{-,1}(\widehat{\eta_{n}})\right]\notag\\
&+\frac{1}{2\pi i}\int_{\Sigma}\frac{(M^{-}(s)G(s))_{2}}{s-z}ds,
\end{align}
whose first-order derivation, one can get
\begin{align}\label{dp-19}
\mu_{-,2}'(z)=&\left(\begin{aligned}
&\frac{i}{z^{2}}u_{+}\\ &~~~0\\
\end{aligned}\right)
-\sum_{n=1}^{2N}\frac{\widehat{V_{n}}(\widehat{z})}{z-\widehat{\eta_{n}}}\left[\mu_{-,1}'(\widehat{\eta_{n}})
+(\widehat{W_{n}}+\frac{2}{z-\widehat{\eta_{n}}})\mu_{-,1}(\widehat{\eta_{n}})\right]\notag\\
&+\frac{1}{2\pi i}\int_{\Sigma}\frac{(M^{-}(s)G(s))_{2}}{(s-z)^{2}}ds.
\end{align}
In addition, we can also get from theorem 5 that
\begin{equation}\label{dp-20}
\mu_{-,2}'(z)=\frac{iu_{+}}{z^{2}}\mu_{-,1}(-\frac{u_{0}^{2}}{z})
-\frac{iu_{0}^{2}u_{+}}{z^{3}}\mu_{-,1}'(-\frac{u_{0}^{2}}{z}).
\end{equation}
When $z=\eta_{s}, s=1,2,3,\cdots,2N,$ we substitution of theorem 5 and equation \eqref{dp-20} into equations \eqref{dp-18} and \eqref{dp-19} to obtain an algebraic system of $4N$ equations. There are $4N$ unknowns for this system, respectively $\mu_{-,1}(\widehat{\eta_{n}})$, $\mu_{-,1}'(\widehat{\eta_{n}})$, $n=1,2,3,\cdots,2N,$ in the form
\begin{align}\label{dp-21}
&~~~\sum_{n=1}^{2N}\widehat{V_{n}}(\eta_{s})\mu_{-,1}'(\widehat{\eta_{n}})
+\sum_{n=1}^{2N}\left[\widehat{V_{n}}(\eta_{s})(\widehat{W_{n}}+\frac{1}{\eta_{s}-\widehat{\eta_{n}}})
+\frac{iu_{+}}{\eta_{s}}\delta_{sn}\right]\mu_{-,1}(\widehat{\eta_{n}})\notag\\
&=-\left(\begin{aligned}
&-\frac{i}{\eta_{s}}u_{+}\\ &~~~~~1\\
\end{aligned}\right)-\frac{1}{2\pi i}\int_{\Sigma}\frac{(M^{-}(s)G(s))_{2}}{s-\eta_{s}}ds,\notag\\
&~~~\sum_{n=1}^{2N}\left(\frac{\widehat{V_{n}}(\eta_{s})}{\eta_{s}-\widehat{\eta_{n}}}
-\frac{iu_{0}^{2}u_{+}}{\eta_{s}^{3}}\delta_{sn}\right)\mu_{-,1}'(\widehat{\eta_{n}})
+\sum_{n=1}^{2N}\left[\frac{\widehat{V_{n}}(\eta_{s})}{\eta_{s}-\widehat{\eta_{n}}}\left(\widehat{W_{n}}
+\frac{2}{\eta_{s}-\widehat{\eta_{n}}}\right)+\frac{iu_{+}}{\eta_{s}^{2}}\delta_{sn}\right]
\mu_{-,1}(\widehat{\eta_{n}})\notag\\
&=\left(\begin{aligned}
&\frac{iu_{+}}{\eta_{s}^{2}}\\ &~~0\\
\end{aligned}\right)+\frac{1}{2\pi i}\int_{\Sigma}\frac{(M^{-}(s)G(s))_{2}}{(s-\eta_{k})^{2}}ds.
\end{align}
 which can obtain $\mu_{-,1}(x,t,\widehat{\eta_{n}})$, $\mu_{-,1}'(x,t,\widehat{\eta_{n}})$, $n=1,2,3,\cdots,2N$. Then we can substituting the results obtained into theorem 5 and equation \eqref{dp-20} to get $\mu_{-,2}(x,t,\eta_{n})$, $\mu_{-,2}'(x,t,\eta_{n})$, $n=1,2,3,\cdots,2N$. Finally, substituting them into equation \eqref{dp-14}, and then substituting equation \eqref{dp-14} into equation \eqref{dp-13} to get the $M_{dp}(x,t,z)$ in terms of the scattering data.

From equation \eqref{dp-13} with equation \eqref{dp-14}, we can get the asymptotic behavior of $M_{dp}(x,t,z)$ as
\begin{equation}\label{dp-22}
M_{dp}(x,t,z)=I+\frac{M^{(1)}_{dp}(x,t)}{z}+O(\frac{1}{z^{2}}),~~~z\rightarrow\infty,
\end{equation}
where
\begin{align}\label{dp-23}
&M^{(1)}_{dp}(x,t)=-i\sigma_{3}U_{+}-\frac{1}{2\pi i}\int_{\Sigma}M^{-}(x,t,s)G(x,t,s)ds\notag\\
&+\sum_{n=1}^{2N}\left[K_{-}(\eta_{n})e^{-2i\theta(\eta_{n})}
\left(\mu_{-,2}'(\eta_{n})+W_{n}\mu_{-,2}(\eta_{n})\right),
K_{+}(\widehat{\eta_{n}})e^{2i\theta(\widehat{\eta_{n}})}
\left(\mu_{-,1}'(\widehat{\eta_{n}})
+\widehat{W_{n}}\mu_{-,1}(\widehat{\eta_{n}})\right)\right].
\end{align}
From equation \eqref{Ip-1}, we can conclude that $M_{dp}(x,t,z)e^{i\theta(x,t,z)\sigma_{3}}$ satisfies the Lax pair \eqref{NLS-3} and \eqref{NLS-4}. In the case of double poles, substituting $M_{dp}(x,t,z)e^{i\theta(x,t,z)\sigma_{3}}$ with equations \eqref{dp-22} and \eqref{dp-23} into the $x$-part of the Lax pair \eqref{NLS-3}, and then the following property of the potential $u(x,t)$ can be found by finding all the coefficients of $z^{0}$:
\begin{align}\label{dp-24}
u(x,t)=&u_{+}+i\sum_{n=1}^{2N}K_{+}(\widehat{\eta_{n}})e^{2i\theta(\widehat{\eta_{n}})}
\left(\mu_{-,11}'(\widehat{\eta_{n}})
+\widehat{W_{n}}\mu_{-,11}(\widehat{\eta_{n}})\right)\notag\\
&-\frac{1}{2\pi}\int_{\Sigma}\left(M^{-}(s)G(s)\right)_{12}ds,
\end{align}
where $K_{+}(\widehat{\eta_{n}})=\frac{2A_{+}(\widehat{\eta_{n}})}{s_{22}''(\widehat{\eta_{n}})}$, $\widehat{W_{n}}=J_{+}(\widehat{\eta_{n}})+2i\theta'(x,t,\widehat{\eta_{n}})$, $J_{+}(\widehat{\eta_{n}})=\frac{B_{+}(\widehat{\eta_{n}})}{A_{+}(\widehat{\eta_{n}})}
-\frac{s_{22}'''(\widehat{\eta_{n}})}{3s_{22}''(\widehat{\eta_{n}})}$, and $\mu_{-,11}'(\widehat{\eta_{n}})$ and $\mu_{-,11}(\widehat{\eta_{n}})$, $n=1,2,\cdots,2N$ are given by equations \eqref{dp-21}.

The proof the above theorem.    $\Box$

\subsection{Trace formula and ``theta" condition}
Next we will derive the trace formula for the double-poles point, where we will find that it is not the same as the ``theta" condition and the trace formula of the simple point. It can also be said that the scattering coefficients $s_{11}$ and $s_{22}$ are expressed by using reflection coefficients and discrete eigenvalues. When the discrete spectral $z_{n}$, $-\frac{u_{0}^{2}}{z^{\ast}_{n}}$ ($\forall n=1,2,\ldots,N$). are the double zeros of $s_{11}$, then the function
\begin{equation}\label{dp-25}
\beta_{1}^{-}(z)=s_{11}(z)\prod_{n=1}^{N}\frac{(z-z^{\ast}_{n})(z+\frac{u_{0}^{2}}{z_{n}})}{(z-z_{n})(z+\frac{u_{0}^{2}}{z^{\ast}_{n}})},
\end{equation}
is analytic in $D^{-}$ and there are no zeros in this area. Similarly, when the discrete spectral $z^{\ast}_{n}$, $-\frac{u_{0}^{2}}{z_{n}}$ ($\forall n=1,2,\ldots,N$). are the double zeros of $s_{22}$, then the function
\begin{equation}\label{dp-26}
\beta_{1}^{+}(z)=s_{22}(z)\prod_{n=1}^{N}\frac{(z-z_{n})(z+\frac{u_{0}^{2}}{z^{\ast}_{n}})}{(z-z^{\ast}_{n})(z+\frac{u_{0}^{2}}{z_{n}})},
\end{equation}
is analytic in $D^{-}$ and there are no zeros in this area. It is natural to find that $\beta_{1}^{-}(z)\beta_{1}^{+}(z)=s_{11}(z)s_{22}(z)$ on the $\Sigma$. And \eqref{Pre-17} hint
\begin{equation}\label{dp-27}
\beta_{1}^{-}(z)\beta_{1}^{+}(z)=\frac{1}{1-\rho(z)\widetilde{\rho}(z)}
=\frac{1}{1+\rho(z)\rho^{\ast}(z^{\ast})},~~~~z\in\Sigma.
\end{equation}
Similarly to the case of simple poles,  we can obtain the trace formulae for the case of double-poles by using the logarithm and Cauchy operators as
\begin{align}\label{dp-28}
&s_{11}(z)=\exp\left[\frac{1}{2\pi i}\int_{\Sigma}\frac{\textmd{ln}(1+\rho(s)\rho^{\ast}(s^{\ast}))}{s-z}ds\right]
\prod_{n=1}^{N}\frac{(z-z_{n})^{2}(z+\frac{u_{0}^{2}}{z^{\ast}_{n}})^{2}}
{(z-z^{\ast}_{n})^{2}(z+\frac{u_{0}^{2}}{z_{n}})^{2}},
~~~~~for~~z\in D^{-},\notag\\
&s_{22}(z)=\exp\left[-\frac{1}{2\pi i}\int_{\Sigma}\frac{\textmd{ln}(1+\rho(s)\rho^{\ast}(s^{\ast}))}{s-z}ds\right]
\prod_{n=1}^{N}\frac{(z-z^{\ast}_{n})^{2}(z+\frac{u_{0}^{2}}{z_{n}})^{2}}
{(z-z_{n})^{2}(z+\frac{u_{0}^{2}}{z^{\ast}_{n}})^{2}},
~~~for~~z\in D^{+}.
\end{align}
Finally,  we want to obtain the asymptotic phase difference of the boundary values $u_{+}$ and $u_{-}$ (also be said to as ``theta" condition) by using the obtained trace formulae. From equation \eqref{Pre-42}, we obtain $s_{11}(z)=-\frac{u_{-}}{u_{+}}$. In equation \eqref{dp-28}, we take $z\rightarrow 0$, and we can get
\begin{equation}\label{dp-29}
\textmd{arg}(-\frac{u_{-}}{u_{+}})=8\sum_{n=1}^{N}\textmd{arg}z_{n}+\frac{1}{2\pi }\int_{\Sigma}\frac{\textmd{ln}(1+\rho(s)\rho^{\ast}(s^{\ast}))}{s}ds,~~~for~~z\rightarrow 0,
\end{equation}
which relates the phase difference between the asymptotic values for the potential to the reflection coefficient and the discrete spectrum.

\subsection{Soliton solutions}
For the case of reflectionless potential, i.e. $\rho(z)=\widehat{\rho(z)}$, we can get the following theorem:

\noindent \textbf{Theorem 9.} \emph{The solution for the reflectionless potential of the NLSLab equation with double poles can be expressed as
\begin{equation}\label{dp-30}
u(x,t)=u_{+}-i\frac{\det\left(\begin{array}{cc}
    \textbf{M} & \textbf{Y}^{\top}\\
    \textbf{B} & 0\\
\end{array}\right)}{\det\textbf{M}},
\end{equation}
where $\textbf{M}=(M^{(sj)})_{2\times2}$, $M^{(sj)}=(m^{(sj)}_{kn})_{2N\times2N}$,
$m^{(11)}_{kn}=\widehat{V_{n}}(\eta_{k})(\widehat{W_{n}}+\frac{1}{\eta_{k}-\widehat{\eta_{n}}})
+\frac{iu_{+}}{\eta_{k}}\delta_{kn}$, $m^{(12)}_{kn}=\widehat{V_{n}}(\eta_{k})$, $m^{(21)}_{kn}=\frac{\widehat{V_{n}}(\eta_{k})}{\eta_{k}-\widehat{\eta_{n}}}\left(\widehat{W_{n}}
+\frac{2}{\eta_{k}-\widehat{\eta_{n}}}\right)+\frac{iu_{+}}{\eta_{k}^{2}}\delta_{kn}$, $m^{(22)}_{kn}=\frac{\widehat{V_{n}}(\eta_{k})}{\eta_{k}-\widehat{\eta_{n}}}
-\frac{iu_{0}^{2}u_{+}}{\eta_{k}^{3}}\delta_{kn}$, $\textbf{B}=(B_{n}^{1},B_{n}^{2})$, $B_{n}^{1}=K_{+}(\widehat{\eta_{n}})e^{2i\theta(\widehat{\eta_{n}})}\widehat{W_{n}}$, $B_{n}^{2}=K_{+}(\widehat{\eta_{n}})e^{2i\theta(\widehat{\eta_{n}})}$, $\textbf{Y}=(Y_{n}^{1},Y_{n}^{2})$, $Y_{n}^{1}=\frac{iu_{+}}{\eta_{n}}$, $Y_{n}^{2}=\frac{iu_{+}}{\eta_{n}^{2}}$.}

For the case of reflectionless potential, the trace formula and theta condition are reduced to the following form
\begin{align}\label{dp-31}
&s_{11}(z)=\prod_{n=1}^{N}\frac{(z-z_{n})^{2}(z+\frac{u_{0}^{2}}{z^{\ast}_{n}})^{2}}
{(z-z^{\ast}_{n})^{2}(z+\frac{u_{0}^{2}}{z_{n}})^{2}},
~~~~~for~~z\in D^{-},\notag\\
&s_{22}(z)=\prod_{n=1}^{N}\frac{(z-z^{\ast}_{n})^{2}(z+\frac{u_{0}^{2}}{z_{n}})^{2}}
{(z-z_{n})^{2}(z+\frac{u_{0}^{2}}{z^{\ast}_{n}})^{2}},
~~~for~~z\in D^{+},
\end{align}
and
\begin{equation}\label{dp-32}
\textmd{arg}(-\frac{u_{-}}{u_{+}})=8\sum_{n=1}^{N}\textmd{arg}z_{n},
\end{equation}
respectively.

Below we will plot the propagation behavior by selecting the appropriate parameters,  which further explain the properties of the solution \eqref{dp-30} and also analyze the influence of parameters on its propagation.

From the Figures 5, we can seen that when the parameter $c=0.1$, the solution appears as the interaction of breather-breather solutions with NZBCs. When the parameter $c$ is gradually increased, the propagation behavior of the solution becomes irregular.

\section{ Conclusions and discussions}
In this work, we mainly investigated the IST of the NLSLab equation with non-zero boundary values. In the case the boundary value is not zero, there are some technical difficulties, which may lead to difficulties in constructing the RH problem. However, for this problem, we solved it by using a Riemann surface. Next, we analyzed the asymptotic Lax pairs to get the Jost function, the scattering matrix and their analyticity and symmetries. This is due to the need to used them to construct the RHP during the inverse scatter transformation. Moreover, we also obtained the asymptotic analysis, residual conditions and theta conditions for discrete spectral points. In solving the inverse scattering process, we obtained some corresponding data on the basis of direct scattering to find the generalized RH problem. Through solving the RH problem, we obtained some soliton solutions of the NLSLab equation, such as stationary solitons, non-stationary solitons, multi-soliton solutions. Finally, we also discussed the double pole case. In this process, we can find the corresponding trace formula, and the theta condition changed accordingly. In addition,  we also selected the corresponding parameters to discuss different parameter solutions, and draw corresponding images to describe the influence of its parameters on the propagation behavior including shape, size and direction etc.
\\

\noindent
{\rotatebox{0}{\includegraphics[width=4.2cm,height=3.5cm,angle=0]{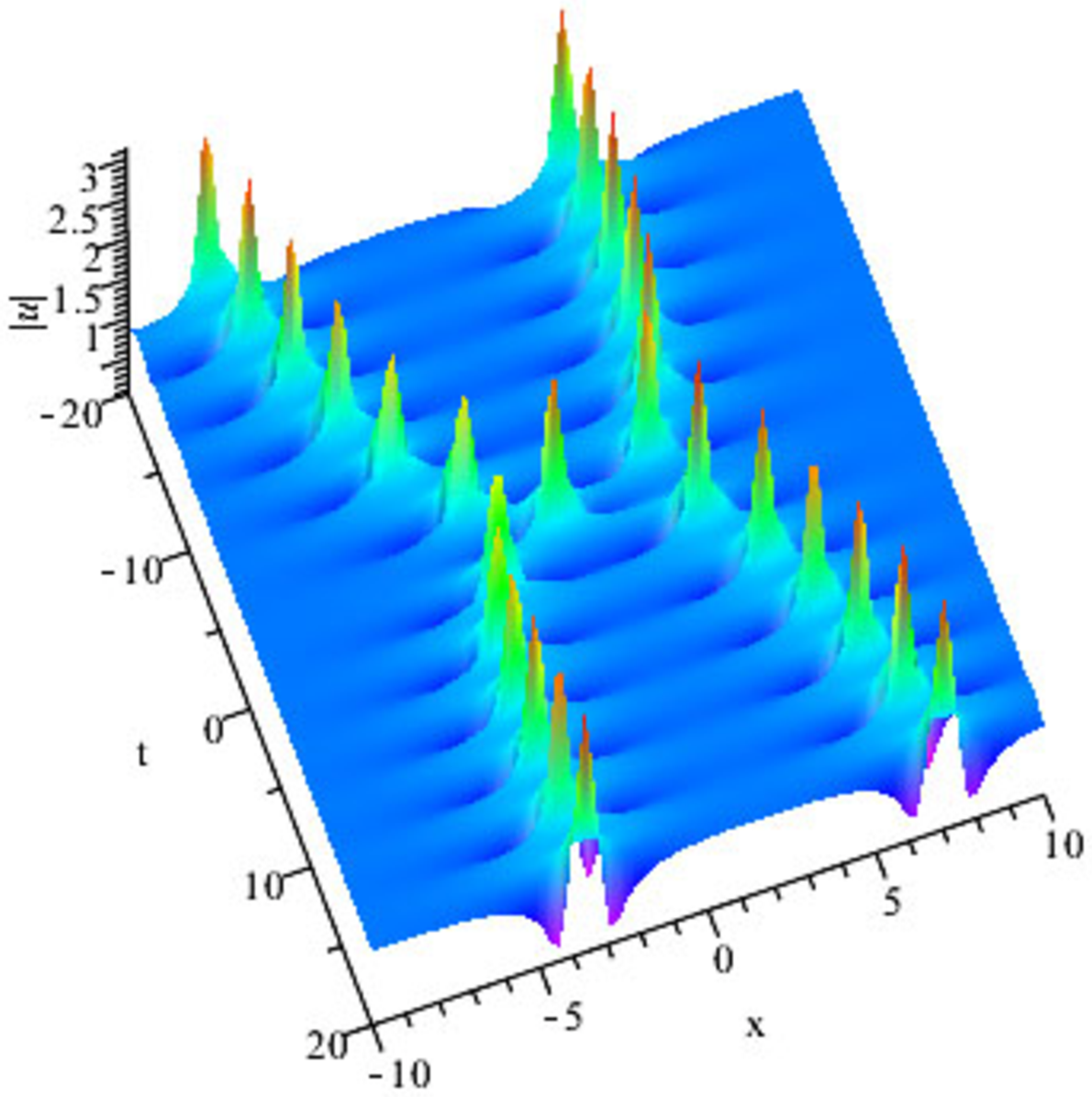}}}
~~~~
{\rotatebox{0}{\includegraphics[width=4.2cm,height=3.5cm,angle=0]{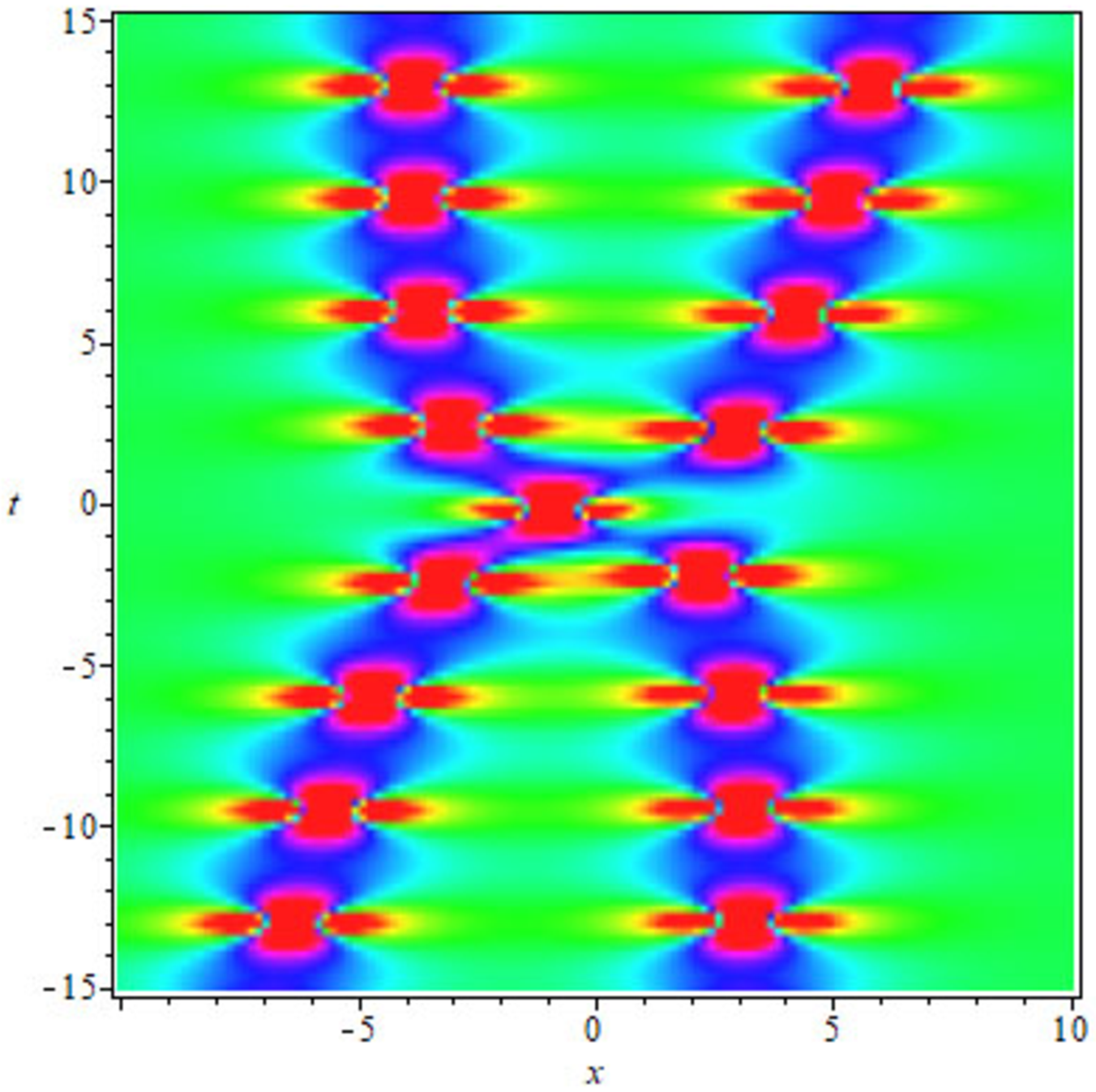}}}
~~~~
{\rotatebox{0}{\includegraphics[width=4.2cm,height=3.5cm,angle=0]{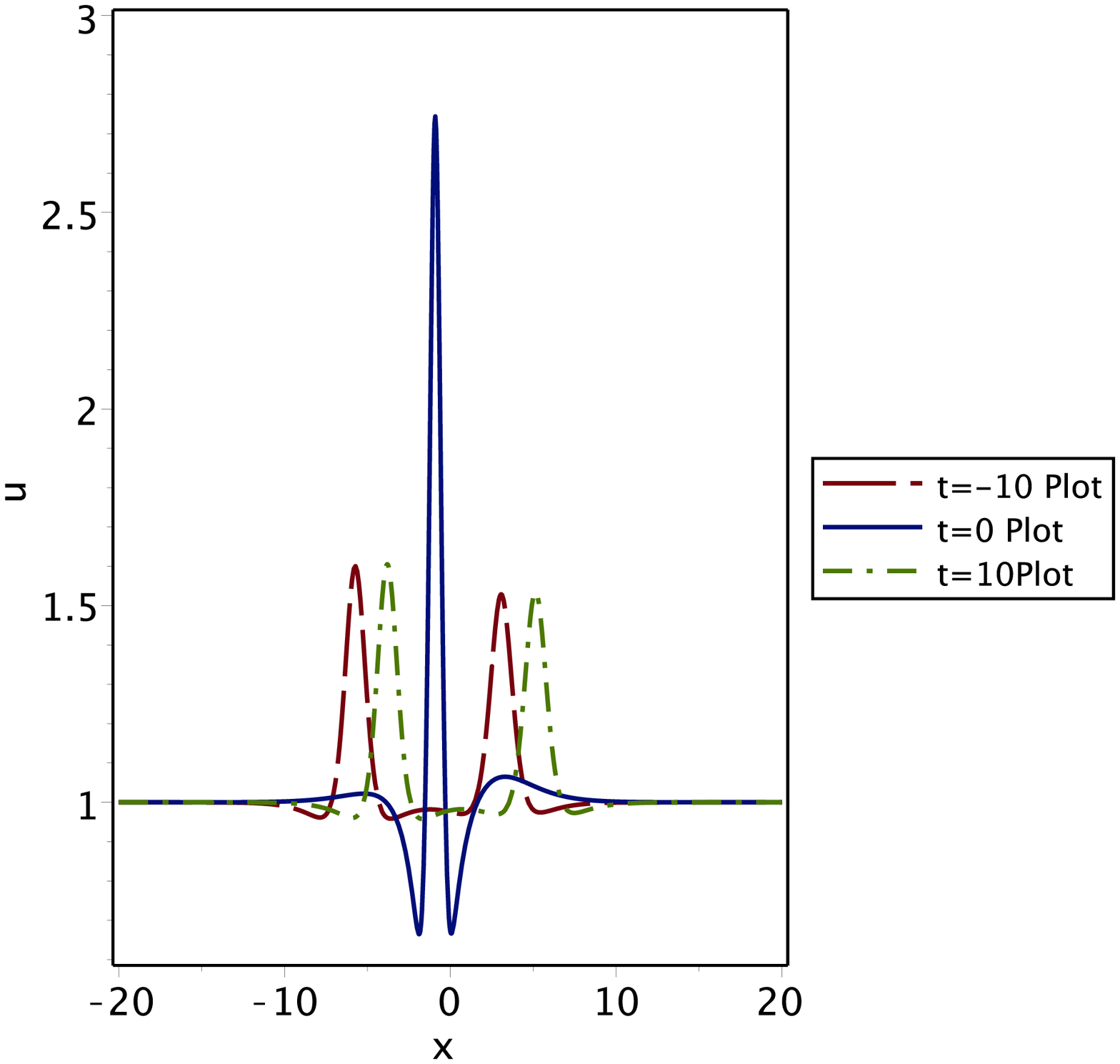}}}
$~~~~~~~~~~~~~~~(\textbf{a1})~~
~~~~~~~~~~~~~~~~~~~~~~~~~~~~~~~~~~~(\textbf{a2})
~~~~~~~~~~~~~~~~~~~~~~~~~~~~~~~~(\textbf{a3})$\\

\noindent
{\rotatebox{0}{\includegraphics[width=4.2cm,height=3.5cm,angle=0]{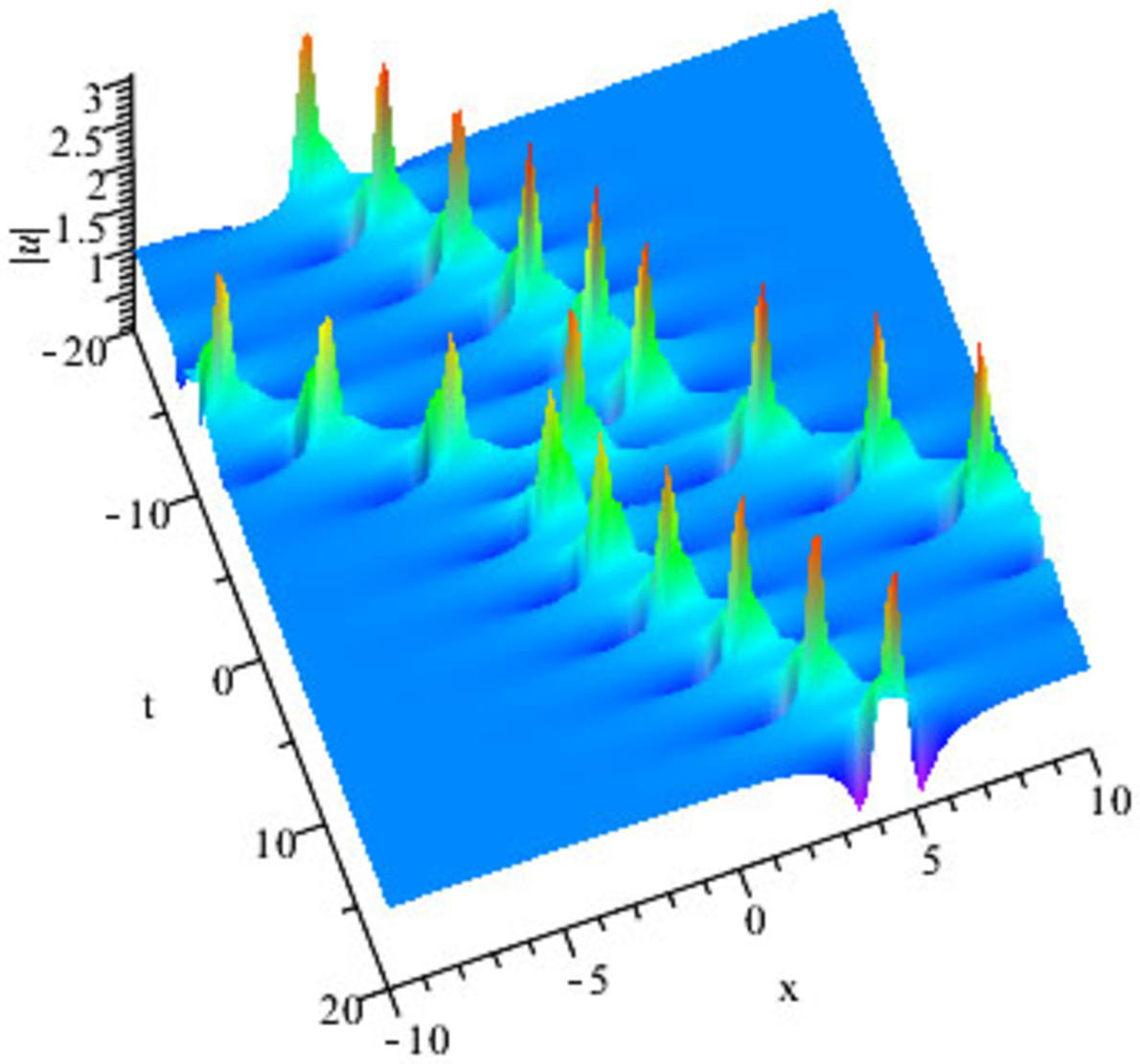}}}
~~~~
{\rotatebox{0}{\includegraphics[width=4.2cm,height=3.5cm,angle=0]{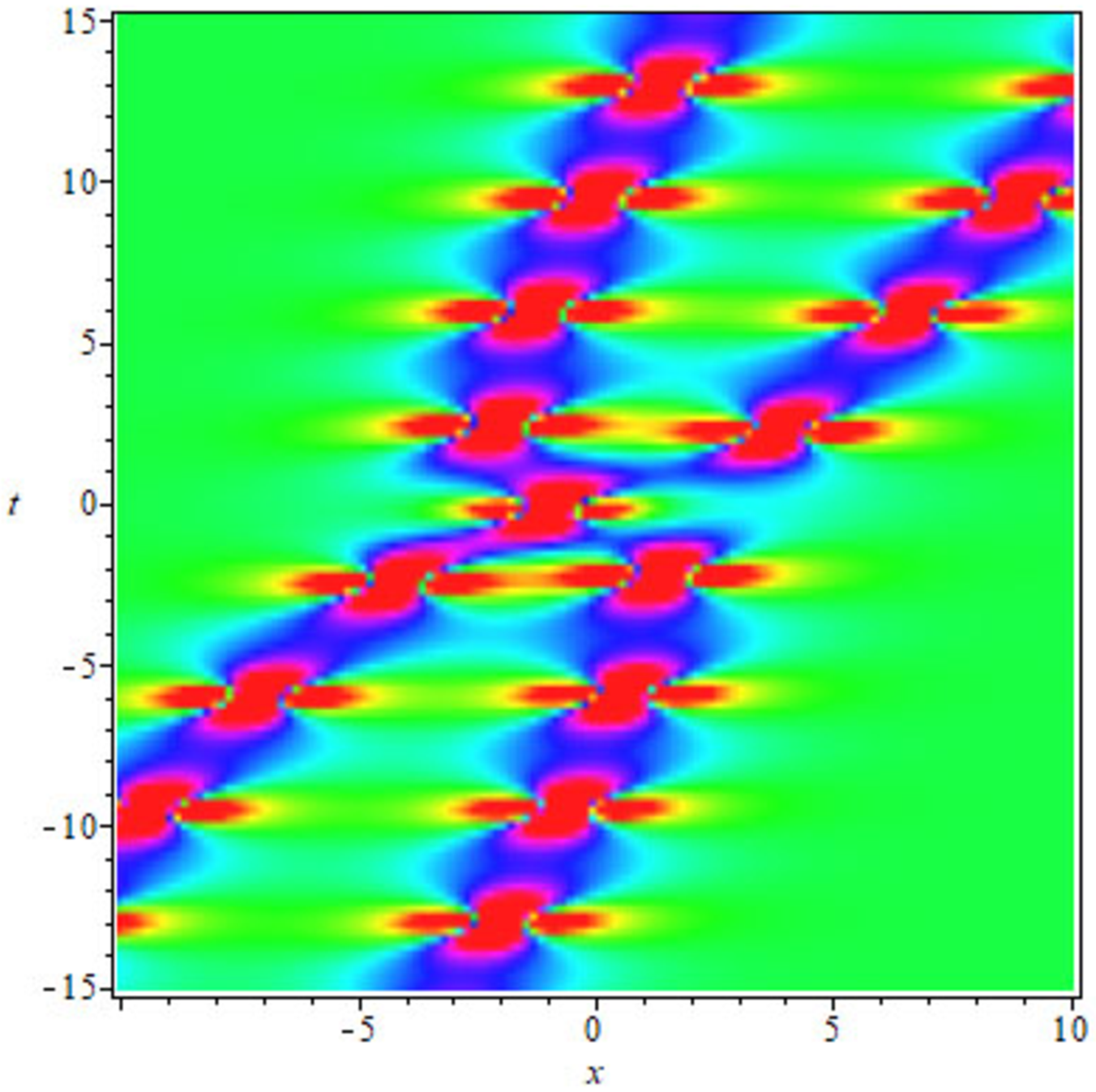}}}
~~~~
{\rotatebox{0}{\includegraphics[width=4.2cm,height=3.5cm,angle=0]{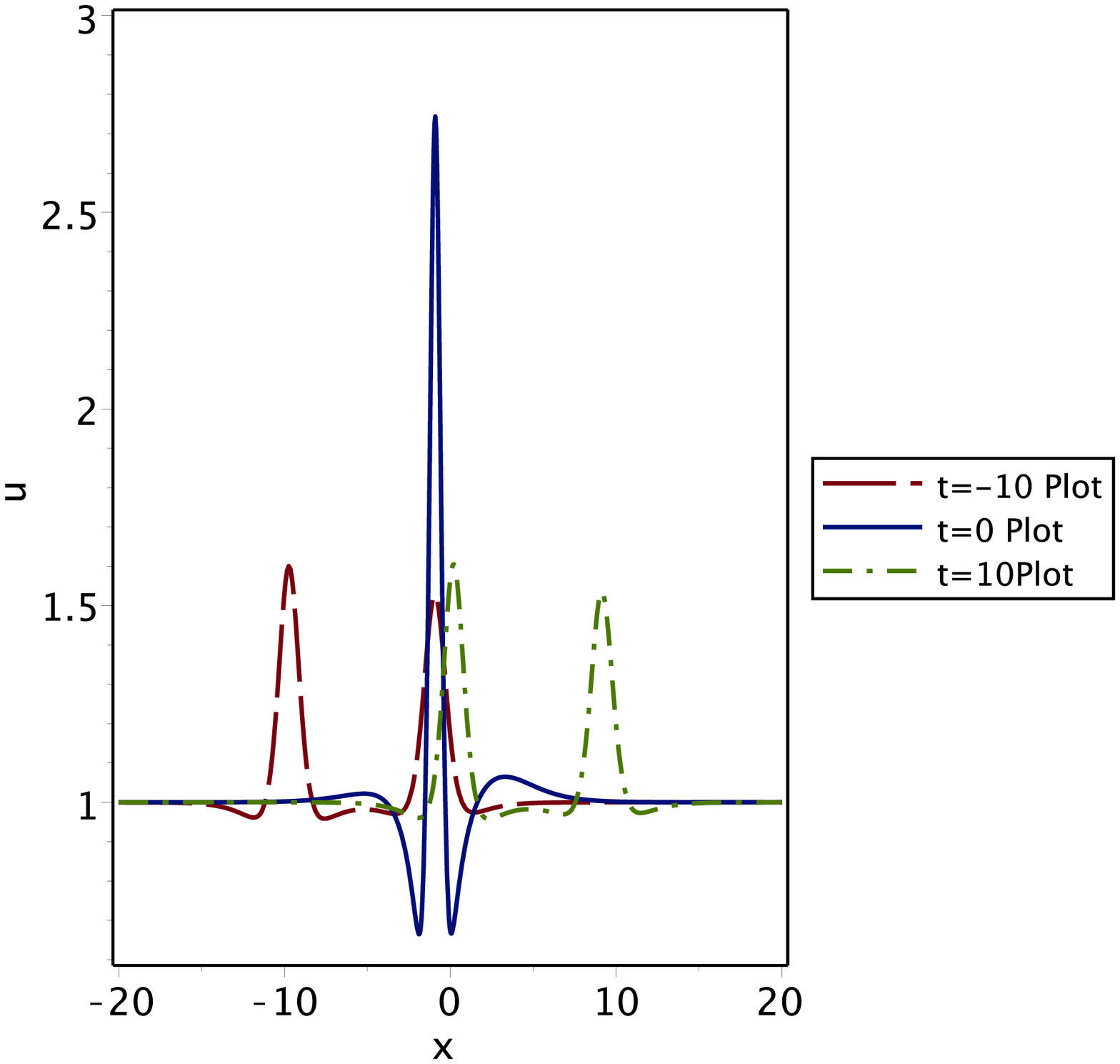}}}
$~~~~~~~~~~~~~~~(\textbf{b1})~~
~~~~~~~~~~~~~~~~~~~~~~~~~~~~~~~~~~~(\textbf{b2})
~~~~~~~~~~~~~~~~~~~~~~~~~~~~~~~~(\textbf{b3})$\\
\noindent
{\rotatebox{0}{\includegraphics[width=4.2cm,height=3.5cm,angle=0]{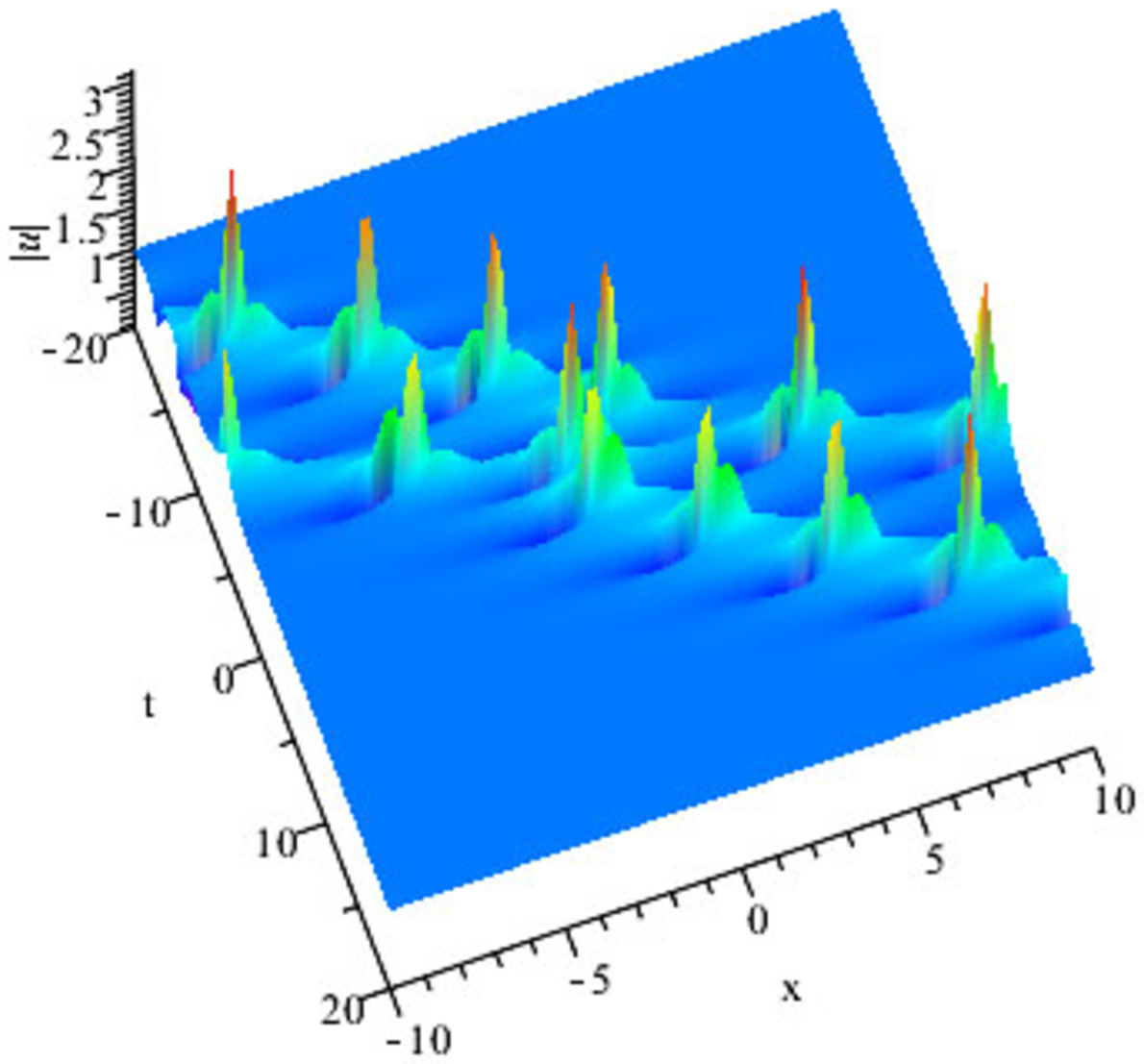}}}
~~~~
{\rotatebox{0}{\includegraphics[width=4.2cm,height=3.5cm,angle=0]{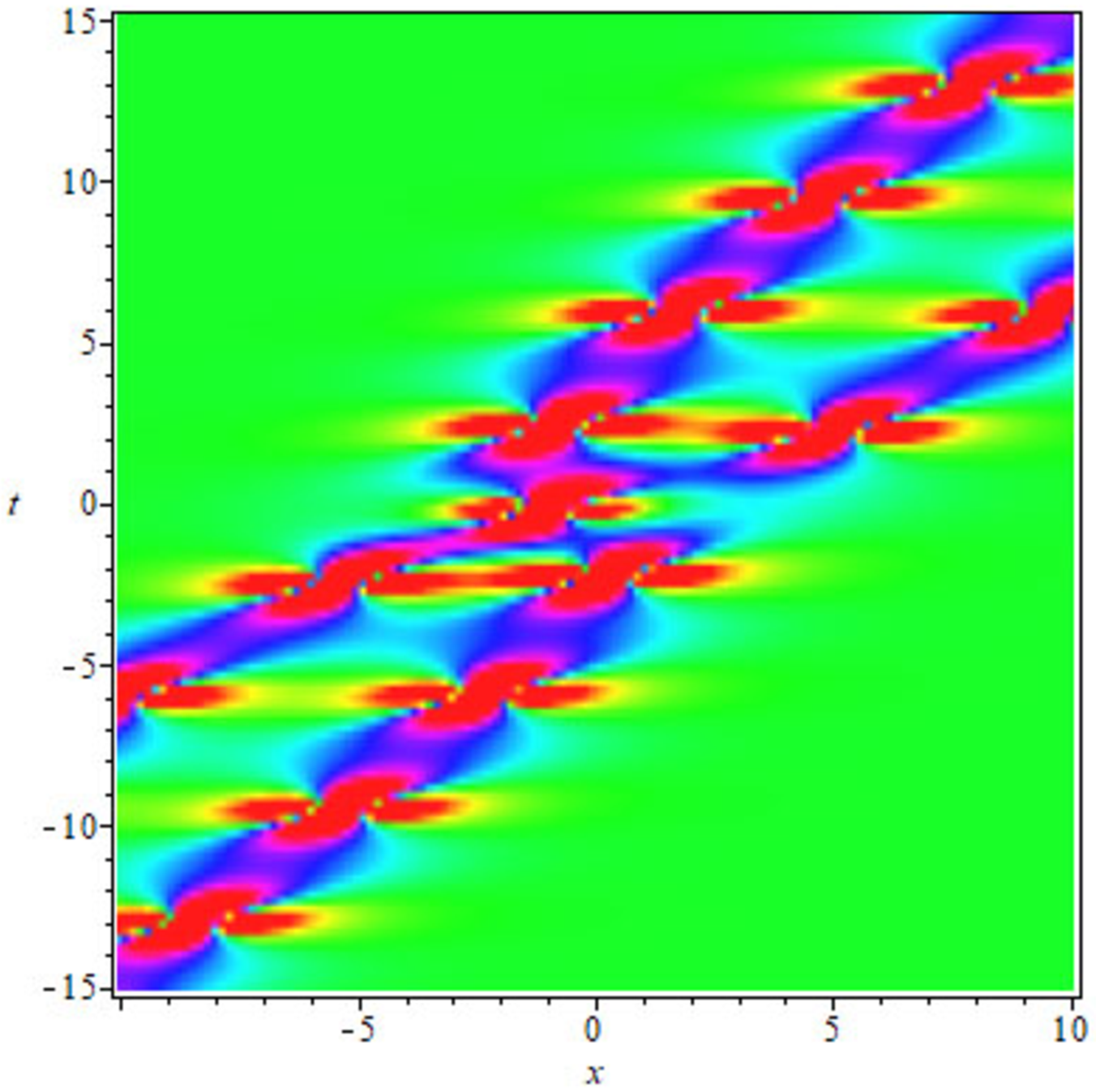}}}
~~~~
{\rotatebox{0}{\includegraphics[width=4.2cm,height=3.5cm,angle=0]{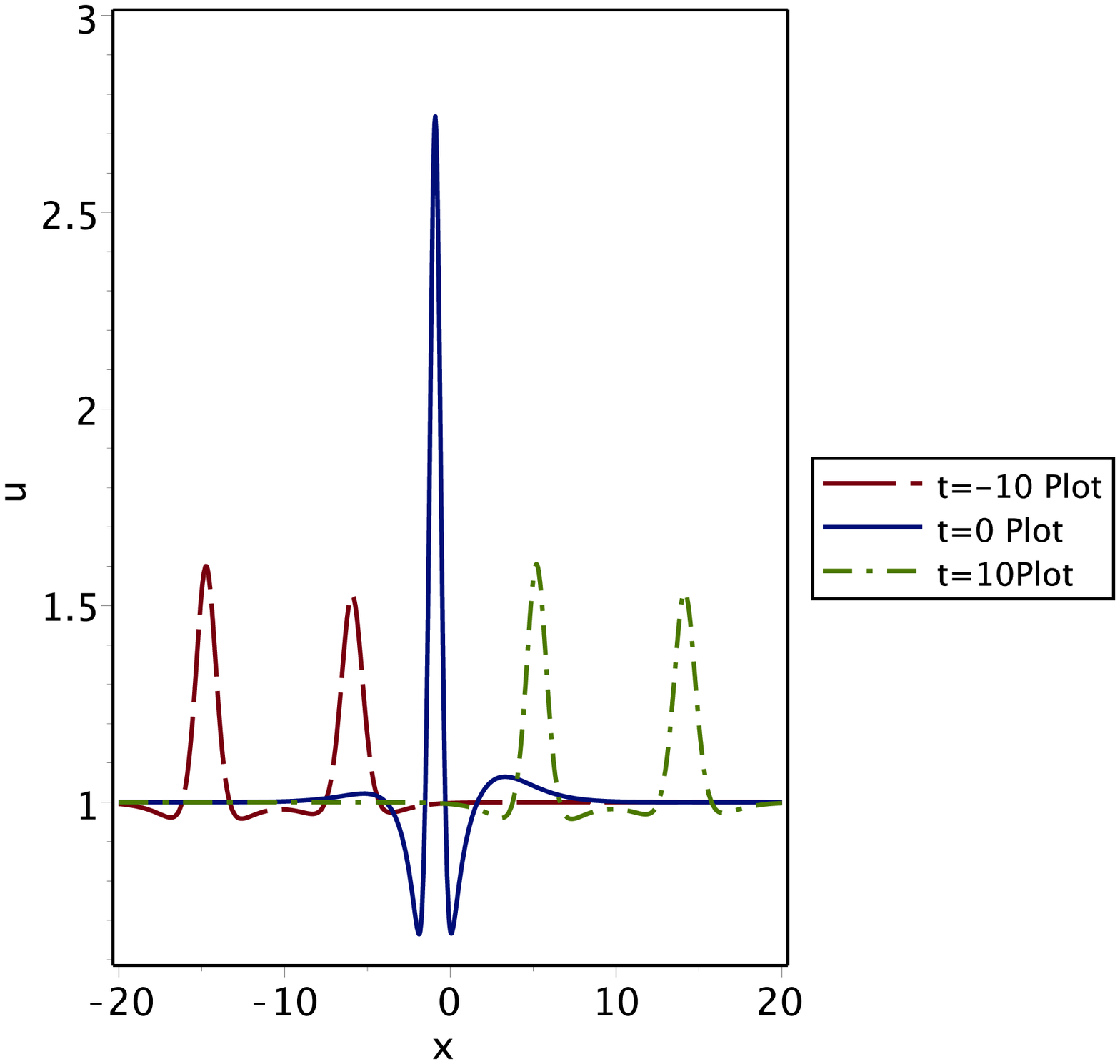}}}
$~~~~~~~~~~~~~~~(\textbf{c1})~~
~~~~~~~~~~~~~~~~~~~~~~~~~~~~~~~~~~~(\textbf{c2})
~~~~~~~~~~~~~~~~~~~~~~~~~~~~~~~~(\textbf{c3})$\\
\noindent { \small \textbf{Figure 5.} (Color online) Double-pole one-soliton solutions $u$ of the NLSLab equation \eqref{NLS-1}  by choosing suitable parameters:
 $N=1, z_{1}=\frac{3i}{2}, u_{0}=1, u_{+}=-1, K_{-}(z_{1})=J_{-}(z_{1})=1$.
($\textbf{(a1)}, \textbf{(a2)},\textbf{(a3)}$): the breather-breather solutions with NZBCs and parameter $c=0.1$;
($\textbf{(b1)}, \textbf{(b2)},\textbf{(b3)}$): the breather-breather solutions with NZBCs and parameter $c=0.5$; ($\textbf{(c1)}, \textbf{(c2)},\textbf{(c3)}$): the breather-breather solutions with NZBCs and parameter $c=1$.}

\section*{Acknowledgements}

This work was supported by the Postgraduate Research and Practice of Educational Reform for Graduate students in CUMT under Grant No. 2019YJSJG046, the Natural Science Foundation of Jiangsu Province under Grant No. BK20181351, the Six Talent Peaks Project in Jiangsu Province under Grant No. JY-059, the Qinglan Project of Jiangsu Province of China, the National Natural Science Foundation of China under Grant No. 11975306, the Fundamental Research Fund for the Central Universities under the Grant Nos. 2019ZDPY07 and 2019QNA35, and the General Financial Grant from the China Postdoctoral Science Foundation under Grant Nos. 2015M570498 and 2017T100413.





\end{document}